\documentclass{article}

\usepackage[utf8]{inputenc}
\usepackage{jheppubedit} 

\usepackage{bbold,graphicx,hyperref}
\usepackage{amsfonts,amsmath,amssymb,amsthm,mathtools}                   
\usepackage{color,xcolor}           
\usepackage{float}         
\usepackage{tikz}        
\usepackage{multirow}        
\usepackage{wrapfig}
\usepackage{braket}        
\usepackage{caption}
\usetikzlibrary{shadings,intersections}
\usetikzlibrary{calc,fadings,decorations.pathreplacing}

\newcommand \mathtikz[1] {\quad \vcenter{\hbox{\tikz{#1}}} \quad}

\newcommand\idA[2] { 
\begin{scope}[xshift=#1,yshift=#2]
\filldraw[fill=white,draw=black] (-0.25,0) rectangle (0.25,-1);
\end{scope}
}

\newcommand\muC[2]{ 
\begin{scope}[xshift=#1,yshift=#2]
\filldraw[left color=lightgray, right color=white] (-0.25,0) to [out=-90,in=180] (0,-0.33) to [in=-90,out=0] (0.25,0) to  (0.75,0) to [in=90,out=-90] (0.25,-1) to [out=-90,in=-90] (-0.25,-1) to [in=-90,out=90] (-0.75,0);
\filldraw[left color=white,right color=lightgray] (-0.5,0) ellipse (0.25 and 0.1);
\filldraw[left color=white,right color=lightgray] (0.5,0) ellipse (0.25 and 0.1);
\draw[dotted] (0.25,-1) arc (0:180:0.25 and 0.1);
\end{scope}
}

\newcommand\muCijk[2]{ 
\begin{scope}[xshift=#1,yshift=#2]
\filldraw[left color=lightgray, right color=white] (-0.25,0) to [out=-90,in=180] (0,-0.33) to [in=-90,out=0] (0.25,0) to  (0.75,0) to [in=90,out=-90] (0.25,-1) to [out=-90,in=-90] (-0.25,-1) to [in=-90,out=90] (-0.75,0);
\filldraw[left color=white,right color=lightgray] (-0.5,0) ellipse (0.25 and 0.1);
\filldraw[left color=white,right color=lightgray] (0.5,0) ellipse (0.25 and 0.1);
\draw[dotted] (0.25,-1) arc (0:180:0.25 and 0.1);
\node at (0,-1.35) {\small $k$};
\node at (-0.5,0.35) {\small $i$};
\node at (0.5,0.35) {\small $j$};
\end{scope}
}

\newcommand{\T}{\scriptstyle \textit{T}}

\newcommand\deltaC[2]{
\begin{scope}[xshift=#1,yshift=#2]
\filldraw[left color=lightgray, right color=white] (-0.25,-1) to [out=90,in=180] (0,-0.66) to [in=90,out=0] (0.25,-1) to [out=-90,in=-90] (0.75,-1) to [in=-90,out=90] (0.25,0) to (-0.25,0) to [in=90,out=-90] (-0.75,-1) to [out=-90,in=-90] (-0.25,-1);
\filldraw[left color=white,right color=lightgray] (0,0) ellipse (0.25 and 0.1);
\draw[dotted] (-0.25,-1) arc (0:180:0.25 and 0.1);
\draw[dotted] (0.75,-1) arc (0:180:0.25 and 0.1);
\end{scope}
}

\newcommand\Int[2]{ 
\begin{scope}[xshift=#1,yshift=#2]
\draw (-0.25,.25) -- (0.25,.25);
\end{scope}
}
\newcommand\muA[2]{ 
\begin{scope}[xshift=#1,yshift=#2]
\draw (-0.75,0) -- (-0.25,0) to [out=-90,in=180] (0,-0.33) to [in=-90,out=0] (0.25,0) -- (0.75,0) to [in=90,out=-90] (0.25,-1);
\draw (-0.25,-1) -- (0.25,-1);
\draw (-0.75,0) to [in=90,out=-90] (-0.25,-1);
\end{scope}
}

\newcommand\pairA[2]{ 
\begin{scope}[xshift=#1,yshift=#2]
\draw (-0.75,0) -- (-0.25,0) to [out=-90,in=180] (0,-0.33) to [in=-90,out=0] (0.25,0) -- (0.75,0) to [out=-90,in=0] (0,-0.83) to [out=180,in=-90] (-0.75,0);
\end{scope}
}

\newcommand\pairAv[2]{ 
\begin{scope}[xshift=#1,yshift=#2]
\draw (-0.75,0) -- (-0.25,0) to [out=-90,in=180] (0,-0.33) to [in=-90,out=0] (0.25,0) -- (0.75,0) to [out=-90,in=0] (0,-0.83) to [out=180,in=-90] (-0.75,0);
\node at (0,0) {\small $e$};
\node at (0,1) {\small $e$};
\node at (0.5,-0.25) {\small $V$};
\end{scope}
}

\newcommand\copairA[2]{ 
\begin{scope}[xshift=#1,yshift=#2]
\draw (-0.75,0) -- (-0.25,0) to [out=90,in=180] (0,0.33) to [in=90,out=0] (0.25,0) -- (0.75,0) to [out=90,in=0] (0,0.83) to [out=180,in=90] (-0.75,0);
\end{scope}
}

\newcommand\pairAred[2]{ 
\begin{scope}[xshift=#1,yshift=#2]
\draw (-0.75,0) -- (-0.25,0) to [out=-90,in=180] (0,-0.33) to [in=-90,out=0] (0.25,0) -- (0.75,0);
\draw [red,thick](0.75,0) to [out=-90,in=0] (0,-0.83) to [out=180,in=-90] (-0.75,0);
\node at (0,1.1) {\small $\beta$};
\end{scope}
}
\newcommand\copairAred[2]{ 
\begin{scope}[xshift=#1,yshift=#2]
\draw (-0.75,0) -- (-0.25,0) to [out=90,in=180] (0,0.33) to [in=90,out=0] (0.25,0) -- (0.75,0);
\draw [red,thick](0.75,0) to [out=90,in=0] (0,0.83) to [out=180,in=90] (-0.75,0);
\end{scope}
}

\newcommand\deltaA[2]{ 
\begin{scope}[xshift=#1,yshift=#2]
\draw (-0.75,-1) -- (-0.25,-1) to [out=90,in=180] (0,-0.66) to [in=90,out=0] (0.25,-1) -- (0.75,-1) to [in=-90,out=90] (0.25,0) -- (-0.25,0) to [in=90,out=-90] (-0.75,-1);
\end{scope}
}

\newcommand\zipper[2]{ 
\begin{scope}[xshift=#1,yshift=#2]
\draw (-0.25,-1) -- (0.25,-1);
\filldraw[right color=white,left color=lightgray] (-0.25,0) to (-0.25,-1) to [out=90,in=225] (0,-0.5) to [out=-45,in=90] (0.25,-1) to (0.25,0);
\filldraw[left color=white,right color=lightgray] (0,0) ellipse (0.25 and 0.1);
\end{scope}
}

\newcommand\mzipper[2]{ 
\begin{scope}[xshift=#1,yshift=#2]
\draw (-0.25,-1) -- (0.25,-1);
\filldraw[right color=white,left color=lightgray] (-0.25,0) to (-0.25,-1) to [out=90,in=225] (0,-0.5) to [out=-45,in=90] (0.25,-1) to (0.25,0);
\filldraw[right color=white,left color=lightgray] (0,0) ellipse (0.25 and 0.1);
\end{scope}
}

\newcommand\cozipper[2]{ 
\begin{scope}[xshift=#1,yshift=#2]
\draw (-0.25,0) -- (0.25,-0);
\filldraw[right color=white,left color=lightgray] (-0.25,-1) to (-0.25,0) to [out=-90,in=135] (0,-0.5) to [out=45,in=-90] (0.25,0) to (0.25,-1) to [in=-90,out=-90] (-0.25,-1);
\draw[dotted] (0.25,-1) arc (0:180:0.25 and 0.1);
\end{scope}
}

\newcommand\epsilonC[2]{ 
\begin{scope}[xshift=#1,yshift=#2]
\filldraw[right color=white,left color=lightgray] (-0.25,0) to [out=-90,in=180] (0,-0.33) to [in=-90,out=0] (0.25,0);
\filldraw[left color=white,right color=lightgray] (0,0) ellipse (0.25 and 0.1);
\end{scope}
}

\newcommand\etaC[2] { 
\begin{scope}[xshift=#1,yshift=#2]
\filldraw[right color=white,left color=lightgray] (-0.25,0) to [out=90,in=180] (0,0.33) to [in=90,out=0] (0.25,0) to [in=-90,out=-90] (-0.25,0);
\draw[dotted] (0.25,0) arc (0:180:0.25 and 0.1);
\end{scope}
}

\newcommand\etaCv[2] { 
\begin{scope}[xshift=#1,yshift=#2]
\filldraw[right color=white,left color=lightgray] (-0.7,0) to [out=90,in=180] (0,0.7) to [in=90,out=0] (0.7,0) to [in=-75,out=-105] (-0.7,0);
\draw[dotted] (0.7,0) arc (0:180:0.7 and 0.3);
\filldraw[color=black] (0,0.7) ellipse (0.05 and 0.05);
\filldraw[color=black] (0,-0.7) ellipse (0.05 and 0.05);
\draw[->, bend right=40] (-0.9,0.2) to (0.9,0.2);
\node at (0,0.9) {\small $e$};
\node at (0,-0.9) {\small $e$};
\end{scope}
}

\newcommand\epsilonCv[2]{ 
\begin{scope}[xshift=#1,yshift=#2]
\filldraw[right color=white,left color=lightgray] (-0.7,0) to [out=-90,in=180] (0,-0.7) to [in=-90,out=0] (0.7,0);
\filldraw[left color=white,right color=lightgray] (0,0) ellipse (0.7 and 0.1);
\end{scope}
}

\newcommand\epsilonA[2] {
\begin{scope}[xshift=#1,yshift=#2]
\draw (-0.25,0) -- (0.25,0);
\draw (-0.25,0) to [out=-90,in=180] (0,-0.33) to [in=-90,out=0] (0.25,0);
\end{scope}
}

\newcommand\etaA[2] {
\begin{scope}[xshift=#1,yshift=#2]
\draw (-0.25,0) -- (0.25,0);
\draw (-0.25,0) to [out=90,in=180] (0,0.33) to [in=90,out=0] (0.25,0);
\end{scope}
}

\newcommand\etaAred[2] {
\begin{scope}[xshift=#1,yshift=#2]
\draw (-0.5,0) -- (0.5,0);
\draw [red,thick](-0.5,0) to [out=90,in=180] (0,0.5) to [in=90,out=0] (0.5,0);
\node at (0,0.7) {\small $\beta$};
\end{scope}
}
\newcommand\epsilonAred[2] {
\begin{scope}[xshift=#1,yshift=#2]
\draw (-0.5,0) -- (0.5,0);
\draw [red,thick](-0.5,0) to [out=-90,in=180] (0,-0.5) to [in=-90,out=0] (0.5,0);
\end{scope}
}

\newcommand\SA[2] { 
\begin{scope}[xshift=#1,yshift=#2]
\draw (-0.25,0) to [out=-90,in=90] (0.25,-1);
\draw[line width=2mm, white] (0.25,0) to [out=-90,in=90] (-0.25,-1);
\draw (-0.25,0) -- (0.25,0) to [out=-90,in=90] (-0.25,-1) -- (0.25,-1);
\end{scope}
}

\newcommand{\pd}{\partial}
\newcommand{\bea}{\begin{eqnarray}}
\newcommand{\eea}{\end{eqnarray}}
\newcommand{\nn}{\nonumber\\}

\DeclareMathOperator{\tr}{Tr}
\renewcommand{\Re}{\operatorname{Re}}
\renewcommand{\Im}{\operatorname{Im}}

\newcommand{\mt}[1]{\textrm{\tiny #1}}

\newcommand{\R}{\mathbb{R}}
\newcommand{\Z}{\mathbb{Z}}
\newcommand{\SL}{\mathrm{SL}}
\newcommand{\PSL}{\mathrm{PSL}}

\newcommand{\U}{\mathrm{U}}

\title{\boldmath A proposal for 3d quantum gravity and its bulk factorization}

\author[a]{Thomas G. Mertens,}
\author[b]{Joan Sim\'on,}
\author[c,d]{Gabriel Wong}

\affiliation[a]{Department of Physics and Astronomy, \\
Ghent University, Krijgslaan, 281-S9, 9000 Gent, Belgium}
\affiliation[b]{School of Mathematics and Maxwell Institute for Mathematical Sciences,\\
University of Edinburgh, Edinburgh EH9 3FD, UK}
\affiliation[c]{Physics Department, Fudan University, Shanghai,China}
\affiliation[d]{Harvard Center of Mathematical Sciences and Applications, USA}

\emailAdd{Thomas.Mertens@ugent.be}
\emailAdd{j.simon@ed.ac.uk}
\emailAdd{gabrielwon@gmail.com}

\date{\today}

\abstract{
Recent progress in AdS/CFT has provided a good understanding of how the bulk spacetime is encoded in the entanglement structure of the boundary CFT. However, little is known about how spacetime emerges directly from the bulk quantum theory. We address this question in an effective 3d quantum theory of pure gravity, which describes the high temperature regime of a holographic CFT.  This theory can be viewed as a $q$-deformation and dimensional uplift of JT gravity.  Using this model, we show that the Bekenstein-Hawking entropy of a two-sided black hole equals the bulk entanglement entropy of gravitational edge modes. In the conventional Chern-Simons description, these black holes correspond to Wilson lines in representations of $\PSL(2,\mathbb{R})\otimes \PSL(2,\mathbb{R}) $.  We show that the correct calculation of gravitational entropy suggests we should interpret the bulk theory as an extended topological quantum field theory associated to the quantum semi-group $\SL^+_{q}(2,\mathbb{R})\otimes \SL^+_{q}(2,\mathbb{R})$. Our calculation suggests an effective description of bulk microstates in terms of collective, anyonic degrees of freedom whose entanglement leads to the emergence of the bulk spacetime. 
}

\begin{document}

\maketitle

\section{Introduction}

In recent years, various lines of investigation have led to the suggestion that black hole entropy arises from the entanglement of spacetime \cite{Srednicki1993,Susskind:1994sm,Maldacena:2001kr,VanRaamsdonk:2010pw,Bianchi:2012ev}. The paradigmatic example supporting this idea comes from the AdS/CFT correspondence \cite{Maldacena:1997re,Aharony:1999ti}, which implies that the two-sided AdS Schwarzschild black hole is dual to an entangled thermofield double (TFD) state of two CFTs living at the disconnected asymptotic boundaries. In this scenario, the Bekenstein-Hawking entropy of the two-sided black hole, given by one quarter of the area of the bifurcation surface, is identified with the entanglement entropy of the TFD state \cite{Maldacena:2001kr}. Moreover, the Ryu-Takayanagi formula generalizes this identification to boundary-anchored extremal surfaces inside a general asymptotically AdS spacetime.  In this case, the area of the extremal surface is dual to the entanglement entropy of spatial subregions of the boundary theory \cite{Ryu:2006bv}. This relation between bulk areas and boundary entanglement entropy is one of the key evidences underpinning the idea that entanglement is responsible for the emergence of a smooth, connected bulk spacetime.
    
We would like to understand bulk emergence directly from the entanglement of the bulk gravitational theory.  For example, in the case of the two-sided black hole, we expect that a single connected bulk geometry can be equivalently described by an entangled sum of one-sided bulk geometries:\footnote{In general not every state in a holographic CFT has a semi-classical dual. But if the bulk theory exists independently, we can still specify a dual state in the bulk Hilbert space.}  this is a bulk manifestation of the ER=EPR paradigm \cite{Maldacena:2013xja}. However, this appealing picture is difficult to verify because it requires a formulation of the bulk gravity Hilbert space as well as its factorization into subregion Hilbert spaces. The former would seem to require solving the bulk string theory, while the latter is difficult to define, even in principle, because the classical notion of a spacetime subregion does not exist in quantum gravity.  We must therefore search for a more fundamental quantum mechanical concept replacing the notion of a subregion.
    
A useful starting point for a bottom-up approach to this problem is provided by three-dimensional pure gravity with a negative cosmological constant.   Here there are black hole solutions even though the bulk degrees of freedom are topological \cite{Banados:1992wn}. In particular, the classical bulk theory can be described by PSL$(2,\mathbb{R})\times $ PSL$(2,\mathbb{R})$ Chern-Simons theory \cite{Witten:1988hc,Achucarro:1986uwr}. For compact gauge groups, the factorization of Chern-Simons theory into subregion Hilbert spaces is well-understood, leading to a well-defined notion of entanglement entropy \cite{Wong:2017pdm}. Therefore we might hope that the Chern-Simons formulation of 3d gravity in the presence of a two-sided black hole will allow a Hilbert space factorization that leads to an entanglement entropy consistent with the Bekenstein-Hawking entropy.\footnote{A different approach was considered in \cite{Benini:2022hzx} by gauging a global 1-form bulk symmetry in a non-abelian Chern-Simons theory, leading to a modular invariant boundary CFT with a factorized partition function on wormhole geometries. We however will not discuss higher topologies in this work.}
    
\begin{figure}[h]
    \centering
    \includegraphics[scale=.35]{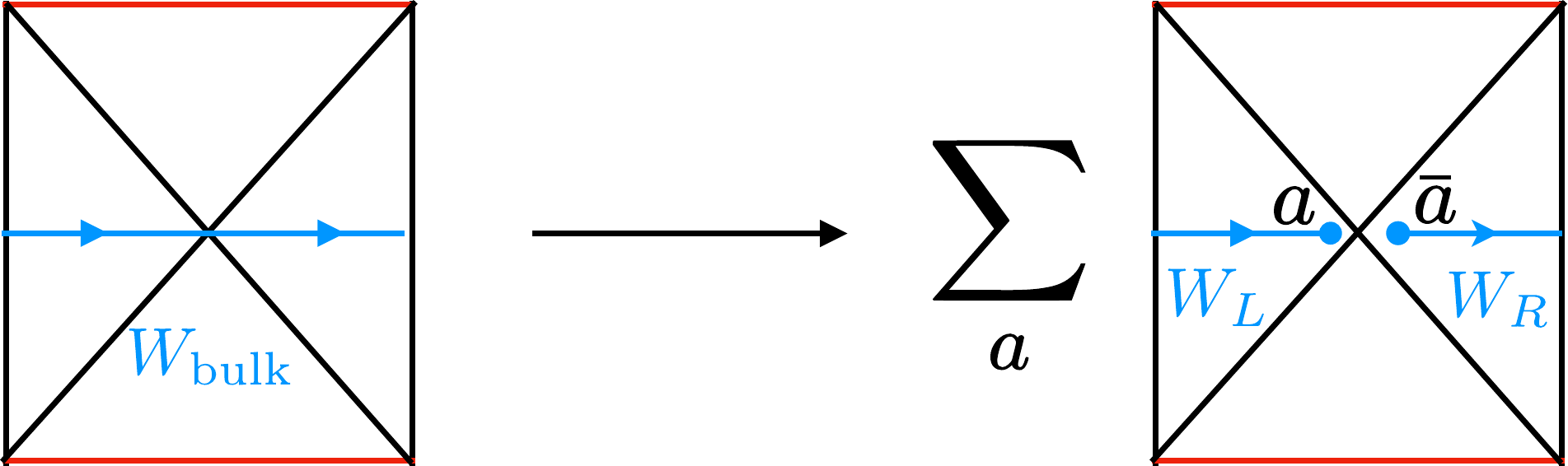}
    \qquad \qquad 
     \includegraphics[scale=.35]{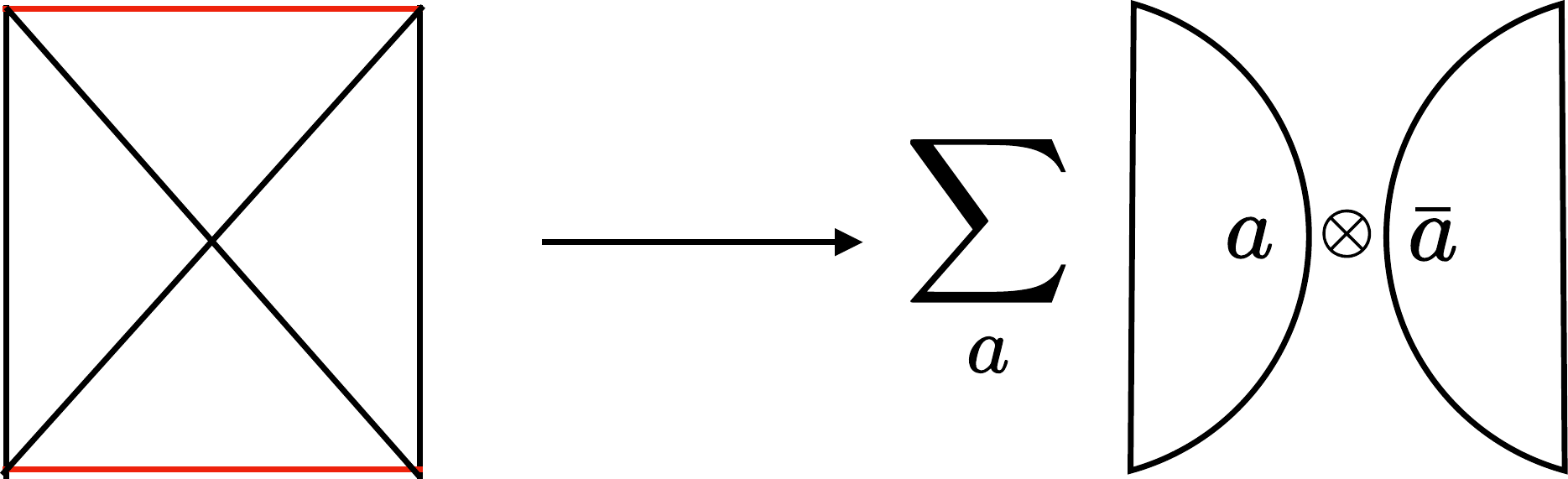}
    \caption{Left: splitting a gauge theory Wilson line $W_{\text{bulk}}$ in left- and right Wilson lines $W_L$ and $W_R$. Right: splitting a gravitational Wilson line disconnects spacetime.}
    \label{U1split}
\end{figure}

To motivate the concepts advocated in this work, consider the bulk Hilbert space $\mathcal{H}_{\text{bulk}}^{\text{gauge}}$ for a pure gauge theory coupled to the fixed background of a two-sided black hole.  As shown in the left of Figure \ref{U1split},  the presence of a Wilson line operator $W_{\text{bulk}}$ crossing the Einstein-Rosen (ER) bridge obstructs a naive factorization of $\mathcal{H}_{\text{bulk}}^{\text{gauge}}$ into two factors corresponding to the left and right wedge.
To split the Wilson line, we must introduce \emph{edge modes}, i.e. charged objects on which Wilson lines in any representation can end in a gauge-invariant way.   As discussed in \cite{Harlow:2015lma}, these charged objects must exist in the microscopic bulk theory, given that the dual CFT does factorize into a tensor product of left and right factors.  However, in the low-energy effective field theory these charges are \emph{confined}: they are entangled to form a singlet under an \emph{edge mode symmetry} $\text{G}_\mt{{S}}$. Explicitly, this singlet condition refers to the factorization
\begin{align}\label{Wsplit}
W_{\text{bulk}} &\to W_{L}W_{R},\nn
    (W_{\text{bulk}})_{bc} &\to \sum_{a} (W_{L})_{b a}(W_{R})_{\bar{a}c},
\end{align}
of the bulk Wilson line, in which the edge mode indices $a$ are always contracted.  Formally,  the factorization map \eqref{Wsplit} defines a co-product on the Hopf algebra of functions on the gauge group.  From the point of view of the low-energy effective field theory, it defines a map into an \emph{extended} Hilbert space:  see e.g.  \cite{Buividovich:2008gq,Casini:2013rba,Donnelly:2014gva,Lin:2018bud,Ghosh:2015iwa,VanAcoleyen:2015ccp} for a sampling of work describing this construction in various contexts.

We would like to generalize this gauge theory set-up to factorize the bulk quantum gravity Hilbert space. This is  heuristically depicted on the right of Figure \ref{U1split} as an entangled sum of one-sided geometries. The essential conceptual shift required in quantum gravity is that rather than factorizing a Wilson line inserted on top of a fixed background, we want to split the Wilson line that makes up the spacetime itself.  In this context, $W_{\text{bulk}}$ represents the wavefunction of a quantum state defining a two-sided bulk geometry. 

Our technical approach will be to take inspiration from 2d JT gravity as a BF gauge theory, and then generalize this to 3d pure gravity and its analogous 3d Chern-Simsons gauge theoretic formulation. This will allow us to interpret black hole entropy as entanglement entropy.

\paragraph{Summary and outline.} 
We begin in \textbf{Section \ref{sec:3dgrav_def}} by defining a model of 3d gravity in which the question of bulk factorization can be addressed. There is a long history and an ongoing debate about how to properly define 3d quantum gravity \cite{Witten:2007kt,Maloney:2007ud,Cotler:2018zff,Maxfield:2020ale,Chandra:2022bqq}. Drawing inspiration from similar approaches to 2d JT gravity, we will define an effective quantum theory of 3d gravity that describes a universal sector relevant for 3d black holes at sufficiently high temperature. From the boundary point of view, this is a theory of ``vacuum Virasoro blocks in the dual channel'':
\begin{equation}
Z(\beta,\mu) = \left|\chi_0(-1/\tau)\right|^2 = \int_0^{+\infty} \hspace{-0.2cm} dp_+ dp_- \, S_{0}{}^{p_+}S_{0}{}^{p_-} \chi_{p_+}(\tau) \chi_{p_-}(\bar{\tau})\,.
\end{equation}
In this sense it is similar in spirit to \cite{Cotler:2018zff}, with a modified boundary condition at infinity that fixes the periodicity along the temporal cycle. The bulk theory involves a sum over (off-shell) black holes of different mass and spin, but does not include a sum over their modular images since we have a distinguished temporal cycle. As a result this is a non-modular invariant theory, dual to a tensor product of chiral and anti-chiral CFTs living at the asymptotic boundary. This theory can be independently derived from gravitational path integral considerations. 
Our model reproduces existing results for JT gravity, but extends it away from (near-)extremality. In fact, we identify a double scaling limit
\begin{equation}
c \gg 1, \qquad \beta/\ell \sim c\,,
\end{equation}
in which the contribution from the descendants is subdominant and the above partition function can be approximated by a doubled JT model
\begin{equation}
   Z(\beta,\mu)  \overset{\beta \gg \ell}{\sim} 
   Z_{\text{JT}}\Big(\frac{b^2\beta}{\ell}(1 + i\mu)\Big)Z_{\text{JT}}\Big(\frac{b^2\beta}{\ell}(1 - i\mu)\Big) .
\end{equation}
Our model can also be extended in the presence of boundary operator insertions, as discussed in Appendix \ref{app:bcor}. \\

In \textbf{Section \ref{sec:review}}, we set the stage for the Hilbert space factorization in our introduced 3d gravity model, by reviewing technical aspects of factorization in gauge theory. In more detail, subsection \ref{sec:algebra}
focuses on the algebraic aspects of the extended Hilbert space formalism, using 2d YM and BF theories as examples, whereas
subsection \ref{sec:path-integral} defines the shrinkable boundary condition that is imposed at the entangling surface. \\

\textbf{Section \ref{sec:facgrav}} then applies this gauge theoretic set of ideas and tools to 2d and 3d gravity. 
2d JT gravity can be formulated as a 2d BF gauge theory. Borrowing factorization methods developed for the latter, we show that factorization in gravity can be implemented by introducing edge states at the entangling surface, tagged by $\SL^+(2,\mathbb{R})$ representation labels for 2d JT gravity. For 3d gravity, we find that consistency with the shrinkable boundary condition leads to a bulk surface (or edge mode) symmetry group given by SL$^+_{q}(2,\mathbb{R})\, \times\, $SL$^+_{q}(2,\mathbb{R})$. The deformation parameter $q$ is related to the cosmological constant as $q=e^{\pi i b^2}$ through the central charge of the dual CFT $c=1 + 6(b+b^{-1})^2$, which equals $\frac{3\ell}{2G_{\text{N}}}$ in the semi-classical regime. As depicted in the right of Figure \ref{U1split}, factorizing this gravitational Wilson line hence allows for a local splitting of the physical spacetime. If we go in the opposite direction, we are probing the emergence of connected spacetime \cite{VanRaamsdonk:2010pw,Maldacena:2013xja} by gluing together the edge states on both sides of the entangling surface. Notice that whereas this picture is readily understood qualitatively from the boundary dual perspective (in e.g. the thermofield double state \cite{VanRaamsdonk:2010pw}) where factorization is manifest, it is highly non-trivial to describe this in technical detail in the bulk gravitational language. The charged objects on the entangling surface in 3d gravity are anyons belonging to the representation category  Rep($\SL^+_{q}(2,\mathbb{R}) \otimes \SL^+_{q}(2,\mathbb{R}))$, and their quantum dimensions give rise to the Bekenstein-Hawking entropy of the two-sided black hole.\footnote{The same claim was made in  \cite{McGough:2013gka}, but there is a problem with the naive application of their arguments. We explain the relevant issues in section \ref{CSdefect}.} \\

Despite the role the Chern-Simons description of 3d gravity plays for us, both technically and inspirationally, we stress the important differences between the gauge theory (CS) and gravity formulations in \textbf{Section \ref{CSdefect}}. In particular, in subsection \ref{s:anyon} we embed our gravitational entanglement entropy formula into a broader discussion of entanglement entropy in QFT in curved backgrounds. In subsection \ref{s:linefactori}, we borrow results from \cite{Guadagnini:1989tj} to give an explicit factorization of Wilson line operators as in \eqref{Wsplit}, with the edge modes transforming under a hidden quantum group symmetry. Finally, in subsection \ref{s:appgaugegra} we show that the gravity path integral excludes conical defect geometries that would ordinarily be included in gauge theory. This ensures the gauge and gravity path integral measures are different and explains why the implementation of the shrinkable boundary condition in gravity forbids the existence of descendants on the entangling surface. The latter is the reason why the bulk entanglement entropy computed in section \ref{sec:facgrav} is finite, matching the Bekenstein-Hawking answer in the semi-classical regime. \\

\textbf{Section \ref{s:exTQFT}} presents a conceptual abstract framework that is capable of jointly describing the earlier gauge and gravity discussions. Subsection \ref{sec:eTQFT} motivates a more abstract perspective on these issues by introducing the main concepts behind the axiomatic formulation of extended topological quantum field theories (TQFTs), see e.g. \cite{Lurie:2009keu}, and linking them to the shrinkable boundary condition in subsection \ref{sec:sh-eTQFT}. 2d gauge theories and 3d pure Chern-Simons (CS) with compact gauge group are revisited in appendix \ref{sec:ex-eTQFT} to illustrate this more abstract framework. The application to gravity is described in subsections \ref{sec:JTtqft} and \ref{sec:3dext}. We show our earlier Euclidean path integral calculations are compatible with the categorical formulation of an extended TQFT where different mathematical objects are assigned, through the computation of the euclidean path integral, to different surfaces of each codimension. In particular, the shrinkable boundary condition at the entangling surface appears as an additional TQFT sewing relation, as shown in \cite{Donnelly:2018ppr}. Given the relevance of the latter to determine the correct factorization map, edge mode symmetry for 3d gravity and the correct entanglement entropy, it suggests our bulk theory should be interpreted as an extended TQFT associated with the quantum semi-group $\SL^+_{q}(2,\mathbb{R}) \otimes \SL^+_{q}(2,\mathbb{R})$. For each chiral copy, we expect this is closely related to the Teichm\"uller TQFT studied in \cite{EllegaardAndersen:2011vps}.
\\

In \textbf{Section \ref{Sec:conclusion}}, we conclude with some different perspectives and suggestive routes for future investigation. \\

\textbf{Appendix} \ref{app:bcor} provides details on a (doubled) JT limit of the boundary correlators. \textbf{Appendix} \ref{app:volume-reg} introduces the relevant finite-volume regularization methods to describe the non-compact groups that appear in low dimensional gravitational models, such as JT and 3d gravity. These methods play an important role to interpret our factorization results in section \ref{sec:facgrav}. \textbf{Appendix} \ref{app:hopf} provides a brief review on Hopf algebras of functions and their deformation leading to the concept of quantum group and its co-representation that appears in subsection \ref{sec:co-rep}. Finally, \textbf{Appendix} \ref{sec:ex-eTQFT} provides some details on pedagogical examples of extended TQFTs.

\section{Universal model for 3d pure gravity}
\label{sec:3dgrav_def}

The first goal of this work is to formulate and interpret a proposal for 3d gravity with negative cosmological constant. Given the success achieved in recent years in 2d JT gravity \cite{Jackiw:1984je,Teitelboim:1983ux,Jensen:2016pah,Maldacena:2016upp,Engelsoy:2016xyb}, it is natural to reconsider the 3d strategy. We build on the universality of the AdS$_3$/CFT$_2$ correspondence at large temperature to propose a partition function that can be independently reproduced by gravitational path integral considerations choosing adequate boundary conditions. We check this proposal reproduces the 2d JT gravity results and, in fact, extends them at low temperature away from extremality.

\subsection{Bulk motivation of our proposal}
\label{sec:bulk_mot}

Before dealing with quantum mechanical considerations, we review the classical bulk connection between 2d JT and 3d gravities, to motive our later proposal. Consider a spherically symmetric ansatz for a metric in 2+1d:
\begin{equation}
\label{mansatz}
    ds^2 = g_{\mu\nu}^{(2)}(x^\mu)dx^\mu dx^\nu + \Phi^2(x^\mu) d\varphi^2 , \qquad \mu,\nu= t,r, \quad \varphi \sim \varphi + 2\pi.
\end{equation}
with coordinates $(t,r,\varphi)$ playing the role of time, radial coordinate, and angular coordinate respectively. For such geometries, the 3d Einstein-Hilbert gravity action with cosmological constant $\Lambda$ reduces to \cite{Achucarro:1993fd}:
\begin{equation}
\label{eqdimred}
S_{\text{EH}} = \frac{1}{16\pi G_{\mt{N}}^{(3)}} \int d^3x \sqrt{-g}(R^{(3)}- \Lambda) \overset{\eqref{mansatz}}{=} \frac{1}{16\pi G_{\mt{N}}^{(2)}} \int d^2x \sqrt{-g} \Phi (R^{(2)}-\Lambda),
\end{equation}
where $G_{\mt{N}}^{(2)} = G_{\mt{N}}^{(3)}/2\pi$. This is the Jackiw-Teitelboim (JT) gravity action \cite{Jackiw:1984je,Teitelboim:1983ux}. Thus, the spherically symmetric $(t,r)$ sector of 3d gravity is directly governed by JT gravity. 

However, in most of the literature on this relation, and in particular in the Schwarzian boundary interpretation \cite{Jensen:2016pah,Maldacena:2016upp,Engelsoy:2016xyb,Stanford:2017thb,Mertens:2022irh}, the JT action is found in the \emph{spatial} $(r,\varphi)$ plane instead, see e.g. \cite{Cotler:2018zff}. To make such link more apparent, it would hence require a reinterpretation of time versus space. Using the language of 2d CFTs, this requires a modular $S$-transformation. This will be key in what follows.

In the rest of this section, we explore this idea allowing us to write down explicit expressions to formulate a concrete proposal for 3d gravity, in parallel with the JT proposal. Quite a few expressions that follow are known in some form in the literature; our main goal here is to present the material in a suggestive form, paralleling our understanding of the JT story.

\subsection{Universality of vacuum character from 2d CFT}
\label{sec:2duniv}

First, we motivate our proposal by describing how it is universally encoded in the high-temperature regime of any holographic CFT (with sufficiently sparse low-energy spectrum).

Consider a discrete microscopic unitary realization of the AdS$_3$/CFT$_2$ duality with modular invariant torus partition function
\begin{equation}
Z(\tau) = \sum_{h,\bar{h}}M_{h,\bar{h}}\, \chi_h(\tau)\chi_{\bar{h}}(\bar{\tau}) =  \sum_{h,\bar{h}}M_{h,\bar{h}}\, \chi_h(-1/\tau)\chi_{\bar{h}}(-1/\bar{\tau})\,,
\label{eq:2dcft-pt}
\end{equation}
where $M_{h,\bar{h}}$ are integers with $M_{0,0}=1$.\footnote{Reference \cite{Collier:2019weq} denoted such a CFT as ``compact".} The set of conformal weights is labeled by $(h,\bar{h})$, or alternatively, by $\Delta = h + \bar{h}$ and $J= h - \bar{h}$.
$\tau = \frac{\beta}{2\pi \ell}(\mu + i)$ is the modular parameter of the boundary torus, $\beta$ is the inverse temperature, $\mu$ is the chemical potential for rotation and $\ell$ is the bulk AdS length. Assume this model only has the Virasoro symmetry and is irrational (we focus on $c>1$). The Virasoro characters can be written as
\begin{equation}
\label{eq:char}
\chi_0(\tau) = \frac{(1-\mathfrak{q})}{\eta(\tau)}\mathfrak{q}^{-\frac{c-1}{24}}, \qquad \chi_h(\tau) = \frac{1}{\eta(\tau)}\mathfrak{q}^{h-\frac{c-1}{24}}, \qquad \eta(\tau) \equiv \mathfrak{q}^{1/24}\prod_{m=1}^{+\infty}(1-\mathfrak{q}^m),
\end{equation}
in terms of the left- and right- modular parameters
\begin{equation}
\mathfrak{q} \equiv e^{2\pi i \tau} = e^{\frac{\beta}{\ell}(i\mu -1)}, \qquad \bar{\mathfrak{q}} \equiv e^{-2\pi i \bar{\tau}} = e^{-\frac{\beta}{\ell}(i\mu + 1)}\,.
\end{equation}
The central charge $c$ can be parameterized as
\begin{equation}
\label{bhcc}
c=1+6(b+b^{-1})^2 \approx \frac{3 \ell}{2 G_{\mt{N}}}\,,
\end{equation}
where the first equality uses standard Virasoro CFT notation (with $c\geq 25$ and hence $b\in \mathbb{R}$) and the second equality uses the Brown-Henneaux holographic formula \cite{Brown:1986nw}, valid in the large $c$ semi-classical regime.

Let us evaluate \eqref{eq:2dcft-pt} at large temperature, i.e. when $\beta/\ell \ll \Delta_{\text{gap}}$, with $\Delta_{\text{gap}}$ the spectral gap between the vacuum and the first excited primary state (see e.g. \cite{Hartman:2014oaa,Dyer:2016pou}). Working in the dual channel, observe the ratio
\begin{equation}
\label{eq:ratio}
\frac{\chi_h(-1/\tau)\chi_{\bar{h}}(-1/\bar{\tau})}{\chi_0(-1/\tau)\chi_0(-1/\bar{\tau})}
= \frac{1}{(1-\tilde{\mathfrak{q}})(1-\bar{\tilde{\mathfrak{q}}})}\tilde{\mathfrak{q}}^h\bar{\tilde{\mathfrak{q}}}^{\bar{h}}, \quad \text{where} \quad
\tilde{\mathfrak{q}} \equiv e^{-2\pi i /\tau} = e^{-4\pi^2 i\frac{\ell}{\beta}\frac{(\mu-i)}{(\mu^2+1)}},
\end{equation}
goes to zero as $\beta/\ell \ll \Delta_{\text{gap}}$. Thus, at large temperature and for a 2d CFT with not too many light primaries (sparse spectrum), the vacuum character in the dual channel dominates the irrational CFT partition function
\begin{equation}
\label{eq:vacdom}
Z(\tau) = \sum_{h,\bar{h}}M_{h,\bar{h}}\, \chi_h(\tau)\chi_{\bar{h}}(\bar{\tau}) \, \approx \,  \left|\chi_0(-1/\tau)\right|^2\,.
\end{equation}
Even if $\beta/\ell \ll \Delta_{\text{gap}}$, since we are not taking the $c\to \infty$ limit (with hence  $c\sim \mathcal{O}(1)$), one retains a quantum theory.
Our main goal is to argue that the resulting partition function $\left|\chi_0(-1/\tau)\right|^2$, capturing the universal vacuum character in the dual channel of any irrational 2d CFT, makes sense on its own and plays the same role as JT gravity does in 2d. Note the role played by the modular $S$-transformation. The dominance of the vacuum module is well-known to only contain the gravitational interactions since it isolates the stress tensor (see e.g. \cite{Fitzpatrick:2015zha,Fitzpatrick:2015dlt,Jackson:2014nla}). Here it dominates in the dual channel instead.

Notice that, due to the exponential suppression in \eqref{eq:ratio}, the vacuum still dominates in \eqref{eq:vacdom} even if $\beta/\ell \approx \Delta_{\text{gap}}$, as long as $\beta(1+\mu^2)/\ell$ is not significantly larger than $\Delta_{\text{gap}}$.\footnote{E.g. $e^{-4\pi^2} \sim 10^{-18}$. This is a purely ``numerical'' suppression, and not a suppression caused by a parametric limit of a ratio of dimensionful parameters.} Having this in mind, we will rewrite the condition on vacuum dominance as 
\begin{equation}
\Delta_{\text{gap}} \, \gtrapprox  \, \frac{\beta}{\ell}.
\end{equation}
This observation will be important in subsection \ref{sec:double-JT}.

\subsection{Grand canonical partition function}
\label{sec:grand-can}

\begin{figure}[h]
\centering
\includegraphics[width=0.85\textwidth]{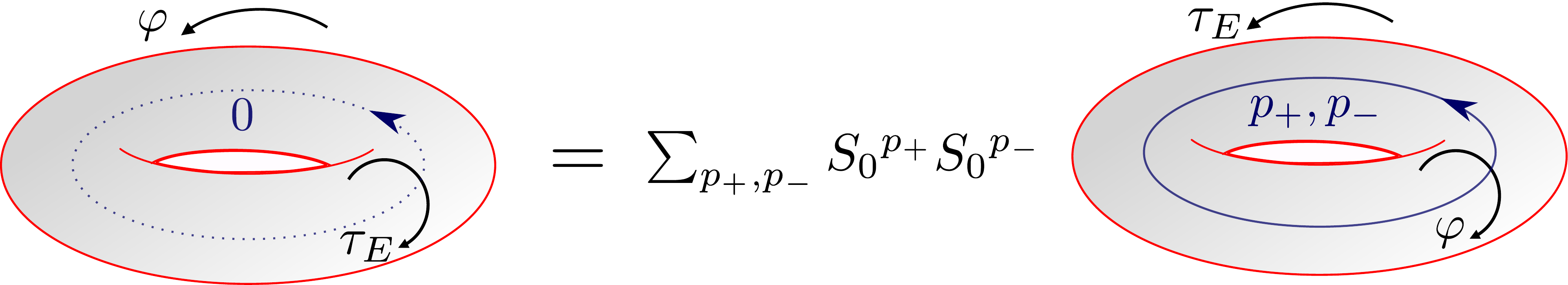}
\caption{Modular $S$-transform of the vacuum character leads to a sum (integral) of black hole states. The blue lines are Wilson loops in the interior of the torus. The labels $(p_+,p_-)$ correspond to defects that have a hyperbolic holonomy around them (as measured on a spatial $\varphi$-slice).}
\label{toruschannel}
\end{figure}

Our proposed partition function, providing a universal description of high-temperature 2d CFT dynamics is:
\begin{equation}
\label{gcpf}
Z(\tau) = \left|\chi_0(-1/\tau)\right|^2 = \int_0^{+\infty} \hspace{-0.2cm} dp_+ dp_- \, S_{0}{}^{p_+}S_{0}{}^{p_-} \chi_{p_+}(\tau) \chi_{p_-}(\bar{\tau}),
\end{equation}
where we parametrized $h = p_+^2 + Q^2/4$ and $\bar{h} = p_-^2 + Q^2/4$, and labeled the characters as such. The RHS has the interpretation as a sum of off-shell (rotating) black hole states labeled by $(p_+,p_-)$, with a specific measure (Figure \ref{toruschannel}). We will come back to this specific interpretation of the gravitational Hilbert space in subsection \ref{s:propgrav} further on.
Using the Virasoro modular $S$-matrix expression
\begin{equation}
S_{ 0}{}^{p_{\pm}} \equiv \text{dim}_q p_\pm = 4\sqrt{2} \sinh ( 2 \pi b p_{\pm} ) \sinh ( 2 \pi b^{-1} p_{\pm} ),
\label{eq:S-modular}
\end{equation}
this can be written as a thermal grand-canonical partition function
\begin{align}
\label{solidtorus}
&Z(\beta,\mu) \equiv \text{Tr}\left[ e^{-\beta H +i\mu \frac{\beta}{\ell} J}\right]  \\
&=\int_0^{+\infty} \hspace{-0.2cm} dp_+ dp_- 32\sinh(2\pi b p_+) \sinh(2\pi b^{-1} p_+)\sinh(2\pi b p_-) \sinh(2\pi b^{-1} p_-) \frac{e^{-\frac{\beta}{\ell}(p_+^2 + p_-^2)} e^{i\mu \frac{\beta}{\ell}(p_+^2 - p_-^2)}}{\left|\eta(\tau)\right|^2}, \nonumber
\end{align}
for (primary) states labeled by the quantum numbers $(p_+,p_-)$, energy $H = \frac{p_+^2 + p_-^2 -\frac{1}{12}}{\ell}$ and angular momentum $J = p_+^2-p_-^2$. 
For a fixed choice of $p_+$ and $p_-$, one has additionally a thermal partition function of boundary gravitons, by series expanding:
\begin{equation}
\frac{1}{\eta(\tau)\mathfrak{q}^{-1/24}} = \frac{1}{\prod_{m=1}^{+\infty}(1-\mathfrak{q}^m)} = \sum_{n=0}^{+\infty}p(n) \mathfrak{q}^n, \qquad p(n) = \# \, \text{partitions of $n$}\,.
\end{equation}
There is an analogous expansion for the anti-holomorphic sector, where the integer $\bar{n}$ would label its descendants. The energy and angular momentum quantum numbers of a generic state are hence:
\begin{equation}
H = \frac{p_+^2 + p_-^2-\frac{1}{12}+n+\bar{n}}{\ell}, \qquad J = p_+^2-p_-^2+n-\bar{n}.
\end{equation}
The density of states in our model \eqref{solidtorus} is continuous, but with no overall IR-divergent volume scaling, which is a result of effectively coarse-graining the microscopics at high temperature. Indeed, our model is a limit of a finite volume boundary partition function \eqref{eq:2dcft-pt},\footnote{Space is a circle $S^1$.} which can hence never obtain an IR divergent volume scaling. Relatedly, a generic state in this Hilbert space does not have integer spin $J$. This is neither required nor expected in our case. Indeed, integer spin $J$ is usually required by modular invariance ($\tau \to \tau +1$), which is lost when approximating the CFT partition function by the vacuum module in \eqref{eq:vacdom}. It is also not expected since this high-temperature regime is a semi-classical regime, where one expects quantization conditions on quantum numbers to be washed out.

The thermal entropy corresponding to \eqref{solidtorus} can be written as:
\begin{align}
\label{eq:BHncl}
&S_{\text{th}} = (1-\beta \partial_\beta) \log Z(\beta,\mu) \nonumber \\
&= -\sum_{n,\bar{n}=0}^{+\infty}\int_0^{+\infty} \hspace{-0.2cm} dp_+ dp_- (\text{dim}_q p_+ \text{dim}_q p_-) p(n) p(\bar{n}) \rho_{n,\bar{n}}(p_+,p_-) \log \rho_{n,\bar{n}}(p_+,p_-) \nonumber \\
&=-\sum_{n,\bar{n}=0}^{+\infty}\int_0^{+\infty} \hspace{-0.2cm} dp_+ dp_- \left[ p_{n,\bar{n}}(p_+,p_-) \log p_{n,\bar{n}}(p_+,p_-) - p_{n,\bar{n}}(p_+,p_-) \log \Big((\text{dim}_q p_+ \text{dim}_q p_-) p(n) p(\bar{n}) \Big) \right],
\end{align}
where we defined
\begin{align}
  \rho_{n,\bar{n}}(p_+,p_-) &= \frac{1}{Z(\beta,\mu)}\,e^{-\frac{\beta}{\ell}(p_+^2 + p_-^2-\frac{1}{12}}+n+\bar{n})\,e^{i\mu \frac{\beta}{\ell}(p_+^2 - p_-^2+n-\bar{n})},
  \end{align}
  and the classical probability distribution
  \begin{align}
  \label{eq:probdis3d}
    p_{n,\bar{n}}(p_+,p_-) &= (\text{dim}_q p_+ \text{dim}_q p_-) p(n) p(\bar{n}) \frac{e^{-\frac{\beta}{\ell}(p_+^2 + p_-^2-\frac{1}{12}+n+\bar{n})}\,e^{i\mu \frac{\beta}{\ell}(p_+^2 - p_-^2+n-\bar{n})}}{Z(\beta,\mu)},
\end{align}

normalized to 1: $$\sum_{n,\bar{n}=0}^{+\infty}\int_0^{+\infty} \hspace{-0.2cm} dp_+ dp_- p_{n,\bar{n}}(p_+,p_-) =1.$$ One of the goals in this paper is to directly interpret this thermal entropy \eqref{eq:BHncl} as a von Neumann entropy at the quantum level. As it stands, its first term equals the classical Shannon entropy of $p_{n,\bar{n}}(p_+,p_-)$, whereas its second term contains the expectation value in this probability distribution of the thermal entropy $\log \Big((\text{dim}_q p_+ \text{dim}_q p_-) p(n) p(\bar{n}) \Big)$ within each such subsector. Notice
the thermal entropy \eqref{eq:BHncl} contains contributions corresponding to the primaries, and to the descendants at the holographic boundary. The descendants at the asymptotic boundary are not part of what we would call the black hole entropy. They are however part of the total holographic entropy in the system. This distinction becomes mute in the limit of semi-classical gravity (to which we turn next).

 Again we emphasize all quantities, including the thermal entropy \eqref{eq:BHncl}, are continuous but do not contain any IR divergent volume factor. This observation will turn out to be crucial later on when we interpret $S_{\text{th}}$ as entanglement entropy.

In the semi-classical high-temperature regime $\beta/\ell \ll 1$, which for the microscopic 2d CFT requires the combined regime $\beta/\ell \ll\text{min} (1,\Delta_{\text{gap}})$, the integral \eqref{solidtorus} is dominated by large values of $p_+$ and $p_-$. This allows us to approximate the density factors by exponentials as:
\begin{equation}
\text{dim}_q p_\pm \approx \sqrt{2} \, e^{2 \pi (b+b^{-1} )p_{\pm}} \sim e^{2\pi \sqrt{\frac{c}{6}h}},
\end{equation}
which is the Cardy scaling at large conformal weight. Next we need to assess the growth of descendants. For an irrational theory for a fixed level $n$, the asymptotic large weight counting is just the number of partitions of the level $n$, scaling as $p(n) \sim e^{2\pi \sqrt{\frac{n}{6}}}$. However, we know from the Cardy scaling that the total number of states in this theory at weight $L_0$ scales as $\rho_{\text{CFT}}(L_0) \sim e^{2\pi \sqrt{\frac{c}{6}L_0}}$. The latter rises much faster when $c>1$. Explicit expressions for the asymptotic density of primary and descendant states at large weight $L_0$ can be computed by a more detailed analysis \cite{Kraus:2016nwo,Datta:2019jeo}:
\begin{equation}
\rho_{\text{primary}}(L_0) \sim e^{2\pi \sqrt{\frac{c-1}{6}L_0}}, \qquad 
\rho_{\text{descendant}}(L_0) \sim e^{2\pi \sqrt{\frac{c}{6(c-1)}L_0}}.
\end{equation}
This means primary states massively overwhelm descendant states for any Cardy computation in a Virasoro theory with $c>1$, and boundary gravitons in this model are subdominant in the semi-classical regime.

If we additionally take the large $c$ limit $(c \sim \frac{\ell}{G_{\mt{N}}}\gg 1)$, or $b\ll 1$, the density of states leads to the thermal entropy matching the Bekenstein-Hawking entropy of rotating BTZ black holes:
\begin{equation}
\label{eq:BHcl}
S_{\text{th}} \, \approx \, \log \left(\text{dim}_q p_+ \text{dim}_q p_-\right) \, \approx \, 2\pi b^{-1} (p_+ + p_-) = \frac{2\pi r_+}{4 G_{\mt{N}}} = S_{\text{BH}}, \qquad r_+ \approx 4b^{-1} G_{\mt{N}}(p_++p_-).
\end{equation}
The interpretation of the BH entropy as the logarithm of the quantum dimension was first observed by H. Verlinde and L. McGough \cite{McGough:2013gka}.\footnote{\label{fn2} They derive this entropy by applying the formulas of topological entanglement entropy as $\log (S_{p_+}{}^{0} S_{p_-}{}^{0})$ to the irrational Virasoro case. A subtlety is that one needs to ad hoc use instead $S_{0}{}^{p_\pm}$ for this identification to work. For rational models, this is not a problem as $S$ is symmetric, but for the irrational Virasoro case, one actually has $S_{p_\pm}{}^{0}=0$ by the modular bootstrap. One of our goals is to precisely understand how to think about \eqref{eq:BHcl}, and more generally \eqref{eq:BHncl}, as entanglement entropy.} \\

This set-up can be immediately extended to CFTs with larger symmetry algebras. Indeed, in that same work \cite{McGough:2013gka}, the matching with the BH entropy was extended to the higher spin case as well. Using the vacuum characters of the $\mathcal{W}_N$ algebra in the dual channel (as in e.g. \cite{Datta:2021efl}), one can reach a similar conclusion. As a final extension, for a boundary SCFT with $\mathcal{N}=1$ supersymmetry, the dominant high-temperature contribution is likewise immediately written down.
Imposing anti-periodicity, respectively periodicity for fermions along the spatial (non-contractible) $\varphi$-circle, one writes for $Z^{\text{NS}}(\beta,\mu)$ and $Z^{\widetilde{\text{NS}}}(\beta,\mu)$ respectively:\footnote{The right hand side contains the super-Virasoro characters:
\begin{equation}
\chi_p^{NS}(\tau) = \sqrt{\frac{\theta_3(\tau)}{\eta(\tau)}}\frac{\mathfrak{q}^{p^2/2}}{\eta(\tau)}, \qquad \chi_p^{R}(\tau) = \sqrt{\frac{\theta_2(\tau)}{2\eta(\tau)}}\frac{\mathfrak{q}^{p^2/2}}{\eta(\tau)}.
\end{equation}
}
\begin{align}
&\int_0^{+\infty} \hspace{-0.2cm} dp_+ dp_- 16 \sinh(\pi b p_+) \sinh(\pi b^{-1} p_+)\sinh(\pi b p_-) \sinh(\pi b^{-1} p_-) \frac{\left|\theta_3(\tau)\right|}{\left|\eta(\tau)\right|^3}e^{-\frac{\beta}{2\ell}(p_+^2 + p_-^2)} e^{i\mu\frac{\beta}{2\ell}(p_+^2 - p_-^2)}, \nonumber \\
&\int_0^{+\infty} \hspace{-0.2cm} dp_+ dp_- 16 \cosh(\pi b p_+) \cosh(\pi b^{-1} p_+)\cosh(\pi b p_-) \cosh(\pi b^{-1}p_-) \frac{\left|\theta_2(\tau)\right|}{\left|\eta(\tau)\right|^3}e^{-\frac{\beta}{2\ell}(p_+^2 + p_-^2)} e^{i\mu\frac{\beta}{2\ell}(p_+^2 - p_-^2)}.
\end{align}
Both of these have the same semi-classical Bekenstein-Hawking growth of states as in the bosonic gravity model.

We emphasize that we do not attempt to find a microscopic model of 3d pure gravity. We are instead aiming for a universal description of 2d irrational CFTs. The result is an effective quantum mechanical model \eqref{solidtorus}, playing a similar role as JT gravity in 1+1d.

\subsection{Proposal for 3d pure gravity}
\label{s:propgrav}
Next we present an independent argument where we directly derive the partition function \eqref{gcpf} from the pure 3d gravitational path integral.
So let us instead start with asymptotically AdS 2+1d pure gravity with Euclidean action
\begin{equation}
I = -\frac{1}{16\pi G_{\mt{N}}}\int_{S^1 \times D^2}d^3x \sqrt{g} (R + 2) -\frac{1}{8\pi G_{\mt{N}}}\oint_{S^1 \times S^1} d^2x \sqrt{h} K\,,
\end{equation}
in the interior of a solid 2-torus $S^1 \times D^2$, including the suitable Gibbons-Hawking-York boundary term on the torus boundary. Following standard arguments \cite{Witten:1988hc,Achucarro:1986uwr}, one writes the first order formulation of this model as $\PSL(2,\R)\times \PSL(2,\R)$ Chern-Simons theory. 
The resulting Chern-Simons models can then, in turn, be rewritten in terms of their boundary WZW description and then finally turned into a combination of left ($L$) and right ($R$) Alekseev-Shatashvili coadjoint orbit actions of the Virasoro group \cite{Alekseev:1988ce,Alekseev:1990mp}, see also \cite{Cotler:2018zff} for a detailed discussion on these steps. 

The Alekseev-Shatashvili actions contain a boundary reparametrization as the degree of freedom. The strategy in \cite{Cotler:2018zff} considered a $\text{Diff }S^1$ reparametrization $f(\tau_E,\varphi)$ on the torus boundary, satisfying  $f(\tau_E,\varphi+2\pi) = f(\tau_E,\varphi) + 2\pi$ and $\partial_\varphi f \geq 0$ along a contractible \textit{spatial} $\varphi$-cycle. This choice describes the boundary graviton fluctuations around the global AdS geometry. Changing the periodicity from $2\pi \to 2\pi \theta$ would then allow for the addition of objects on a spatial slice of global AdS$_3$, such as massive particles, represented as conical defects. The resulting model again describes the boundary gravitons on such a background. In this sense, their analysis is perturbative, since it concerns the fluctuations around a given background.

Our choice of boundary condition will be different: we impose instead the above periodicity along the \emph{temporal} cycle, explicitly implementing a swap between time and space cycles, making the reparametrization at fixed \emph{spatial} coordinate an element of $\text{Diff }S^1$ instead. 
Given the perspective advocated in this work, such geometric action description of \eqref{gcpf} would not be interpreted as perturbative, as defined above. 
Our boundary condition is also different from the choice of boundary conditions imposed in \cite{Coussaert:1995zp},  which is summing over possible choices of defect (with flat measure in $\lambda$ allowing one to recombine both geometric actions into the 2d Liouville CFT). We comment on this difference further on.

Let us be more explicit while considering more general constant-representative orbits with seed element $b_0$. We shall revert to the vacuum orbit at the end of the discussion. On a 2d Euclidean torus space with coordinates $({\T},\sigma)$,\footnote{These will turn out to be rescaled versions of our usual coordinates ($\tau_E,\varphi$), see further around equation \eqref{eq:casee}.} we evaluate the path integral
\begin{align}
\label{pspi}
\int
\left[\mathcal{D}f_L\right]\left[\mathcal{D}f_R\right] \text{Pf}(\omega_L) \text{Pf}(\omega_R)\, e^{- I_{\text{orbit}}^L[f_L] - I_{\text{orbit}}^R[f_R]}\,,
\end{align}
where the symplectic measure is
\begin{equation}
\omega = \frac{c}{48\pi}\int_0^{2\pi} d{\T} \left(\frac{\delta \ddot{f} \wedge \delta \dot{f}}{\dot{f}^2} + b_0 \delta \dot{f} \wedge  \delta f\right),
\end{equation}
and the Alekseev-Shatashvili geometric actions \cite{Alekseev:1988ce,Alekseev:1990mp} are
\begin{equation}
\label{as}
I_{\text{AS,orbit}}^{L,R}[f] =  \pm i \int d\sigma d{\T} \left[\frac{c}{48\pi}\frac{f'}{\dot{f}}\left(\frac{\dddot{f}}{\dot{f}}- 2 \left(\frac{\ddot{f}}{\dot{f}}\right)^2 \right) - b_0 \dot{f}f'\right]\,,
\end{equation}
where $\dot{} \equiv \frac{d}{d\T}$ and ${}^\prime \equiv \frac{d}{d\sigma}$. Notice once more space $\sigma$ and time ${\T}$ were swapped, following \cite{Mertens:2018fds}, compared to the standard discussions. The $L$-sector has the top sign and the $R$-sector the bottom sign. These are complex conjugates of each other. We parameterize the orbit parameter as $b_0 \equiv - \frac{c}{48\pi} \theta^2$. Geometrically, this corresponds to having a periodicity $2\pi\theta$ along the temporal cycle.

Adding the Schwarzian action as the Hamiltonian weighing different configurations, as done in \cite{Cotler:2018zff}, and after integration by parts in the time direction, the resulting orbit action used in \eqref{pspi} is equivalent to
\begin{align}
\label{as2}
I_{\text{orbit}}^L[f] &=  -\int d\sigma d{\T} \left(-\frac{ic}{48\pi}\frac{\dot{f'}\ddot{f}}{\dot{f}^2} - ib_0 \dot{f}f' +  \frac{c}{24\pi}\left\{\tan  \frac{\theta}{2}f,\tau\right\}\right) \\
&= \frac{c}{24\pi} \int d\sigma d{\T} \left(\frac{\partial \dot{f}\ddot{f}}{\dot{f}^2} - \theta^2 \dot{f} \partial f\right),
\label{as3}
\end{align}
where $\partial = \frac{1}{2}(\partial_{\T} + i\partial_\sigma)$. The $R$-action is the complex conjugate where $\partial \to \bar{\partial} \equiv \frac{1}{2}(\partial_{\T} - i\partial_\sigma) $. This action was written in this form (with time and space swapped) in \cite{Cotler:2018zff}, see also \cite{Henneaux:2019sjx} for an analysis from the two boundary phase space perspective. 

Since the only difference is swapping the time $\T$ and space $\sigma$ cycles, the integration domain $D$ consists of all reparametrization functions $f_{L,R}$ satisfying $\dot{f}_{L,R} \geq 0$ and the twisted periodicity constraints
\begin{equation}
\label{twper}
\begin{cases}
f({\T} + 2\pi,\sigma) &= f({\T},\sigma) + 2\pi, \\
f({\T} + 2\pi \Re(\tau), \sigma + 2 \pi \Im(\tau)) &= f({\T},\sigma),
\end{cases}
\end{equation}
modulo independent SL$(2,\mathbb{R})$ M\"obius transformations
\begin{equation}
\label{mob}
\tan  \frac{\theta}{2}f \to \frac{a\tan  \frac{\theta}{2}f+b}{c\tan  \frac{\theta}{2}f+d}, \qquad ad-bc=1,
\end{equation}
where $a,b,c,d$ are now allowed to depend on $\sigma$, since both reparametrization sectors, i.e. $f_L$ and $f_R$, do not communicate directly. Using \eqref{as3}, one can check that
\begin{equation}
f_{\text{cl}}({\T},\sigma) = {\T} - \frac{\Re(\tau)}{\Im(\tau)}\sigma
\end{equation}
is a classical saddle solution with on-shell action
\begin{equation}
I_{\text{on-shell}} = \frac{c}{12} \pi i \tau \theta^2.
\end{equation}
In order to match with our 3d gravity proposal, we must work in the dual channel. This requires us to interpret the dependence on the torus modular parameter in terms of $\tau \to -1/\tau \equiv \tau^\prime$. Hence,
\begin{equation}
\Re(\tau^\prime) = -\frac{2\pi \mu}{\beta(1+\mu^2)}, \quad \Im(\tau^\prime) = \frac{2\pi}{\beta(1+\mu^2)},
\end{equation}
so that the classical solution is simply $f_{\text{cl}}({\T},\sigma) = {\T} +\mu\sigma$. After suitable rescalings of ${\T}$ and $\sigma$ (and $f$) to obtain a standard periodicity of $\beta$ and $2\pi$ for the rescaled coordinates $\tau_\mt{E}$ and $\varphi$ respectively, one can rewrite the identification \eqref{twper} as:
\begin{equation}
\begin{cases}
\label{eq:casee}
f(\tau_\mt{E} + \beta,\varphi) &= f(\tau_\mt{E},\varphi) + \beta, \\
f(\tau_\mt{E} + 2\pi \frac{\mu}{1+\mu^2}, \varphi + 2\pi) &= f(\tau_\mt{E},\varphi),
\end{cases}
\end{equation}
with $\mu \neq 0$ corresponding to a non-trivial twisting. In the rescaled coordinates $(\tau_\mt{E},\varphi)$, the classical saddle solution becomes
\begin{equation}
f_{\text{cl}}(\tau_\mt{E},\varphi) = \tau_\mt{E} - \frac{\mu}{1+\mu^2}\varphi,
\end{equation}
or in terms of the Poincar\'e time coordinate:
\begin{equation}
F(\tau_\mt{E},\sigma) = \tan \left[\frac{\pi \theta}{\beta} \left( \tau_\mt{E} - \frac{\mu}{1+\mu^2}\varphi\right)\right].
\end{equation}
The resulting on-shell action is the same as before, but still evaluated on $\tau^\prime = -1/\tau$
\begin{equation}
I_{\text{on-shell}} = -\frac{c}{12} \pi i \theta^2 \frac{2\pi}{\beta(\mu + i)}\,.
\end{equation}

The classical saddle solution captures an Euclidean BTZ black hole, when expressed in terms of the second order metric formulation.\footnote{The classical saddle determines the stress tensor expectation value from which one can reconstruct the metric, either directly or through the Chern-Simons gauge connection.} For example, for $\mu=0$, one would recover the Euclidean non-rotating BTZ metric
\begin{equation}
ds^2 =\ell^2 \left[\left(\frac{2\pi}{\beta}\right)^2\sinh^2\rho d\tau_E^2 + d\rho^2 + \cosh^2\rho d\varphi^2\right], \qquad \tau_E \sim \tau_E + \beta, \quad \varphi \sim \varphi + 2\pi,
\end{equation}
with event horizon at $\rho=0$ and the holographic boundary at $\rho \to +\infty$.

These types of path integrals \eqref{pspi} are famously one-loop exact and evaluate precisely to the Virasoro characters for different primaries depending on the choice of $\theta$ \cite{Alekseev:1988ce,Alekseev:1990mp}. Indeed, the saddle point approximation reproduces the $\mathfrak{q}^{h-c/24}$ part of the character, whereas the one-loop determinant, coming from the fluctuations, captures the contribution from the descendants, i.e. the boundary gravitons. 

The above discussion applies for any $\theta$. In particular, when $\theta=1$, it leads to the solid torus partition function \eqref{gcpf}. Hence starting from a pure gravity perspective, one automatically lands on our model \eqref{solidtorus}, recovering the same partition function we encountered earlier for holographic irrational 2d CFTs in the high temperature regime $\beta/\ell \ll \Delta_{\text{gap}}$.

The geometric action \eqref{as3} has a chiral Virasoro symmetry algebra, and the combination in \eqref{pspi} has a separate left- and right-moving Virasoro algebra, as dictated by the asymptotic AdS (or Brown-Henneaux) boundary conditions \cite{Brown:1986nw}, but it is not a full CFT; it is a tensor product of two chiral CFTs instead:
\begin{equation}
\text{3d pure gravity} \, = \text{chiral CFT}_{\mt{L}} \otimes \text{chiral CFT}_{\mt{R}}.
\label{eq:3dfactor}
\end{equation}
The difference with a full CFT is that the above models are clearly not modular invariant. In particular, the coadjoint orbit action \eqref{as3}  has a unique saddle point, as pointed out above.
Indeed, if we specify the temperature and chemical potential at the boundary, there is a unique BTZ black hole that fits this description with on-shell values of mass and angular momentum, $M(\beta,\mu)$ and $J(\beta,\mu)$. In the set-up of A. Maloney and E. Witten \cite{Maloney:2007ud}, only a boundary torus is specified. In that case, an entire $\SL(2,\Z)$ family of saddle solutions exist. Here we specify more: we choose the interpretation of the two cycles as time and space. This uniquely fixes a single BTZ black hole saddle.

\begin{figure}[h]
\centering
\includegraphics[width=0.6\textwidth]{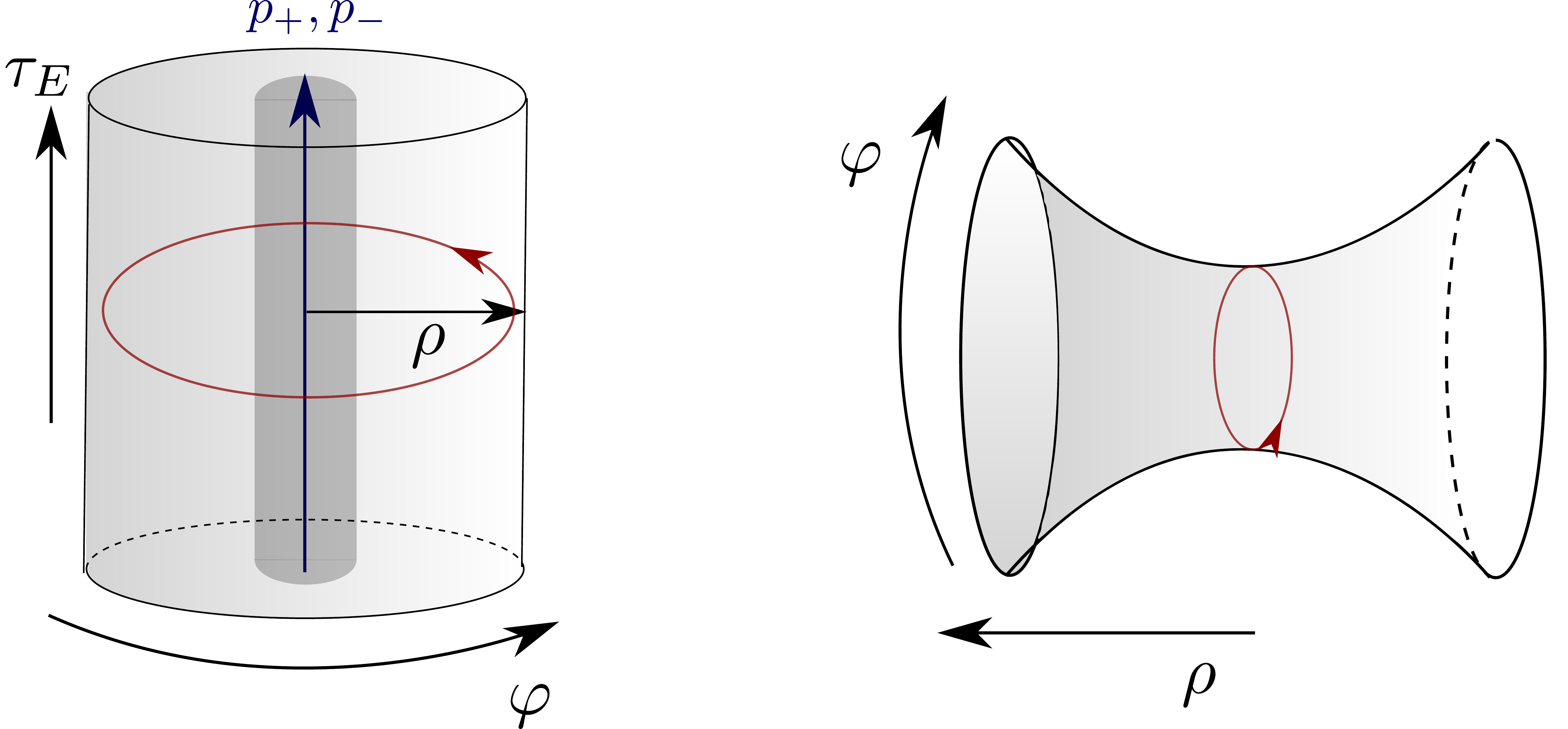}
\caption{Left: Black hole state defined by specifying its hyperbolic monodromy around a spatial slice, as measured by the red Wilson loop. Right: Spatial slice of the classical geometry containing a spatial wormhole.}
\label{hilbertgrav}
\end{figure}

Let us next characterize the states appearing in our gravitational Hilbert space in \eqref{gcpf}, keeping in mind the right hand side of Figure \ref{toruschannel}. On a solid cylinder, there is a unique classical solution of 3d gravity with specified hyperbolic monodromies $(p_+,p_-)$ (as measured with a linking Wilson loop in the fundamental representation): the BTZ black hole with these mass $H$ and angular momentum $J$ parameters (Figure \ref{hilbertgrav}). 

Geometrically, its spatial slice contains a spatial wormhole (Einstein-Rosen bridge):
\begin{equation}
ds^2_{\text{spatial}} = R^2\frac{d\rho^2 + d\varphi^2}{\cos^2(R\rho/\ell)}, \qquad R^2 = 8 H \ell^2,
\end{equation}
which opens up close to the hyperbolic defect at $\rho=0$.\footnote{This is identical to macroscopic ``punctures'' on the Liouville worldsheet \cite{Seiberg:1990eb}.} Since the classical phase space (two copies of Teichm\"uller space of the 2d hyperbolic cylinder) with these fixed holonomies is zero-dimensional, there is a unique quantum state associated with it. We conclude that the Hilbert space states in \eqref{solidtorus} are to be interpreted as quantized versions of the rotating BTZ black holes.

Finally, we note that the global AdS$_3$ vacuum is not contained as a state in the Hilbert space of our model \eqref{gcpf}, since the vacuum character is not included on the RHS of \eqref{gcpf}. This feature caused some confusion in the past when attempting to interpret 2d Liouville CFT as the boundary dual of pure gravity. This is because of a notorious mismatch between the Liouville CFT Cardy formula (with effective central charge $c_{\text{eff}}\equiv c - 24 h_{\text{min}}=1$ using $h_{\text{min}} = (c-1)/24$ for the Liouville CFT lowest primary which is at the bottom of the continuum \cite{Kutasov:1990sv}), and the Bekenstein-Hawking formula \cite{Strominger:1997eq,Carlip:2005zn,Carlip:1998qw,Martinec:1998wm} that uses instead the Brown-Henneaux central charge $c$ \eqref{bhcc}, see also \cite{Donnay:2016iyk} for a pedagogical discussion. In short, Liouville CFT has far too few high-energy states to explain the black hole degeneracy, which has led some authors to attempt to include additional states into the spectrum.
Indeed, the 2d Liouville CFT torus partition function (in any channel) is\footnote{There is an infinite volume factor $V_\phi$ from the Liouville zero-mode, that is not generated from the bulk gravitational path integral. So this factor has to be removed. The factor of $2$ originates simply from our choice throughout this work of limiting the $p_\pm$ integrals to positive values.}
\begin{equation}
\label{eq:lioupf}
  \frac{Z_{\text{Liouville}}(\tau)}{V_\phi} = 2 \int_0^{+\infty} \hspace{-0.2cm} dp\,  \chi_p(\tau) \chi_p(\bar{\tau}).
\end{equation}
Comparing to our result \eqref{gcpf}, there are two differences: (1) Liouville CFT \eqref{eq:lioupf} does not have twisted primaries in its spectrum, gravitationally interpretable as having no rotating black hole states, (2) the spectral density is flat for Liouville CFT. In fact, the partition function \eqref{eq:lioupf} is the same as that of the free boson CFT, whose asymptotic growth is determined by the Cardy formula with $c=1$ indeed. The distinction with our proposal can now be appreciated as follows. The asymptotic Cardy growth is usually computed by the dominance of the lowest weight character in the dual channel. For Liouville CFT, by modular invariance, this is the module $h=\bar{h}=(c-1)/24$ at the bottom of the continuum. For our proposal, by construction, this is the vacuum character at $h=\bar{h}=0$. We hence observe that in this language in our case $c_{\text{eff}}=c$, in spite of also not including the vacuum character on the right hand side of \eqref{gcpf}.\footnote{Our conclusion mirrors the case of 2d JT gravity, where the AdS$_2$ global vacuum is also not contained within the Hilbert space of the model. In both cases, the lowest-energy state is at the bottom of the continuum.} \\

\begin{figure}[h]
\centering
\includegraphics[width=0.85\textwidth]{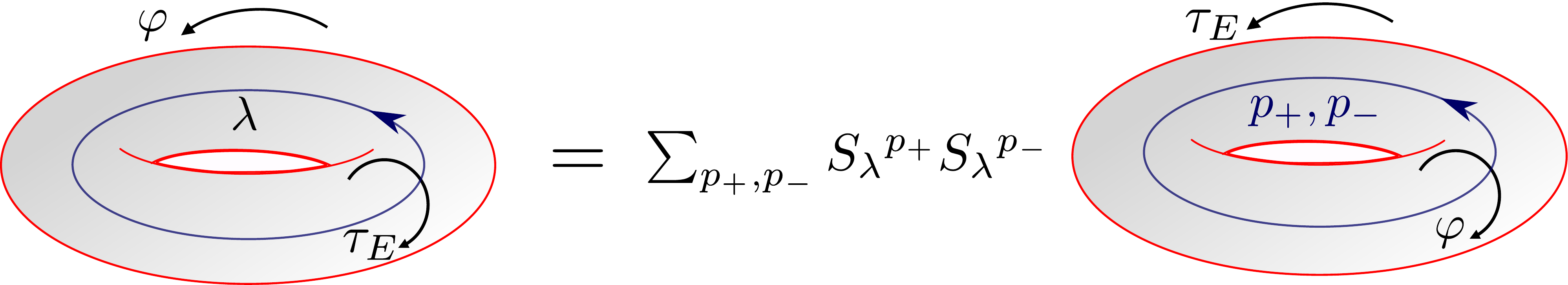}
\caption{Modular $S$-transform of the non-vacuum character.}
\label{toruschannel2}
\end{figure}

We can enrich the proposal \eqref{solidtorus} further by inserting additional defects or Wilson loops in the interior of the solid torus. Inserting a Wilson loop along the (non-contractible) spatial $\varphi$-direction in the $(\lambda,\lambda')$ continuous representation of $\SL(2,\R)\times \SL(2,\R)$,\footnote{Where $h=\lambda^2/2+Q^2/4$ to relate to the notation of equation \eqref{eq:char}.} just changes the Virasoro characters into:
\begin{align}
\label{defecttr}
Z_{\text{trumpet}} &= \chi_{\lambda}\left(-\frac{1}{\tau}\right)\chi_{\lambda'}\left(-\frac{1}{\bar{\tau}}\right) \\
&= \int_0^{+\infty} \hspace{-0.2cm} dp_+ dp_- 4\cos(2\pi \lambda p_+)\cos(2\pi \lambda' p_-) \frac{e^{-\beta(p_+^2 + p_-^2)} e^{i\mu\beta(p_+^2 - p_-^2)}}{\left|\eta(\tau)\right|^2}. \nonumber
\end{align}
The parameter $\lambda = i \theta$ with $\theta$ defined above. Insertion of such a Wilson loop modifies the holonomy along the Euclidean time direction $\tau_E$, and implements defect insertions (Figure \ref{toruschannel2}).

This amplitude is also required to glue tori together into more complicated surfaces. In this work, this amplitude will play a role in determining the correct edge state Hilbert space later on.

\subsection{Application: dimensional reduction to JT gravity squared}
\label{sec:double-JT}

Since our motivation was to find a conceptual 3d generalization of the 2d JT model and its Schwarzian description, it is natural to discuss how precisely JT gravity is encoded within our set-up.

It is well known that near-extremality freezes one of the two holomorphic sectors of the 2d CFT \cite{Strominger:1998yg,Balasubramanian:2003kq,Balasubramanian:2009bg}. It was shown in more detail in \cite{Ghosh:2019rcj} how the Schwarzian model emerges within a universal 2d CFT sector describing the low-energy excitations in such near-extremal situation. This is an explicit realization of the more general statement that JT gravity governs the near-horizon near-extremal dynamics of all higher-dimensionsal black holes that have an AdS$_2$ near-horizon region.\footnote{See e.g. \cite{Nayak:2018qej,Iliesiu:2020qvm,Castro:2021csm} for some recent discussion on these statements.} What we are aiming for here is a statement that is outside of this near-extremal near-horizon regime, i.e. for generic values of $\mu$, but holding specifically for 3d gravity.

We already revisited the high temperature and large $c$ limit of our partition function \eqref{solidtorus}, reproducing the classical BTZ black hole thermodynamics. Let us consider instead its low-temperature regime, where the partition function \eqref{solidtorus} is expected to be dominated by small momentum values $p_+$ and $p_-$. In fact, working in the double-scaling regime
\begin{equation}
c \gg 1, \qquad \beta/\ell \sim c\,,
\label{eq:double-scale}
\end{equation}
the contribution from the descendants is subdominant and the full partition function \eqref{solidtorus} can be approximated by
\begin{align}
\label{doubleJT}
Z(\beta,\mu)  &\overset{\beta \gg \ell}{\approx} 32 e^{\frac{\beta}{12\ell}} (2\pi b^3)^2 \int_0^{+\infty} \hspace{-0.2cm} dp_+ dp_- p_+ \sinh(2\pi p_+) p_-\sinh(2\pi p_-) e^{-\frac{b^2\beta}{\ell}(p_+^2 + p_-^2)} e^{i\mu \frac{b^2\beta}{\ell}(p_+^2 - p_-^2)} \nonumber \\
&= 32e^{\frac{\beta}{12\ell}} (2\pi^3 b^3)^2 \, Z_{\text{JT}}\Big(\frac{b^2\beta}{\ell}(1 + i\mu)\Big)Z_{\text{JT}}\Big(\frac{b^2\beta}{\ell}(1 - i\mu)\Big),
\end{align}
where we explicitly used $b^2 \approx 4G_{\mt{N}}/\ell \ll 1$, which is indeed compatible with $\beta/\ell \sim c$. This particular scaling limit was studied extensively in \cite{Mertens:2017mtv} in order to reproduce Schwarzian QM from Liouville CFT. Notice this double-scaling limit gets contributions from all energies, ending up with a \textit{quantum} theory \eqref{doubleJT} that structurally matches the JT gravity spectral form factor.

In order to achieve this regime in a microscopic 2d CFT, one must combine this scaling regime \eqref{eq:double-scale} with the vacuum character dominance of subsection \ref{sec:2duniv}: $\beta/\ell \ll \Delta_{\text{gap}}$. However, these conditions together seem impossible to satisfy when accommodating the bootstrap conjecture / result for the gap $\Delta_{\text{gap}} \leq c/12$ \cite{Hellerman:2009bu,Hartman:2019pcd}.\footnote{We thank Alexandre Belin for pointing this out.}
However, at the end of subsection \ref{sec:2duniv} we observed that the regime where the vacuum dominates the dual channel is larger than at first sight expected, since even if only $\Delta_{\text{gap}} \gtrapprox \beta/\ell$, contributions from other primaries are suppressed. In the end, this implies our result \eqref{doubleJT} is a good approximation for not-too-small JT temperatures $T_{\text{JT}}$ where $\beta_{\text{JT}} = \frac{b^2\beta}{\ell}$.\footnote{This argument is a bit finicky, so as a particular example, set $c=10000$, $\Delta_{\text{gap}} = c/12$, $\mu=0$ and $\beta_{\text{JT}}=1$. We get the ratio \eqref{eq:ratio} $\approx 5\cdot 10^{-6}$, meaning contributions from other primaries are indeed heavily suppressed. At the same time, one directly checks numerically that $\frac{\eqref{solidtorus}}{\eqref{doubleJT}} \approx 1.00005$, so the solid torus partition function is approximated very well (up to $0.005\%$) by the doubled JT partition function. This single datapoint then shows that there is a dynamic range of the parameters where the doubled JT approximation holds, assuming the spectral gap $\Delta_{\text{gap}}$ is not much smaller than its maximally allowed value $c/12$.} Turning on the chemical potential $\mu$, the condition is more precisely: $\Delta_{\text{gap}} \gtrapprox \beta(1+\mu^2)/\ell$. For real $\mu$, this leads to a shrinking of the validity window of the doubled JT regime. For imaginary $\mu$ on the other hand, the window of validity enlarges all the way up to the extremal case $\mu=i$ where the spectral gap condition would always be satisfied. This extremal situation is the one analyzed in \cite{Ghosh:2019rcj}. Our focus on the other hand is deliberately on the opposite regime.

When coming instead from our 3d pure gravity proposal of subsection \ref{s:propgrav}, other primaries (next to the vacuum) are absent altogether, and the doubled JT result \eqref{doubleJT} can be trusted for arbitrary small temperatures.

The expression \eqref{doubleJT} is an interesting result on its own right, which we expect to have implications for the calculation of other 3d observables in the regime \eqref{eq:double-scale}. Appendix \ref{app:bcor} elaborates on this remark. We emphasize that we did not tune the chemical potential for rotation $\mu$, which allows us to explore physics outside of the near-extremal regime considered in \cite{Ghosh:2019rcj}. \\

We complement the above discussion with an analysis of the bulk metric fluctuations accounting for the different states in this model. The low-temperature and large $c$ double-scaled limit \eqref{eq:double-scale} removes all descendants from the spectrum. Geometrically, this feature should be captured by removing all fluctuations along the angular $\varphi$-direction. Let us hence consider a general ansatz for a $\varphi$-independent metric
\begin{equation}
\label{mansatzgen}
    ds^2 = g_{\mu\nu}^{(2)}(x^\mu)dx^\mu dx^\nu + \Phi^2(x^\mu) (d\varphi + V_\nu(x^\mu)dx^\nu)^2, \qquad \mu,\nu= \tau_E,r, \quad \varphi \sim \varphi + 2\pi\,.
\end{equation}
Notice the latter still involves a one-form $V_\nu(x^\mu)dx^\nu$ which is capable of describing rotating configurations. In the second-order formulation of 3d gravity, this metric ansatz results in a 2d model of JT gravity coupled to a Maxwell term \cite{Strominger:1998yg}. For our purposes however, the first order formulation is far more direct to find a doubled JT partition function. Working in the Chern-Simons formulation of 3d gravity, this ansatz is equivalent to specifying the gauge connections $A = \frac{e}{\ell} + \omega$ and $\bar{A} = \frac{e}{\ell} - \omega$. The specific details will not be needed though. All we need is the property $\partial_\varphi A = \partial_\varphi \bar{A}= 0$. Denoting $B \equiv A_\varphi$, the resulting ansatz reduces a single Chern-Simons action into the BF action
\begin{equation}
\label{eq:csdimree}
\frac{\ell}{64\pi G_{\mt{N}}^{(3)}} \int d^3x \, \text{Tr} \left( A \wedge dA + \frac{2}{3} A \wedge A \wedge A\right) \,\, \to \,\,\frac{2\pi\ell}{32\pi G_{\mt{N}}^{(3)}} \int d^2x \, \text{Tr} \left( BF\right) + I_{\text{bdy}}.
\end{equation}
Projecting onto $\varphi$-independent fluctuations can be motivated from the CS perspective as follows. By conformal invariance the limit of large $\beta/\ell$ is equivalent to that of small angular $\varphi$ periodicity. Since the 3d topology is of the form $S^1 \times D^2$ where $S^1$ is the angular $\varphi$ direction (see Figure \ref{toruschannel}), this limit projects on the KK zero-mode in the angular $\varphi$-direction. In order for the resulting action not to vanish, we simultaneously need to scale the prefactor of the CS action $\sim \ell/G_{\mt{N}}^{(3)}\sim c$ to go to infinity in a double-scaled way. This is precisely encoded in \eqref{eq:double-scale}.
Thus, the full 3d gravity action functional becomes, setting $G_{\mt{N}}^{(3)} = 2 \pi \ell G_{\mt{N}}^{(2)}$:
\begin{equation}
\label{BFmBF}
\frac{1}{32\pi G_{\mt{N}}^{(2)}} \int d^2x \left[\text{Tr} \left( B F\right) - \text{Tr} \left( \bar{B}  \bar{F}\right)\right] + I_{\text{bdy}}.
\end{equation}
Notice this is the difference of two JT actions. Along a contour where $B \in i \mathbb{R}$, the result of these independent path integrations yield complex conjugate results, and directly lead to the doubled JT partition function of \eqref{doubleJT}.

To avoid confusion, we note that setting $\mu=0$ leads to a non-rotating BTZ saddle, but is not sufficient to remove all rotating states; it still leads to a doubled JT model, accommodating rotating off-shell black hole states.

Finally, even stronger, if we were to set $V_\nu=0$ in \eqref{mansatzgen} by hand, restricting to only non-rotating geometries, also off-shell, the Chern-Simons gauge fields have a specific form:
\begin{alignat}{3}
A_{\tau_E} d\tau_E + A_r dr = &&\, (e^1,e^2,\omega^{12}), \qquad
A_\varphi d\varphi &= (\omega^{23},0,e^3), \\
\bar{A}_{\tau_E} d\tau_E + \bar{A}_r dr = && \, (e^1,e^2,-\omega^{12}), \qquad
\bar{A}_\varphi d\varphi &= (-\omega^{23},0,e^3),
\end{alignat}
in terms of the dreibein $e$ and spin connection $\omega$. Both BF models in \eqref{BFmBF} are equal but opposite, and the total bulk action equals
\begin{equation}
\frac{1}{16\pi G_{\mt{N}}^{(2)}} \int d^2x \, \text{Tr} \left( B F\right) + I_{\text{bdy}},
\end{equation}
which is just the gauge-theoretical version of \eqref{eqdimred}.

We can further extend this doubled JT regime of 3d gravity to include boundary correlation functions. Whereas we believe this is an important discussion that gives strength to the current arguments, the methods used are somewhat orthogonal to the main text and we therefore defer it to Appendix \ref{app:bcor}.

\subsection{Summary}
We provided a proposal for an effective 3d quantum theory of pure gravity based on the universal high-temperature dynamics of 2d holographic CFTs. From such boundary perspective, this is a theory of ``vacuum Virasoro blocks in the dual channel'' with partition function given in \eqref{solidtorus}. Our effective model mimics similar claims in 2d JT gravity, but operates far away from extremality.  In fact, we identified a low temperature scaling limit \eqref{eq:double-scale}, governed by a doubled JT model, which is of independent interest. Our partition function \eqref{solidtorus} can be independently derived from gravitational path integral considerations leading to a geometric Alekseev-Shatashvili coadjoint orbit type-action \cite{Alekseev:1988ce,Alekseev:1990mp} by a choice of boundary conditions differing from the ones in \cite{Cotler:2018zff}. Because of the latter, our partition function has a single saddle point, for a fixed value of the chemical potentials, since our theory is not modular invariant. 

It is possible to extend our proposal with boundary operator insertions (see Appendix \ref{app:bcor}) and Wilson loop defects. 

The thermal entropy of the quantum black hole \eqref{eq:BHncl} matches the BTZ black hole entropy \eqref{eq:BHcl} in the semi-classical regime. In the remainder of this work, we will further investigate the structural similarities between JT gravity and 3d gravity on the one hand, and the structural differences between gauge theory and gravity on the other hand, with the specific goal of explaining
the black hole entropy in \eqref{eq:BHncl} as gravitational entanglement entropy, that is, understanding factorization of 3d gravity in terms of gravitational edge states.

\section{Factorization in gauge theory}
\label{sec:review}

Section \ref{sec:3dgrav_def} provided an effective model and Hilbert space for 3d gravity in which we can in principle study whether black hole entropy can be understood as gravitational entanglement entropy. However, gravity is a gauge theory and gauge theories are known to have physical Hilbert spaces that are non-factorizable. Since 2d JT gravity can be directly formulated as a 2d BF gauge theory, this section reviews how the issue of factorization can be handled in gauge theory, using 2d YM and BF theories as specific examples, before applying these tools to 2d JT and 3d gravity in the next section.

\subsection{Gauge theory and factorization}
\label{sec:algebra}

A defining feature of gauge theory is the presence of non-local degrees of freedom represented by Wilson lines.  The presence of Wilson lines passing through any entangling surface implies the physical Hilbert space $\mathcal{H}_{\text{physical}}$ does not factorize into subregions 
\begin{equation}
  \mathcal{H}_{\text{physical}} \neq \mathcal{H}_{\mt{V}} \otimes \mathcal{H}_{\bar{\mt{V}}}.
\end{equation}

However, factorization can be achieved by an extension of the Hilbert space. One introduces a regulator splitting the entangling surface $S$ into two pieces. This separates a Cauchy slice $\Sigma$ into two spatial subregions $V$ and $\bar{V}$, each with its own separate boundary. Gauge invariance requires the addition of \textit{edge mode} degrees of freedom on $\pd V$ and $\pd \bar{V}$ transforming under large gauge transformations which we will denote by $G_\mt{S}$.\footnote{Alternatively, one may quantize the theory for a given set of "boundary conditions" at $S$ and sum over all such possible choices, leading to a schematic decomposition 
\begin{equation}
  \mathcal{H}_{\text{physical}} = \bigoplus_\alpha \mathcal{H}_{\mt{V},\alpha} \otimes \mathcal{H}_{\bar{\mt{V}},\alpha}
\end{equation}
in superselection sectors. This perspective has been employed mainly in the literature on edge states in Maxwell theory \cite{Donnelly:2014fua,Donnelly:2015hxa,Blommaert:2018rsf}, with the superselection label $\alpha = E_\perp$. } Hence, edge modes carry charges under this group, allowing the construction of new Wilson lines ending on the boundary of each subregion. This is the physical mechanism underlying the factorization of the original Wilson line operators. We shall refer to $G_\mt{S}$ as the surface or edge mode symmetry.

After introducing the edge modes, $\mathcal{H}_{\text{physical}}$ can be embedded into a tensor product Hilbert space via the factorization map
\begin{align}\label{fact}
    i:\mathcal{H}_{\text{physical}} \hookrightarrow \mathcal{H}_{V}\otimes \mathcal{H}_{\bar{V}}.
\end{align}
The reduced density matrix on $V$ and its entanglement entropy is then defined in the extended Hilbert space $\mathcal{H}_{V}\otimes \mathcal{H}_{\bar{V}}$. The image of   $\mathcal{H}_{\text{physical}}$ under the factorization map is a  fusion product of $\mathcal{H}_{V}$ and $ \mathcal{H}_{\bar{V}}$  called the entangling product  \cite{Donnelly:2016auv}:
\begin{align}
    \mathcal{H}_{\text{physical}} \, \simeq \, \mathcal{H}_{V}\otimes_{\text{G}_{S}} \mathcal{H}_{\bar{V}},
\label{eq:fusion}
\end{align}
where $\otimes_{\text{G}_{S}}$ refers to a quotient by a diagonal action of the surface group symmetry $\text{G}_{S}$ on $\mathcal{H}_{V} \otimes \mathcal{H}_{\bar{V}}$:
 \begin{align} 
    \ket{v}\otimes \ket{w} \sim \ket{v \cdot g}\otimes \ket{g^{-1} \cdot w},\qquad \forall\,g \in \text{G}_{S}\,
 \end{align} 
 Here $v \cdot g$ denotes the right action of G$_{S}$ and similarly for $w$.  We will be more explicit below.
 
In gauge theories, the quotient can be interpreted as a projection onto the singlet sector of $\mathcal{H}_{V}\otimes \mathcal{H}_{\bar{V}}$ under $\text{G}_\mt{S}$, which imposes the Gauss' law constraint. In the math literature, the entangling product is referred to as the relative tensor product of modules, which are just Hilbert spaces equipped with an action of the symmetry group $\text{G}_{S}$.  This is the notion of factorization which we shall pursue in bulk 3d gravity.  In the next section, we illustrate these ideas in the context of two-dimensional gauge theories.

\subsubsection{Example: 2d gauge theory}
\label{sec:2d-gauge}

Consider a 2d gauge theory on an interval I$\,= \left\{x\in \mathbb{R}\, | \,x\in [x_1,\,x_2] \right\}$ ($\times \text{time}$), such as 2d Yang-Mills theory with action $S=\frac{1}{4g_\mt{YM}} \int_{\mathcal{M}} \text{Tr} F\wedge *F $, or the 2d BF model $S= - \int_{\mathcal{M}} \text{Tr} BF + \frac{1}{2} \oint_{\partial \mathcal{M}} B A_t$, both sharing the same Hilbert space structure. We focus on 2d BF theory from here on.

The interval has two endpoints where boundary conditions must be chosen.  For convenience, take the boundary condition $A_t-B\vert_{\partial \mathcal{M}}=0$ at both ends for 2d BF theory, although we will consider more general boundary conditions later. Applying Gauss' law in the bulk implies the physical Hilbert space $\mathcal{H}_{\text{physical}}$ consists of square integrable functions on the compact gauge group $\text{G}$
\begin{align}
    \mathcal{H}_{\text{physical}}= \text{L}^{2}(\text{G})\,.
\end{align}
A complete basis for this Hilbert space is provided by the Peter-Weyl theorem, which gives the spectral decomposition of $\text{L}^{2}(\text{G})$ in terms of the group representations $\mathcal{P}_{R}$:
\begin{align}\label{pw}
     \text{L}^{2}(\text{G}) = \bigoplus_{R}  \, \mathcal{P}_{R} \otimes \mathcal{P}_{R}^\star .
\end{align}
Here, the label $R$ ranges over all representations of $\text{G}$. This notation corresponds to the following interpretation: the left-right regular representation is a representation of $G \otimes G$ obtained by the action:
\begin{equation}
\label{lrregular}
f(g) \to f(h_L g h_R^{-1}), \qquad f \in \text{L}^{2}(\text{G}), \qquad (h_L,h_R) \in G \otimes G.
\end{equation}
It maps a square-integrable function $f$ into another one, and hence forms a representation of $G \otimes G$. Famously, decomposing this into irreducible representations gives precisely the above symmetric decomposition \eqref{pw} where all $\mathcal{P}_{R}$ appear precisely once.

The Peter-Weyl theorem is the simplest example of how information about a group can be reconstructed from its representations. For example, when $\text{G}= \U(1)$, its content just corresponds to the Fourier transform with modes $e^{i n \varphi}$, mapping the 1d representations, labeled by $n \in \mathbb{Z}$, to functions on $\U(1)$.  The classical statement of the Tannaka-Krein duality (see e.g. \cite{Chari:1994pz}) goes a bit farther to say that given the Clebsch-Gordan coefficients and the fusion rules for a set of representations of $\text{G}$, one can reconstruct the compact group $\text{G}$ itself.  A generalization of this duality to  quantum groups \cite{majid_1995} will be particularly useful when we consider surface symmetries for 3d gravity.  

Concretely, the Peter-Weyl theorem implies that a basis of states for the interval Hilbert space is given by the matrix elements in all irreducible representations $R$ of $\text{G}$
\begin{equation}
\label{eq:PW-2dYM}
  \left\{ |R,a,b\rangle\,, \quad a,b = 1,2, \dots \text{dim R}\right\}.
\end{equation}
Each normalized basis state $\ket{R,a,b}$ has a wave function on the group manifold given by 
\begin{equation}
  \langle g |R,a,b\rangle = \sqrt{\text{dim R}}\,R_{ab}(g) ,
\label{eq:G-wave}  
\end{equation}
where  $R_{ab}(g)$ are the matrix elements in the representation $R$ satisfying the orthogonality relation 
\begin{align}\label{ortho}
    \int d g\,   R_{ab}(g)  R^{*}_{cd}(g) =  \frac{V_\text{G}}{\text{dim R}}\, \delta_{bd} \delta_{ac}\,,
\end{align}
where $V_G$ is the volume of the gauge group G. In the context of JT gravity, the gauge group is non-compact and the set of representations of interest will be continuous. The above discussion still applies, with the normalization $(\dim R)$ on the right hand side defining the Plancherel measure $d\mu(R)= (\dim R)\, dR $. This can be viewed as a regularized dimension for the irreducible representation $R$, and gives rise to the continuous completeness relation 
\begin{align}
  \delta(g-g') = \sum_{ab} \int d\mu (R)\,\, R_{ab}(g)R^{*}_{ab}(g')\,.
\end{align}
The indices $a$ and $b$ may or may not be continuous as well. This depends on the particular choice of basis made within each representation.

\paragraph{Factorization and fusion.}
Let us characterize the edge modes and their transformation properties under the surface symmetry $\text{G}_{\mt{S}}$ in more detail. At the boundary of a spatial region, would-be gauge transformations are promoted to physical symmetries. We can view the ungauged physical degrees of freedom as the edge modes of the physical boundary.\footnote{This is equivalent to adding Stueckelberg fields.}  In the case of an interval $\text{I}$ with boundary condition $A_t-B\vert_{\partial \mathcal{M}}=0$ at each end, these large gauge transformations are just a copy of the gauge group G at each endpoint, which we identify with the surface symmetry $\text{G}_{\mt{S}}$. They act on the states in the interval by left and right multiplication at the left and right endpoint, respectively. In terms of the group basis $\ket{g}$, this action is given by
\begin{equation}\label{LR}
  \ket{g} \to \ket{h_{L}^{-1}g}\,, \quad\quad \ket{g} \to \ket{g h_{R}}\,.
\end{equation}

Let us split the original Cauchy surface (interval) into $V=[x_1,\,y-\epsilon]$ and  $\bar{V}=[y+ \epsilon ,\,x_2] $. If we use the $A_t-B\vert_{\partial \mathcal{M}}=0$ boundary condition at the regulated entangling surface, then the extended Hilbert space on the two intervals is given by $\text{L}^{2}(\text{G}) \otimes \text{L}^{2}(\text{G})$. The surface symmetry at the split entangling surface is $\text{G}_{\mt{S}}=\text{G}$, acting by right multiplication on $V$ and vice versa for $\bar{V} $.  The next step is to define a factorization map 
\eqref{fact}. In this case, there is a natural choice arising from the structure of $\text{L}^{2}(\text{G})$ as a Hopf algebra. This means that in addition to the pointwise multiplication rule for functions,  $\text{L}^{2}(\text{G})$ also has a co-multiplication:
\begin{align}
i: \text{L}^{2}(G) &\to \text{L}^{2}(G)\otimes \text{L}^{2}(G), \nn
i \ket{g} &=\frac{1}{\sqrt{V_\text{G}}} \int_{G} dg_{1}dg_{2}\, \delta({g_1\cdot g_2,g})\ket{g_{1}}\otimes \ket{g_{2} } ,
\label{eq:Ifact-2}  
\end{align}
which defines the factorization map.  This map is an isometry because it has an adjoint  $i^*\left(\ket{g_{1}}\otimes \ket{g_{2}}\right)= \frac{1}{\sqrt{V_\text{G}}}\ket{g_{1} g_{2}} $  that fuses back the split intervals in the sense that $i^* \circ i=\mathbb{1}$, where one uses left- or right-invariance of the Haar measure to prove this statement.

Within $\text{L}^{2}(\text{G})\otimes \text{L}^{2}(\text{G}) $, the physical Hilbert space is recovered by taking a quotient with respect to the diagonal action of $\text{G}_{\mt{S}}=\text{G}$:
\begin{align}
    \ket{g_{1}} \to \ket{ g_{1} h}, \qquad 
    \ket{g_{2}} \to \ket{ h^{-1} g_{2}}.
\end{align}
This quotient corresponds to the entangling product  \eqref{eq:fusion}, which is isomorphic to the original Hilbert space:
\begin{align}\label{eq:entp}
  \mathcal{H}_\text{physical} \simeq \text{L}^{2}(\text{G}) \otimes_{\text{G}} \text{L}^{2}(\text{G}).
  \end{align}
Explicitly, elements of the quotient are equivalence classes $ 
 ( \ket{g_{1}}, \ket{g_{2}})  \sim (\ket{g_{1} h}, \ket{h^{-1} g_{2}} ) $.
 In the context of 3d gravity, $\text{G}_{\mt{S}}$ is deformed to a quantum group, which also has a co-product that we will use to define a factorization map in the same way.

 \paragraph{Wilson lines and Factorization in the representation basis.} The representation basis may provide a more intuitive description of the Hilbert space factorization.  In this basis, the factorized wavefunctions are  obtained from pulling back $R_{ab}(g)$ through the group multiplication map $G \times G \to G$, which is dual to the co-product \eqref{eq:Ifact-2}
\begin{align}
\label{eq:lfact} 
    i :  \text{L}^{2}(G) &\to \text{L}^{2}(G) \otimes \text{L}^{2}(G), \\
    \langle g|R,a,b\rangle &\to \frac{1}{\sqrt{V_\text{G}}}\langle g_1\cdot g_2| R,a,b\rangle = \frac{\sqrt{\text{dim R}}}{\sqrt{V_\text{G}}}\,R_{ab}(g_1\cdot g_2) = \frac{1}{\sqrt{V_\text{G}}\sqrt{\text{dim R}}}\,\sum_c \langle g_1|R,a,c\rangle \langle g_2|R,c,b\rangle, \nonumber   
\end{align}
where we used the defining property of the representation matrix $R_{ab}(g_1\cdot g_2) = \sum_c\,R_{ac}(g_1)R_{cb}(g_2)$. 
In the representation basis, the entanglement edge modes are labeled by the right index of the wavefunctions on the left interval and vice versa on the right interval.  \eqref{eq:lfact} explicitly shows how these edge modes (labeled by $c$) are entangled in the physical state. We can give a more local picture of factorization by 
expressing the group elements $g\in \text{G}$ as Wilson lines of the original connection $A_{\mu}(x)$:
\begin{align}
    R_{ab}(g) = \text{P} \exp \left(i\,\int_{I} A\right)_{ab}\,.
\end{align}
Then the factorization map \eqref{eq:lfact}, just corresponds to splitting the Wilson line in each representation $R$ of the gauge group: 
\begin{align}
\text{P} \exp \left(i\,\int_{I} A\right)_{ab}  = \sum_c \text{P} \exp \left(i\,\int_{V} A\right) _{ac} \text{P} \exp \left(i\,\int_{\bar{V}} A\right)_{cb}. 
\end{align}

\subsection{The shrinkable boundary condition}
\label{sec:path-integral}

Not all factorization maps \eqref{fact} define a meaningful notion of entanglement.  In the absence of constraints, the embedding of $\mathcal{H}_{\text{physical}}$ into an extended Hilbert space can have arbitrary edge mode degeneracies leading to an arbitrary amount of entanglement.  Arguably, \textit{locality} is the most natural and important physical constraint to be imposed.

\paragraph{Stretched entangling surface and the shrinkability condition.}
Locality can be imposed by defining the factorization map via a Euclidean path integral that splits a Cauchy slice into two, as shown in the figure below. 
\begin{align}\label{split}
\begin{tikzpicture}[scale=0.6, baseline={([yshift=-0.1cm]current bounding box.center)}]
\draw[thick,->] (0,1.5) -- (0,0);
\node at (-1,0.3) {\small time};
\end{tikzpicture} \hspace{0.2cm}
   \mathtikz{\deltaA{0cm}{0cm} ;\draw (0cm,-.9 cm) node {\footnotesize $e$ };\draw (.5 cm,-1.2 cm) node {$V$ };\draw (-.5 cm,-1.2 cm) node {$\bar{V}$ }}.   
\end{align}
 
Mathematically, the spacetime process described by \eqref{split} is a cobordism, i.e. a manifold interpolating between an ingoing and an outgoing boundary. This cobordism introduces a \emph{stretched} entangling surface $S_{\epsilon}$, which is the surface traced out by the Euclidean modular evolution of the entanglement boundary for $V$ (or $\bar{V}$).  This leads to a codimension-1 boundary placed at a distance $\epsilon$ from the original entangling surface $S$; this is depicted by the semi-circular arc connecting the entanglement boundaries for $V$ and $\bar{V}$ in \eqref{split}.

To define the corresponding path integral, an entanglement boundary condition on $S_{\epsilon}$ must be chosen. This is denoted by $e$ in \eqref{split}.  Since $S_{\epsilon}$ is not a physical boundary, we choose $e$ to preserve the cross-boundary correlations of the initial  non-factorized state, in the limit $\epsilon \to 0$. This constraint is captured by the \textit{shrinkable} boundary condition, which says that 
\begin{align}\label{shrink}
\lim_{\epsilon \to 0}
    \mathtikz { \etaA{0cm}{0cm}\deltaA{0cm}{0cm} \draw (0cm,-1cm) node {\footnotesize $e$ } ;\muA{0cm}{-1cm}; \epsilonA{0cm}{-2cm};\draw (.5 cm,-1.2 cm) node {$V$ }} =\lim_{\epsilon \to 0} \mathtikz { \copairA{0cm}{0cm} \pairA{0}{0} \draw (0cm,0cm) node {\footnotesize $e$ }; \draw (.5 cm,-0.2 cm) node {$V$ }} = \mathtikz{\etaA{0cm}{0cm} \epsilonA{0cm}{0cm}}
\end{align}
The right hand side is the disk path integral that computes the norm squared of the Hartle-Hawking state, defined by cutting the geometry at a moment of time reflection symmetry. Equation \eqref{shrink} says that when $\epsilon \to 0$, the disk path integral is equal to the annulus path integral arising from sequentially splitting and fusing the Cauchy slice. In the context of an entanglement calculation, the annulus is viewed as the trace of a reduced density matrix $\rho_{\mt{V}}$ supported on the subregion Hilbert space $\mathcal{H}_{\mt{V}}$. Hence, equation \eqref{shrink} matches the path integral description of the defining relation:
\begin{align}
    Z_{\text{Disk}} = \tr_{\mt{V}} \rho_{\mt{V}}\,.
\end{align} 
Notice the shrinkable boundary condition $e$ provides a statistical interpretation to the path integral $Z_{\text{Disk}}$ on a spacetime with a shrinking (interior) thermal circle.

While the stretched entangling surface is naively of codimension one, the gluing of the two subregions $V$ and $\bar{V}$ really occurs at a codimension-2 surface, a fact which becomes manifest in the shrinking limit $\epsilon \to 0$.  It is important to distinguish this from the gluing of spacetimes along a proper codimension-1 surface. In Lorentzian signature, modular flow would evolve $\pd V$ and $\pd \bar{V}$ into a stretched Rindler horizon which approaches an infinite redshift surface as $\epsilon \to 0$. This implies the edge modes surviving the $\epsilon \to 0$ limit must have zero modular frequency, a key feature that is absent when gluing along a codimension-1 surface. As discussed in \cite{Blommaert:2018oue}, the static nature of the edge modes in 4d Maxwell theory is also crucial for the (subregion) state counting interpretation of the Euclidean path integral.

\paragraph{Shrinkability as a completeness relation.}
Shrinkability can be viewed as a completeness relation which states that summing over a complete set of edge modes in a subregion $V$ recovers the path integral on the contractible spacetime. Crucially, this condition determines both the symmetry $\text{G}_{\mt{S}}$ and the required spectrum of representations that closes up the stretched entangling surface.

It is instructive to remember how this works in some of the understood theories. For example, in 2d BF theory and Chern-Simons theory, the shrinkable boundary condition sets to zero both the component $A_{\tau_E}-A_{\varphi}$ of the gauge field and $\text{P}\exp \oint d\tau_E A_{\tau_E} = 1$ along the $\tau_E$ direction of the stretched entangling surface, allowing us to shrink it down to a point.\footnote{In the 2d BF theory case, it should be understood that $A_{\varphi}=B$ by dimensional reduction of the 3d Chern-Simons action, as discussed around equation \eqref{eq:csdimree}.}
\begin{itemize}
\item In 2d BF gauge theories, quantization satisfying this boundary condition leads to edge modes transforming in all representations of the gauge group $\text{G}$.  We can see this by applying the shrinkable boundary condition to the sphere:
\begin{align}
       \mathtikz{\epsilonCv{0}{0}\etaCv{0}{0}}&=
         \mathtikz{\pairAv{0}{0} \copairA{0}{0}}\label{ZS2}
\end{align}
 This implies that  $\sum_{R} (\dim R)^{2} e^{-4\pi \epsilon C(R)} = \tr_{V} \rho_{V}$.
 To reproduce the degeneracy $(\dim R)^{2}$ factor on the left hand side, we need a subregion Hilbert space with $(\dim R)^{2}$ edge modes for each representation $R$. This gives a physical motivation for the Peter-Weyl theorem: the shrinkable boundary condition 
 \begin{equation}
 A_{\tau_E}-B\vert_{\partial\mathcal{M}}=0, \qquad \text{P}\exp \oint d\tau_E A_{\tau_E} = 1,
  \end{equation}
  gives $\text{L}^{2}(\text{G})$
 as the subregion Hilbert space, while satisfying \eqref{ZS2} requires this Hilbert space to have $(\dim R)^{2}$ degenerate states for each $R$, as stated by the Peter-Weyl theorem applied to this discussion (see \eqref{eq:PW-2dYM}). The large gauge transformations are given by the gauge group $G$, which is identified with the surface symmetry $G_{\mt{S}}$.  The path integral factorization map \eqref{split} agrees with the co-product defined in \eqref{eq:lfact}.

\item In Chern-Simons theory with compact gauge group $\text{G}$, the shrinkable boundary condition gives rise to CFT edge modes \cite{Wong:2017pdm}. In this case, the entangling surface is a circle, so the surface symmetry group $\text{G}_{\mt{S}}$ is the loop group of $\text{G}$ since these are the large gauge transformations.  The reason for the appearance of these CFT edge modes is exactly the same as the usual holographic duality of Chern-Simons theory, where the stretched entangling surface is treated as the holographic boundary of a subregion.  
\end{itemize}

\subsection{Towards factorization in gravity}

Any attempt to borrow the lessons from the 2d gauge theory factorization to 3d gravity requires considering, at least, two extra problems:
\begin{enumerate}
\item if one succeeds in defining a Hilbert space for a bulk subregion, one expects to deal with different sets of boundary conditions at different boundaries: Brown-Henneaux boundary conditions at the holographic boundary and the shrinkable boundary condition at the entangling surface.  
\item both 2d JT and 3d gravity involves \textit{non-compact} gauge groups, whose representation theory is more complex.
\end{enumerate}

The first problem, and still within a 2d gauge theory context, is handled by considering other boundary conditions, labeled by a subgroup $H$.  This corresponds to setting the gauge field along $H$ to zero at the (right) boundary.  The interval $\hspace{-0.25cm}\mathtikz{\Int{0}{0};\node at (-.35,.25) {\small $e$};\node at (.37,.28)  {\small $H$}}\hspace{-0.25cm}$ is then assigned to the Hilbert space 
\begin{align} 
\mathcal{H}_{e H}= \bigoplus_{R} \mathcal{P}_{R}\otimes \mathcal{P}_{R,0},  
\end{align} 
where $\mathcal{P}_{R,0}$ denotes a projection of $ \mathcal{P}_{R}$ onto a state invariant under the right action of the subgroup $H$. The Peter-Weyl theorem implies that this projection gives the space of $\text{L}^2$-functions on the coset: 
\begin{align}\label{coset}
\text{L}^{2}(\text{G/H})= \bigoplus_{R} \mathcal{P}_{R}\otimes \mathcal{P}_{R,0}.
\end{align}
This coset decomposition will be relevant for gravity when assigning a Hilbert space to the subregion between the entanglement surface and the holographic boundary.

Regarding the second problem, and keeping the discussion general for now, one expects the unitary tempered irreducible representations for non-compact groups to play a relevant role. Since one important lesson to extract from the 2d gauge theory examples is that the shrinkable boundary condition requires not only the knowledge of a symmetry group, but also the spectrum of representations appearing in the Peter-Weyl decomposition, we would expect to need to specify their Plancherel measure $d \mu (R)$ for non-compact groups. That is, the shrinkable boundary condition would imply a Peter-Weyl theorem taking the form
\begin{align}
\text{L}^{2}(\text{G}) = \int_{\oplus_{R} }   \, d \mu (R) \, \mathcal{P}_{R} \otimes \mathcal{P}_{R}^{*}\,,
\end{align}
where we view $\text{G}$ as a configuration space variable in a bulk subregion, and with an analogous expression to \eqref{coset} for the coset discussion. We will provide explicit expressions of this kind for both 2d JT and 3d gravity in section \ref{sec:facgrav}.

\section{Factorization in gravity} 
\label{sec:facgrav}

Let us next apply the ideas reviewed in section \ref{sec:review} to gravity. We will first provide a general discussion, stressing the differences between gauge theory and gravity when attempting to implement the shrinkable boundary condition. Then, we shall frame the known JT story in the language developed in section \ref{sec:review}, to finally present new results on 3d pure gravity within this framework.

\subsection{Motivation: shrinkable boundary condition in gravity}
\label{sec:sh_gravity}
In essence, the \emph{shrinkable boundary condition} allows us to interpret gravitational entropy as bulk entanglement entropy. In the Euclidean computation of Gibbons and Hawking \cite{Gibbons:1976ue}, gravitational entropy is defined via the semiclassical evaluation of
\begin{align}\label{Sbeta}
    S=( 1-\beta \pd_{\beta})|_{\beta_{H}} \log Z(\beta),
\end{align}
where $Z(\beta)$ is an Euclidean path integral on a spacetime with a thermal circle of length $\beta$ at infinity, and $\beta_{H}$ is the Hawking temperature.   The black hole entropy arises from the tree level evaluation of $\log Z(\beta)$ on a Euclidean saddle with the topology of a cigar in the directions transverse to the horizon. Cutting out a small disk from the tip of the cigar and inserting a shrinkable boundary condition there, our discussion in section \ref{sec:path-integral} suggests the resulting annulus path integral can be viewed as a trace over a subregion Hilbert space:
\begin{align} \label{Ztr}
    Z(\beta) = \tr_{\mt{V}} e^{-\beta H_{\mt{V}}},
\end{align}
where $H_{\mt{V}}$ generates rotations around the entangling surface, and is identified with the modular Hamiltonian for the Hartle-Hawking state defined by cutting the cigar in half.   In Lorentzian signature, this corresponds to a boost in the spacetime exterior to the black hole.  The gravitational entropy \eqref{Sbeta} can then be written as an entanglement entropy
\begin{align}
    S= -\tr \rho_{\mt{V}} \log \rho_{\mt{V}} ,\qquad \rho_{\mt{V}}= e^{-\beta H_{\mt{V}}}.
\end{align}
This naive argument is insufficient in gravity, because the shrinkable boundary condition is non-local along the Euclidean time circle  \cite{Jafferis:2019wkd}.   
The reason is that in contrast with the analogous QFT computation, where the temperature variation in \eqref{Sbeta} would insert a conical singularity at the entangling surface, the gravity calculation only involves saddle point geometries which are smooth in the bulk.   To ensure there is no conical singularity at the entangling surface,  the shrinkable boundary condition (at arbitrary $\beta$) must fix the conical angle around the tip to be $2\pi$.  This constraint cannot be imposed by a local boundary condition, since it requires the integral of the spin connection around the stretched entangling surface to be $2 \pi$.\footnote{This is clear in 2d, since the metric around the conical singularity is locally of the form $ds^2 = dr^2 + r^2 d\theta^2$, where $\theta \sim \theta + 2 \pi (1-\alpha)$. This metric has a spin connection $\omega = d\theta$, so that we get $\oint_{S^1} \omega = 2\pi (1-\alpha)$. Avoiding a conical singularity ($\alpha=0$) hence means indeed the advertized property. The same calculation applies in higher dimensions, if we restrict to a normal plane transverse to a point on the codimension-2 entangling surface.} 

Following \cite{Jafferis:2019wkd}, one approach is to quantize the subregion theory on $V$  using a local boundary condition. A gravity path integral with such a local boundary condition defines a naive ``cutting map" $\mathcal{J}:\mathcal{H}_{\text{physical}} \to \mathcal{H}_{V} \otimes \mathcal{H}_{\bar{V}}$, which must then be supplemented with a defect operator $\sqrt{D}$ on $\mathcal{H}_{V}$ that implements the $2\pi$  cone angle.\footnote{In the BF formulation of JT gravity, the defect operator imposes the constraint that the Chern class of the gauge bundle is equal to one; in the gravity language, this corresponds to the fact that the boundary Schwarzian has winding number one around the boundary circle.} The correct embedding of the Hartle-Hawking state into the extended Hilbert space is then given by
\begin{align}\label{sdj}
    i = \sqrt{D} \mathcal{J} .
\end{align}
This approach has the advantage of making a direct connection to the horizon area operator, since this generates translations in the conical angle around the entangling surface-we will comment more on this in the conclusion. However, it does not give an interpretation of the black hole entropy in terms of entanglement entropy of edge modes.  This is because the shrinkable condition in the presence of the defect operator means
\begin{align}\label{TrD}
    Z(\beta)= \tr ( D\,e^{- \beta H_{\mt{V}} }), \qquad   S= \tr(\log D\rho_{V})  - \tr \rho_{V} \log \rho_{V},
\end{align}
where $H_{\mt{V}}$ is the modular Hamiltonian defined with the naive local boundary condition, and where $\rho_{V}= D e^{-\beta H_{V}}$ is the reduced density matrix defined by the factorization map \eqref{sdj}.\footnote{Equivalently, if we define $\tilde{\rho}_{V} = e^{-\beta H_{V}}$ , we can write the entropy as
\begin{align}
    S= -\tr D \tilde{\rho}_{V} 
    \log \tilde{\rho}_{V}.
\end{align}
Here $\tilde{\rho}_{V} $ is a reduced density matrix obtained from a factorization map with a local boundary condition.}
The defect insertion in \eqref{TrD} modifies the trace in order to obtain the appropriate density of states.
The black hole entropy is then given by the first term, i.e. the expectation value of the defect operator, and does not have a manifest state counting interpretation. 

In the next two sections, we take a more abstract approach. Rather than insisting on a local boundary condition which must be supplemented by a defect insertion, we simply apply the shrinkable boundary condition criteria to identify the appropriate gravitational edge modes.

\subsection{Bulk factorization and entropy in 2d JT gravity} 
\label{sec:JT}

We review and expand the work of \cite{Blommaert:2018iqz}, which we interpret as giving a factorization of the  JT gravity Hilbert space into a  tensor product 
$\mathcal{H}_{V}\otimes_{ G_{S}}\mathcal{H}_{\bar{V}}$ of modules.\footnote{In the mathematical definition of the relative tensor product of modules $\mathcal{H}_{V}\otimes_{G_{S}}\mathcal{H}_{\bar{V}}$, $G_{S}$ is usually taken to be the group algebra, which just means we allow for the addition of group elements. } This factorization was defined using the formulation of JT gravity as an ``$\SL^+(2,\mathbb{R})$ BF gauge theory", and led to a derivation of black hole entropy as entanglement entropy. Our main goal is to learn the subtleties of factorization in the gravitational context, before generalizing to the more difficult problem in the 3d pure gravity context in the next subsection.

\begin{figure}[h]
\centering
\includegraphics[scale=.4]{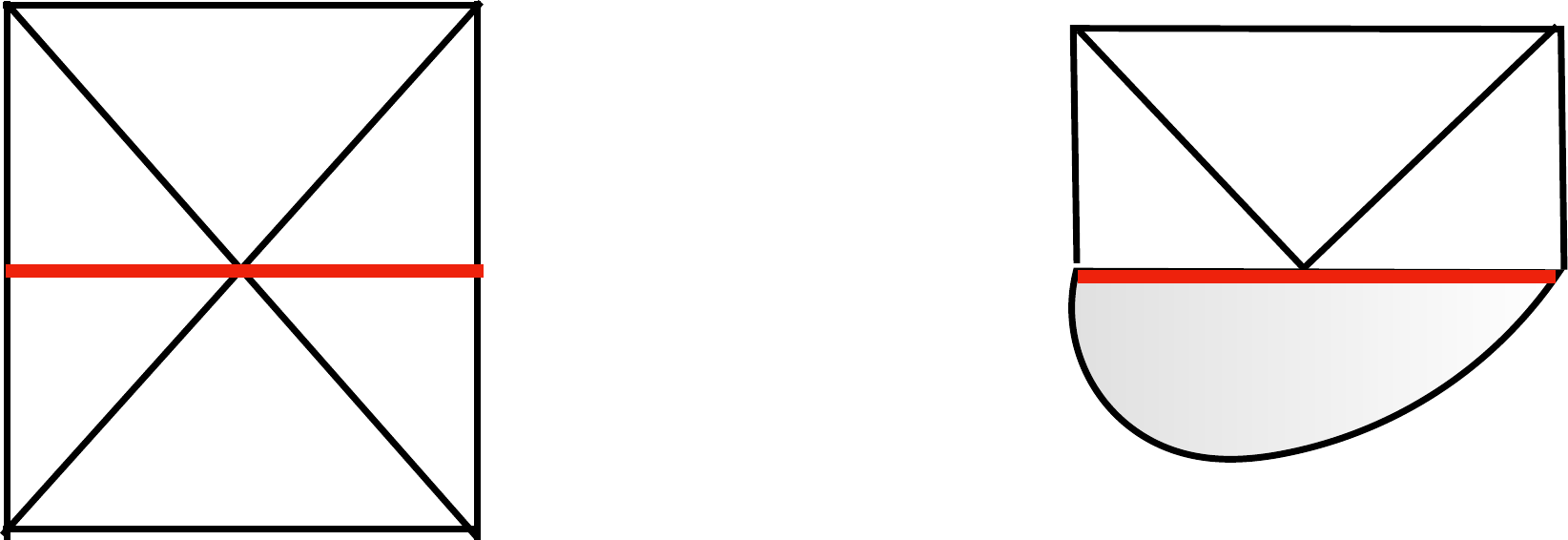}
\caption{The left figure shows the Penrose diagram for the two-sided AdS$_2$ geometry, with the red time slice denoting the wormhole. On the right, we show the Lorentzian evolution of the resulting Hartle-Hawking state.}
\label{BTZ}
\end{figure}

\paragraph{Shrinkable boundary condition and the edge mode density of states.}
Consider the Lorentzian two-sided AdS$_2$ geometry, as shown in the left of Figure \ref{BTZ}.

The time slice on which the bulk Hilbert space is supported is a wormhole connecting the two asymptotic boundaries, corresponding to the red interval in Figure \ref{BTZ}.  We denote this two-sided Hilbert space by $\mathcal{H}_{\mathfrak{i}_{L} \mathfrak{i}_{R}}$, with $\mathfrak{i}_{L}, \mathfrak{i}_{R}$ labeling the asymptotic boundaries that satisfy holographic gravitational boundary conditions.  
A basis for this ``interval" Hilbert space is given by the energy eigenstates
\begin{align}\label{kii}
    \ket{k\, \mathfrak{i}_{L}\,\mathfrak{i}_{R}},\quad k\in \mathbb{R}^+\, ,
\end{align}
where $k$ is a momentum variable related to the energy as $E=k^2$ in units where the Schwarzian coupling coefficient $C=1/2$ \cite{Jensen:2016pah,Maldacena:2016upp,Engelsoy:2016xyb,Stanford:2017thb}.

The momentum $k$ can also be viewed as labeling the representation of a Wilson line threading the wormhole and ending on the two asymptotic boundaries. Indeed, as in 2d YM or BF theory, the corresponding wavefunctions have a group theoretic interpretation as representation matrix elements of the underlying group structure, in our case the semi-group $\SL^+(2,\mathbb{R})$: 
\begin{align}
\label{eq:nn}
     \braket{g|k, \mathfrak{i}_{L}\mathfrak{i}_{R}}= \sqrt{ k\sinh 2\pi k } \,\,R^{k}_{\mathfrak{i}_{L}\mathfrak{i}_{R} }(g),\quad g \in \SL^+(2,\mathbb{R}),
 \end{align}
with $k$ labeling a representation in the continuous principal series and $R^{k}_{ab}(g)$ a representation matrix.  The fixed indices $\mathfrak{i}_{L}\,\mathfrak{i}_{R}$ are associated to the left- respectively right boundary, and are fixed by asymptotic gravitational boundary conditions which directly descend from the 3d Brown-Henneaux boundary conditions \cite{Brown:1986nw,Bershadsky:1989mf}. These are in essence coset boundary conditions described below equation \eqref{coset} in terms of the parabolic generators of $\SL(2,\mathbb{R})$. However, instead of being invariant under left- and right-multiplication by these parabolic one-parameter subgroups of $\SL(2,\mathbb{R})$, the wavefunction transforms by an irrelevant factor. We will nevertheless refer to these as coset boundary conditions. We refer to \cite{Blommaert:2018iqz} for more details on this gravitational coset. The normalization factor of \eqref{eq:nn} is important, and is explained below. \\

Our goal is to define a factorization map into the extended Hilbert space:
\begin{align}\label{jtfact}
    \mathcal{H}_{\mathfrak{i}_{L} \mathfrak{i}_{R}} \hookrightarrow \mathcal{H}_{\mathfrak{i}_{L} e}\otimes \mathcal{H}_{ e\mathfrak{i}_{R}},
\end{align}
where $\mathcal{H}_{\mathfrak{i}_{L} e} $ and $ \mathcal{H}_{ e\mathfrak{i}_{R}}$ denote one-side Hilbert spaces supported on "intervals" with one endpoint at asymptotic infinity, and the other on the bulk entangling surface. There is no assumption about the locality of the boundary condition at the entangling surface.

Let us appeal to the shrinkable boundary condition to determine the subregion Hilbert spaces and the factorization map \eqref{jtfact}. Consider therefore the Hartle-Hawking state 
\begin{align}
\label{eq:HH}
   \mathtikz{\etaA{0}{0}}=  \ket{\text{HH}_\beta}=  \int_{0}^{+\infty}  d k\,  \sqrt{k \sinh 2 \pi k}\,e^{-\frac{\beta C(k)}{2}} \ket{k\, \mathfrak{i}_{L}\,\mathfrak{i}_{R}}\,, \quad \quad \text{with} \quad
    C(k)= k^2\, ,
\end{align}
which can be prepared by the Euclidean path integral on a half disk with boundary length $\beta/2$. The Hartle-Hawking state $\ket{\text{HH}_\beta} \in \mathcal{H}_{\mathfrak{i}_{L} \mathfrak{i}_{R}}$ is defined to satisfy
\begin{align}
    Z_{\text{disk} }(\beta) = \mathtikz{\etaA{0cm}{0cm} \epsilonA{0cm}{0cm}} \equiv \braket{\text{HH}_\beta|\text{HH}_\beta},
\end{align}
where the disk partition function is:
\begin{equation}
    Z_{\text{disk} }(\beta) = \int_{0}^{+\infty}  d k\,(k \sinh 2 \pi k)\, e^{-\beta C(k)}.
\label{eq:JT-disk}
\end{equation}

Our desired factorization map $i$, when applied to $\ket{\text{HH}_\beta}$, produces a half annulus. 
\begin{align}
    \mathtikz{ \etaA{0}{0}   } \to   \mathtikz{ \etaA{0}{0} \deltaA{0}{0}  ;\draw (0cm,-1cm) node {\footnotesize $e$} }
\end{align} 
The shrinkability condition then says that $ Z_{\text{disk} }(\beta)$ is equal to the  $\epsilon \to 0$ limit of the full annulus:
\begin{align}\label{ShrinkableBC}
\mathtikz{\etaAred{0cm}{0cm} \epsilonAred{0cm}{0cm}}&= \lim_{\epsilon \to 0} \mathtikz { \copairAred{0cm}{0cm} \pairAred{0}{0} \draw (0cm,0cm) node {\footnotesize $e$ }; \draw (.5 cm,-0.2 cm) node {$V$ }}
\end{align}

\begin{figure}[h]
\centering
\includegraphics[width=0.4\textwidth]{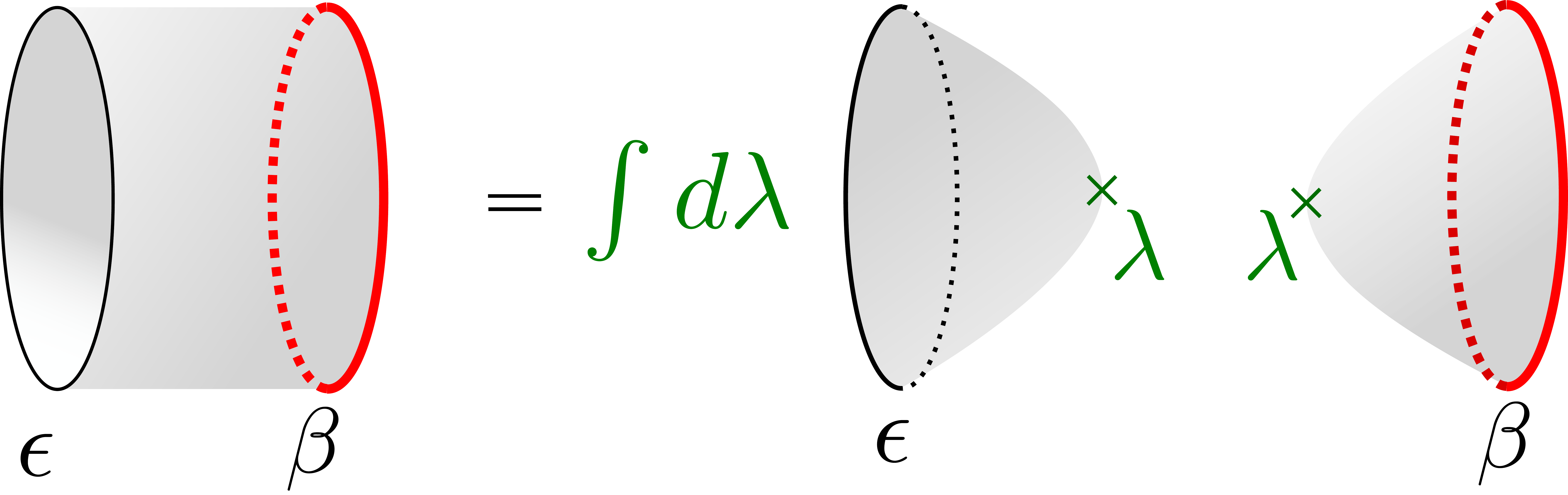}
\caption{At finite $\epsilon$, the two-boundary amplitude in BF or YM theory can be computed by gluing defect amplitudes together. The holographic boundary is drawn in red color.}
\label{DecompAnn}
\end{figure}

When $\epsilon$ is finite, the annulus path integral can be computed in the ``closed string" channel as a two boundary amplitude. In order to appreciate where the density of states is coming from, it is convenient to insert a complete set of intermediate states $\mathbb{1} = \int d \lambda \ket{\lambda}\bra{\lambda}$  which can be equivalently viewed as defect insertions as shown in the right of Figure \ref{DecompAnn}.

This decomposition can then be written as
\begin{equation}
  Z(\epsilon ,\beta) =\int d\lambda\,Z_\mt{inner}(\epsilon,\lambda)\,Z_\mt{outer}(\beta,\lambda)\,.
\label{eq:gluingJT}
\end{equation}
Each building block partition function has now only a single boundary, of either length $\epsilon$ or $\beta$, respectively. The parameter $\lambda$ can be interpreted as the geodesic length of the gluing circle, but this interpretation will not be needed here. Importantly, such single-boundary BF models can be described on their respective boundaries in terms of quantum mechanics on the group manifold of interest.

Let's compute the different blocks in detail. In Figure \ref{DecompAnn}, the red exterior boundary is at asymptotic infinity, and satisfies asymptotic gravitational boundary conditions. Thus, this ``outer" path integral produces a partition function given by \cite{Mertens:2019tcm,Stanford:2017thb}:
\begin{equation}
    Z_\mt{outer}(\beta,\lambda)\equiv \int_{0}^{+\infty} dk\, \cos (2 \pi \lambda k)\, e^{-\beta k^2}\,,
\label{Zouter}
\end{equation}
where $\cos (2 \pi \lambda k)=\braket{\lambda|k}$ represents the wavefunction of a boundary state $\ket{\lambda}$ whereas the wavefunction of the asymptotic boundary state is a constant (independent of $k$). 
On the other hand, the inner path integral $Z_{\text{inner}}$ is an amplitude between $\ket{\lambda}$ and the \emph{entanglement boundary state} $\ket{e}$, which ``caps off" the spacetime when $\epsilon \to 0$.  This takes the form 
\begin{equation}
   Z_\mt{inner}(\epsilon,\lambda)\equiv \braket{e|\exp^{-H_{\text{closed}}}|\lambda} \nn
   =\int_{0}^{+\infty} dk\, \braket{e|k}\, \cos (2\pi \lambda k)\, e^{-\epsilon k^2}.
\label{eq:edge}
\end{equation}
Gluing the two partition functions together by integrating over $\lambda$ in \eqref{eq:gluingJT} then gives
\begin{align}
    Z(\epsilon, \beta)= \int_{0}^{+\infty} dk \braket{e|k} e^{-(\epsilon +\beta) k^{2} }.
\end{align}
The shrinkability condition $Z_{\text{disk}}= \lim_{\epsilon \to 0}  Z(\epsilon, \beta)$ then implies that $\braket{e|k}= k \sinh 2 \pi k$. We thus conclude that the partition function for the inner disk is
\begin{align}
\label{Zinner}
    Z_{\text{inner}}(\epsilon, \lambda)= \int_0^{+\infty} d k \, (k \sinh 2 \pi k) 
\cos (2 \pi \lambda k)\,e^{-\epsilon k^{2}}.
\end{align}
Comparing \eqref{Zinner} with \eqref{Zouter}, we can infer that $(k \sinh 2 \pi k)$ plays the role of a density of states originating from the inner entangling boundary, since the $\cos(2\pi \lambda k)$ factor is here merely for the purposes of the gluing. Such density counts the (modular) zero-energy edge modes at fixed $k$ localized at the entangling surface.

A technical comment is in order concerning our gluing procedure \eqref{eq:gluingJT}. The actual gluing integral one should use both in BF models and in JT gravity, contains a twist variable as well. In the case of Teichm\"uller space, the range of the twist variable is infinite and global diffeomorphisms are not modded out. One can show that the degree of freedom of the twist variable leads in this case to an overall multiplicative volume factor $V_C$ on the right hand side of \eqref{eq:gluingJT} (see e.g. appendix D of \cite{Blommaert:2018iqz}). $V_C$ can also be interpreted as the regularized volume of the maximal torus as we clarify further below. Given the TQFT nature of our gluing procedure, we note that Teichm\"uller space gluing is indeed the natural way to implement the shrinkability constraint. With this in mind, we correct our expression for the $e$-brane boundary state by compensating accordingly:
\begin{equation}
\label{eq:eoverlap}
\braket{e|k}= \frac{1}{V_C} k \sinh 2 \pi k.
\end{equation}
 
\paragraph{Peter-Weyl Theorem and the subregion Hilbert space.} 

The density of states associated to the boundary label $e$ is the Plancherel measure for $\SL^+(2,\mathbb{R})$ as defined by the spectral decomposition \cite{Ponsot:1999uf}:
\begin{align}\label{L2JT}
\text{L}^{2}(\SL^+(2,\mathbb{R})) = \int_{\oplus_{k \geq 0} }   \, (k  \sinh 2\pi k )\, \mathcal{P}_{k} \otimes \mathcal{P}_{k}\,.
\end{align}
Here, $\SL^+(2,\mathbb{R})$ is the semi-subgroup of $\SL(2,\mathbb{R})$ matrices with positive matrix elements and  $\mathcal{P}_{k}$ denotes a continuous series representation of $\SL(2,\mathbb{R})$.  This rather non-trivial generalization of the Peter-Weyl theorem arises from a $q \to 1$ limit of an analogous equation for $\SL^+_{q}(2,\mathbb{R})$ (see eq \eqref{L2}), which originated from a series of studies on Liouville theory and quantum Teichm\"uller space \cite{Ponsot:1999uf}.  It defines a ``complete" set of representations for $\SL^+(2,\mathbb{R})$ given by the continuous series representations with measure $k \sinh  2\pi k$.  As in the ordinary Peter-Weyl theorem, the right hand side of \eqref{L2JT} is the decomposition of the regular representation of $\SL^+(2,\mathbb{R})$, in which $\SL^+(2,\mathbb{R})$ acts on itself by left and right multiplication.

The shrinkable boundary condition implies that the subregion Hilbert space $\mathcal{H}_{e \mathfrak{i}_{R} }$ is given by 
\begin{align}\label{PWJT}
    \mathcal{H}_{e \mathfrak{i}_{R} }= \int_{\oplus_{k \geq 0} }   \, (k  \sinh 2\pi k )\, \mathcal{P}_{k} \otimes \mathcal{P}_{k,\mathfrak{i}_{R}}.
\end{align}
Here $\mathcal{P}_{k,\mathfrak{i}_{R}}$ denotes the subspace which is invariant with respect to right multiplication by a parabolic subgroup of  $\SL^+(2,\mathbb{R})$, up to a phase: this corresponds to fixing the right index of its representation matrix to $\mathfrak{i}_{R}$.  The representations $\mathcal{P}_{k} $ correspond to the unrestricted left indices which describe edge modes on the bulk entangling surface, each with regularized dimension  
$\dim k =k  \sinh 2 \pi k$. Comparing \eqref{PWJT} to \eqref{L2JT}, implies we can characterize  the subregion Hilbert space  as: 
\begin{align} \label{L2coset}
 \mathcal{H}_{e \mathfrak{i}_{R} }= \text{L}^{2}(\SL^+(2,\mathbb{R})/\sim),
\end{align} 
where the symbol $\sim$ denotes the right coset. This generalizes \eqref{L2JT} in a way analogous to the ordinary gauge theory generalization \eqref{coset} of the Peter-Weyl theorem. 

\paragraph{Factorization map.}
As in ordinary gauge theory, the factorization map is defined via the co-product on the $\text{L}^{2}$-space of functions on $\SL^+(2,\mathbb{R})$ and its cosets.  Since the Hilbert spaces $\mathcal{H}_{e\mathfrak{i}_{R}}$, $\mathcal{H}_{\mathfrak{i}_{L}e}$ and $\mathcal{H}_{\mathfrak{i}_{L}\mathfrak{i}_{R}}$ are all subspaces of $\mathcal{H}_{ee}= \text{L}^{2}(\SL^+(2,\mathbb{R}))$, it suffices to define the factorization map there and then restrict as appropriate. 
The co-product defines the factorization map in the group basis to be
\begin{align}
 \mathtikz{\deltaA{0cm}{0cm} ;\draw (0cm,-.9 cm) node {\footnotesize $e$ }}:
\ket{g} \to \frac{1}{\sqrt{V_\text{G}}} \int_G d g_{1} \int_G dg_{2} \, \delta(g_{1}g_{2}, g)   \ket{g_{1}}\otimes \ket{g_{2} } ,
\end{align} 
where $g,g_{1},g_{2}$ should be restricted to appropriate cosets of $\SL^+(2,\mathbb{R})$ if one or both of the boundaries are asymptotic boundaries.

As in our review of 2d gauge theory in section \ref{sec:2d-gauge}, the entanglement of edge modes becomes manifest in the representation basis, whose wavefunctions on the $\SL^+(2,\mathbb{R})$ manifold are given by 
\begin{align}
     \braket{g|k, s_{1},s_{2}}&= \sqrt{ k\sinh 2\pi k } \,\,R^{k}_{s_{1}s_{2} }(g),
     \end{align} 
with normalization determined by the Plancherel measure:
\begin{align}
\int_{\SL^+(2,\mathbb{R})} dg\,\, R^{k}_{s_{1} s_{2}}(g) R^{k'*}_{s_{3} s_{4}}(g)  &=\frac{ \delta(k-k') \delta(s_{1}-s_{3} ) \delta (s_{2}-s_{4})}{k \sinh 2\pi k} .
\end{align} 
In this basis, the factorization map is just a continuous generalization of \eqref{eq:lfact}
\begin{align}
\label{rfact}    
  i : \mathcal{H}_{ee} &\to \mathcal{H}_{ee} \otimes \mathcal{H}_{ee} ,\\
    \langle g|k,s_{1},s_{2}\rangle &
    \to  
    \frac{1}{\sqrt{V_\text{G}}}\frac{1}{\sqrt{k \sinh 2\pi k}}\, \int_{\mathbb{R}} ds\, \langle g_1|k,s_{1},s\rangle \langle g_2|k,s,s_{2}\rangle, \nonumber
\end{align}
where $s$ labels the entangled edge modes, $V_\text{G}=\int_G dg$ is the (regularized) volume of the (semi)group manifold
and the prefactor is determined by the normalization.\footnote{One might interpret the volume regulator as precluding a strict factorization of the Hartle-Hawking state. We prefer to read it as showing us how to make sense of gravitationally factorized states. Analogous comments hold for the 3d case later on, though we will not be as explicit about the analogs of these volume factors.} Notice that applying the factorization map \eqref{rfact} to the Hartle-Hawking state \eqref{eq:HH} 
\begin{align}\label{hhfact}
    i\ket{\text{HH}_{\beta}  }= \frac{1}{\sqrt{V_\text{G}}}\int_{0}^{+\infty}  d k\, \int_{\mathbb{R}} ds  \,e^{-\beta C(k)/2} \ket{k\, \mathfrak{i}_{L},s} \ket{k\,s,\mathfrak{i}_{R}},
\end{align}
gives the usual thermofield double state. Indeed, a thermal density matrix is obtained after tracing over the left region
\begin{align}
\label{rhob}
    \rho (\beta) \equiv \text{Tr}_{\mathcal{H}_{\mathfrak{i}_{L} e}} \Big(i\ket{\text{HH}_{\beta}  } \bra{\text{HH}_{\beta}}i^* \Big) = \frac{1}{V_\text{G}}\int_{0}^{+\infty} d k\, \int_{\mathbb{R}} ds \, e^{-\beta C(k)}\ket{k\,\mathfrak{i}_{R},s}\bra{k\,\mathfrak{i}_{R},s} .
\end{align}
Notice that, crucially, the $\sqrt{k \sinh 2\pi k} $ factor from the $\ket{\text{HH}_{\beta}}$ wavefunction cancels the same factor appearing in the denominator of  \eqref{rfact}.\footnote{Had we followed \cite{Jafferis:2019wkd} and used the co-product factorization map for the universal cover of $\SL(2,\mathbb{R})$, these factors would not have cancelled.
}
We can explicitly check that
  \begin{equation}
  \label{eq:checkisomet}
 \bra{\text{HH}_{\beta}}i^* i\ket{\text{HH}_{\beta}  } =  \braket{\text{HH}_{\beta} \vert \text{HH}_{\beta}  }.
 \end{equation}
 Indeed, to compute the overlap between \eqref{hhfact} and its adjoint, we use the formal equalities 
 \begin{equation}
 \label{eq:formal}
 \delta(k-k) = \frac{V_C}{2\pi},\qquad \int_{\mathbb{R}}ds = \frac{2\pi V_\text{G}}{V_C} k \sinh 2\pi k,
 \end{equation}
 where $V_C$ is the volume of a maximal torus, computable as $V_C \equiv \int_{G, g \sim h g h^{-1}}dg$, the volume of the conjugacy class elements obtained by identifying group elements according to $g \sim h g h^{-1}$. These identities can be derived by comparing the compact and non-compact group orthogonality and character orthogonality relations as reviewed in Appendix \ref{app:volume-reg}.
 
 To check the factorization map \eqref{hhfact} is compatible with the shrinkable boundary condition  starting with $\rho (\beta)$ \eqref{rhob} and using the above formulas, we observe the projector onto the irreducible representation $k$ of $\SL^+(2,\mathbb{R})$ has a non-trivial trace, given by the  Plancherel measure. Thus,
\begin{align}
\label{eq:pfjt}
    \tr_{\mathcal{H}_{e \mathfrak{i}_{R} }} \rho (\beta) &= \frac{1}{V} \int_{0}^{+\infty} dk \tr \Big( \int_{-\infty}^{+\infty} ds \ket{k\,\mathfrak{i}_{R},s}\bra{k\,\mathfrak{i}_{R},s}\Big) e^{-\beta k^{2}} =\int_{0}^{+\infty} dk \, \left(k\sinh 2\pi k\right)\, e^{-\beta k^{2}},
\end{align}
which manifestly equals $Z_{\text{disk}}(\beta)$.

After normalizing $\rho \equiv \rho(\beta)/Z_{\text{disk}}(\beta)$, the von Neumann entropy $S=-\text{Tr} \rho \log \rho$ takes the form:
\begin{equation}
\label{eq:JT_vN}
\boxed{S= -\int_{0}^{+\infty}  d k\,   P(k) \log P(k) \, + \, \int_{0}^{+\infty}  d k\, P(k) \log \dim k \, + \, \log \frac{2\pi V_\text{G}}{V_C}
}
\end{equation}
where we defined a classical probability distribution: 
\begin{equation*}
\dim k = k \sinh 2\pi k , \qquad
P(k) =\dim k\,\frac{e^{-\beta C(k)} }{Z_{\text{disk}}(\beta) } .
\end{equation*}

There is an interesting list of points to stress regarding this result:
\begin{itemize}
\item The last term is an artifact of having infinite-dimensional representations. Indeed, it combines with the second term into $\int_{0}^{+\infty}  d k\, P(k) \log\left(\int_{\mathbb{R}} ds\right)$. Crucially, it does not depend on the state of the system and we can easily renormalize it by subtracting it. Note also that it has nothing to do with the short-range UV divergent term in any entanglement entropy in a continuum theory. This can be appreciated by the fact that it vanishes both for compact gauge groups and for the abelian non-compact gauge group $\mathbb{R}$, something that is not true for the short-range UV divergent term. Upon dropping this last term, \eqref{eq:JT_vN} takes the standard gauge theory form \cite{Lin:2018xkj}.
\item $P(k)$ defines a classical probability distribution, whose classical Shannon entropy equals the first term in \eqref{eq:JT_vN}. The second term is the averaged entropy within a fixed $k$-sector. This form of the entropy corresponds to a block-diagonal density matrix 
\begin{equation}
\rho = \bigoplus_k P(k) \rho_k, \qquad \text{where  } P(k) =\dim k\,\frac{e^{-\beta C(k)} }{Z_{\text{disk}}(\beta) }, \qquad \rho_k = \int_{\mathbb{R}} ds \, \frac{\ket{k\,\mathfrak{i}_{R},s}\bra{k\,\mathfrak{i}_{R},s}}{V_\text{G} \dim k},
\end{equation}
with $\int_0^{+\infty} dk P(k) = 1$ and $\text{Tr} \rho_k = 1$. Even more so, in parallel with the compact case, the density matrix $\rho$ can be written as a fully diagonal matrix:
\begin{align}
\label{eq:diagrho}
\rho = \bigoplus_{k,s} P(k,s) \rho_{k,s}, \qquad \text{where  }
P(k,s) = \frac{V_C}{2\pi V_\text{G}} \,\frac{e^{-\beta C(k)} }{Z_{\text{disk}}(\beta)}, \qquad \rho_{k,s} =  \frac{2\pi \ket{k\,\mathfrak{i}_{R},s}\bra{k\,\mathfrak{i}_{R},s}}{V_C},
\end{align}
with $\int_0^{+\infty} dk \int_\mathbb{R} ds \, P(k,s) = 1$ and $\text{Tr} \rho_{k,s }= 1$. In this language, the von Neumann entropy \eqref{eq:JT_vN} is entirely classical Shannon entropy:
\begin{align}
S= -\int_{0}^{+\infty}  d k\, \int_\mathbb{R} ds\,  P(k,s) \log P(k,s).
\end{align}
\item If we start out with JT gravity plus the Einstein-Hilbert action $- \frac{S_0}{4\pi} (\int R + 2 \oint K)$, as arises in the near-horizon near-extremal regime of higher-dimensional black hole physics, or as a ``regulator'' of the sum over wormhole corrections \cite{Saad:2019lba}, the density of states would be $\rho(k) = e^{S_0} \, k \sinh 2\pi k$, with $S_0$ the (naive) extremal black hole entropy. Factorizing such a system only requires a small modification in the reduced density matrix:
\begin{align}
\label{rhobn}
    \rho_{N} (\beta)  = \frac{1}{N} \sum_{a=1}^{N}\frac{1}{V_\text{G}}\int_{0}^{+\infty} d k\, \int_{\mathbb{R}} ds \, e^{-\beta C(k)}\ket{k\,\mathfrak{i}_{R},s,a}\bra{k\,\mathfrak{i}_{R},s,a},
\end{align}
where the additional labels $a=1\hdots N$ encode the ``extremal'' microstates, and $N = e^{S_0}$ is assumed to be an integer in this notation. The resulting von Neumann entropy of $\rho_{N}(\beta)/Z_{\text{disk}}(\beta)$ then just gets further shifted by $S_0$:
\begin{align}
S&= - \int_{0}^{+\infty}  d k\,   P(k) \log P(k) \, + \, \int_{0}^{+\infty}  d k\, P(k) \log \dim k \, + \, \log \frac{2\pi V_\text{G}}{V_C} +S_0.
\end{align}
For fixed value of $k$, for both labels $s$ and $a$ the state \eqref{rhobn} is maximally mixed leading indeed to an entropy that is just the log of the dimension of the respective Hilbert spaces: $\log \frac{2\pi V}{V_C} \dim k$ and $S_0$. These additive shifts to the entropy are non-universal and depend on the regulator (here finite-volume regularization of the group manifold). We believe it is a specific case of the more general ambiguity of entropy in type II$_\infty$ von Neumann algebras \cite{Witten:2021unn}. The latter has a trace operation, but there is an ambiguity in that trace (existence of an automorphism in the operator algebra), which at the level of the entanglement entropy translates into an arbitrary additive constant.
\end{itemize}

A stationary phase evaluation of the integral over $k$ in \eqref{eq:JT_vN}, fixing a single value $k^*$ in the distribution $P(k)$, gives the semi-classical entanglement entropy 
\begin{equation}
    S \, \approx \, \log k^* \sinh 2 \pi k^* \, \approx \, \frac{\phi_{h}}{4G_{\mt{N}}},
 \end{equation}
in agreement with the BH entropy of the classical JT black hole. 

\paragraph{Relation to the  classical phase space approach.}
The fact that only representations of $\SL^+(2,\mathbb{R})$ appears in the subregion Hilbert space shares an interesting resemblence with the classical formulation of gravitational edge modes given in \cite{Donnelly:2016auv}. It was shown there that the gravitational surface symmetry group $\text{G}_\mt{S}$ associated with an entangling surface $S$ is given by:\footnote{In pure gravity theory in higher dimensions, there is also a $(\mathbb{R}^{2})^{S}$ factor which describes the transverse deformations of $S$. See \cite{Geiller:2017xad,Geiller:2017whh,Freidel:2020xyx} for some specific relevant work in 3d and \cite{Ciambelli:2021vnn,Ciambelli:2021nmv} for interesting related work.}
\begin{align}
    G_\text{S} = \text{Diff}(\text{S})  \ltimes  \SL(2,\mathbb{R})^{\mt{S}},
\end{align}
where $\text{Diff}(\text{S})$ are diffeomorphisms mapping the entangling surface $S$ onto itself, while $\SL(2,\mathbb{R})^{\mt{S}}$ are independent linear transformations of the normal plane to the entangling surface, including the boosts, at every point of $S$. In JT gravity, the entangling surface is just a point, so the surface symmetry reduces to $G_\text{S} = \SL(2,\mathbb{R})$.  However, it was noted in that same work \cite{Donnelly:2016auv} that the $\SL(2,\mathbb{R})$ Casimir in this case is always negative, because it corresponds to the determinant of a Lorentzian metric normal to $S$. Within the  $\SL(2,\mathbb{R})$ representations appearing in the Plancherel decomposition (see e.g. \cite{VK}), this condition selects out solely the continuous principal series representations of $\SL(2,\mathbb{R})$, which is consistent with the spectral decomposition \eqref{L2JT}.

\subsection{Bulk factorization and entropy in 3d gravity}
\label{sec:3dbulk}

The main goal of this section is to provide a derivation of the black hole entropy of the quantum gravitational 3d black hole in terms of the bulk entanglement entropy across the horizon.  To do so, we shall define bulk factorization following the same approach just described for JT gravity.  The essential new ingredient is a $q$-deformed surface symmetry  $\SL^+_{q}(2,\mathbb{R})$ at the bulk entangling surface, where the deformation parameter $q$ is related to the cosmological constant. We will find that solving the shrinkable boundary condition implies that the boundary category associated to a bulk entangling surface is the representation category of $\SL^+_{q}(2,\mathbb{R})$, in direct analogy with JT gravity.  This determines the bulk factorization map, which leads to a state counting interpretation of black hole entropy in terms of quantum group edge modes. 

\subsubsection{Review: the bulk phase space}

It is convenient to start our discussion in 3d gravity by reviewing its bulk classical phase space.

\paragraph{Euclidean geometries.}
A general Euclidean solution to Einstein's equations with a negative cosmological constant is given by the metric \cite{Banados:1998gg}:
\begin{equation}
    ds^2 = \ell^2\left[\mathcal{L}^{+}(w) dw^2 + \mathcal{L}^{-}(\bar{w})d\bar{w}^2 + d\rho^2 - \left(e^{2\rho} + \mathcal{L}^{+}(w) \mathcal{L}^{-}(\bar{w})\,e^{-2\rho}\right)dw\,d\bar{w}\right],
\label{3dmet}
\end{equation}
where $w= \tau_E +i \varphi$ is a boundary coordinate and the functions $\mathcal{L}^{\pm}$ can be identified with the left and right moving Virasoro stress tensor components. Equivalently, in the Chern-Simons formulation, they parametrize the degrees of freedom of the boundary theory which arise from applying the bulk-boundary correspondence satisfying AdS$_3$ boundary conditions. Either way, these functions parameterize the classical phase space.  

In particular, the Euclidean BTZ black hole corresponds to having only the zero modes $\mathcal{L}^{+}_{0},\mathcal{L}^{-}_{0}$. These are related to the inner and outer horizons of the black hole by $\mathcal{L}_{0}^{\pm} = (r_+ \pm r_-)^2/(4\ell^2)$, where $\ell$ is the AdS radius, or to the mass and the angular momentum by 
\begin{align}\label{BHC}
    M \ell = \mathcal{L}^{+}_{0} +\mathcal{L}^{-}_{0}=  \frac{ r_{+}^2 +r_{-}^2 }{8 \ell^{2}} , \qquad
    J =  \mathcal{L}^{+}_{0} - \mathcal{L}^{-}_{0}=\frac{r_{+}r_{-}}{4\ell}.
\end{align}

\paragraph{Phase space of Lorentzian, 2-sided geometries.}
The above Ba\~{n}ados metric \eqref{3dmet} describes a single-boundary Euclidean geometry. In Lorentzian signature however, by continuation past any horizons, one reaches a two-sided configuration whose spatial topology is that of an Einstein-Rosen bridge. Our 3d gravity phase space consists of classical solutions with the topology of the two-sided Lorentzian BTZ black hole. 

Here we describe this phase space following \cite{Henneaux:2019sjx,Banerjee:2022jnv}. In the presence of two asymptotic boundaries, there are now two pairs of stress tensors describing boundary degrees of freedom, which we denote by
\begin{align}
        \mathcal{L}^+_{L}(w_{L}^+), \,\,
        \mathcal{L}^-_L(w_{L}^-), \,\,\mathcal{L}^+_{R}(w_{R}^+), \,\,  \mathcal{L}^-_{R}(w_{R}^-).
\end{align}
The subscripts $L,R$ label the left and right boundaries which are responsible for the doubling of the degrees of freedom relative to the Euclidean section, while $\pm$ labels the two chiral stress tensor components, as before. These degrees of freedom are not independent: they are correlated through their zero modes. For example, BTZ black holes satisfy
\begin{align}
         \frac{1}{2\pi} \oint d\varphi \, \mathcal{L}^+_{L} = \frac{1}{2\pi} \oint d\varphi \,  \mathcal{L}^+_{R} =\frac{1}{2}\left(M\ell+J\right)\,, \qquad \frac{1}{2\pi} \oint d\varphi \,  \mathcal{L}^-_{L} = \frac{1}{2\pi} \oint d\varphi \,  \mathcal{L}^-_{R} =\frac{1}{2}\left(M\ell-J\right).
\end{align}
Thus, their zero-modes are equal and should be matched with the pair $\mathcal{L}^{+}_{0},\mathcal{L}^{-}_{0}$ giving rise to the black hole charges as in equations \eqref{BHC}. 

The Chern-Simons formulation in terms of bulk gauge connections $A^\pm$ gives a Wilson line description of the coupling between the phase space degrees of freedom of the two sides.
Here $\mathcal{L}^{+}_{L,R}$ and  $\mathcal{L}^{-}_{L,R}$ are the boundary degrees of freedom that remain after imposing AdS$_3$ boundary conditions on the gauge fields $A^+$ and $A^-$, respectively. Denoting the $\SL(2,\mathbb{R})$ generators by $L_{0},L_{\pm}$, these AdS$_3$ boundary conditions imply
\begin{align}
  A^{+}_r = 0\,, \quad A^{+}_\varphi &= L_- + \mathcal{L}^+_R(t,\varphi)\,L_+ \quad \text{(right boundary)}\, , \\
  A^{+}_r = 0\,, \quad A^{+}_\varphi &= L_+ + \mathcal{L}^+_L(t,\varphi)\,L_- \quad \text{(left boundary)}\,.
\end{align}
Besides $\mathcal{L}^+_{L,R}$, the phase space has one extra degree of freedom: the radial Wilson line $W_\mathcal{C}$
\begin{equation}
  W_\mathcal{C} \equiv {\cal P} \exp\left[-\int_\mt{L}^\mt{R} A^{+}_r(\varphi=0,r)\,dr\right],
\end{equation}
linking the holonomy on the two asymptotic boundaries 
\begin{equation}
  {\cal P}\exp\left[-\oint_\mt{R} (L_- + \mathcal{L}^+_R(\varphi)\,L_+)\,d\varphi\right] = W_\mathcal{C}\,{\cal P}\exp\left[-\oint_\mt{L} (L_+ + \mathcal{L}^+_L(\varphi)\,L_-)\,d\varphi\right]\,{W_\mathcal{C}}^{-1}.
\label{eq:C-link}
\end{equation}
This is why the phase space does not factorize into two independent boundary theories. This shared holonomy between the two boundaries can be described by a single quantum mechanical degree of freedom $p_{+}$, parameterizing the Wilson loop around $\varphi$ at arbitrary $r$. To see this, assume the Wilson loop is in the hyperbolic conjugacy class of SL$(2,\mathbb{R})$. It can then be written as
\begin{align}
   \tr {\cal P}\exp\left[-\oint d\varphi A^{+}_{\varphi}\right]&= \tr \exp( p_{+} L_{0} ),
\end{align}
i.e. in terms of a real proportionality factor $p_{+}$, as claimed.\footnote{This phase space variable is denoted $k_0$ in \cite{Henneaux:2019sjx,Banerjee:2022jnv}.} Since the above discussion holds for both chiralities, we conclude the Hilbert space is described by the non-zero modes of all four $\mathcal{L}^\pm_{L}, \mathcal{L}^\pm_{R}$, combined with a common zero-mode $p_\pm$ for each chirality.

\subsubsection{The two-sided Hilbert space and the Hartle-Hawking state} 

The 3d analog of the interval Hilbert space in JT is a Hilbert space supported on a spatial annulus with two circular boundaries.  For example, the two-sided Hilbert space is supported on the Einstein-Rosen bridge connecting the two asymptotic boundaries, while the one-sided Hilbert space has one boundary at asymptotic infinity, and one at the bifurcation surface. Following the notation in JT gravity, we denote the two-sided Hilbert spaces by $\mathcal{H}_{\mathfrak{i}_{L} \mathfrak{i}_{R}}$, with the indices $\mathfrak{i}_{L}, \mathfrak{i}_{R}$ labeling the asymptotic AdS$_3$ boundary conditions.  

Quantization of the stress tensor pair $\mathcal{L}^\pm_{L}, \mathcal{L}^\pm_{R}$ leads to boundary Hilbert spaces labeled by shared chiral primaries $p_{\pm}$, each supporting an infinite tower of descendants. The two-sided Hilbert space is given by
\begin{align}
        \mathcal{H}_{\mathfrak{i}_{L} \mathfrak{i}_{R}}
        =  \text{span}\Big\{ \ket{ p_{+} \mathfrak{i}_{L}\mathfrak{i}_{R};\, m_{L}  m_{R}} \otimes\ket{p_{-} \mathfrak{i}_{L}\mathfrak{i}_{R}; n_{L}n_{R} } \Big\},
    \end{align} 
where $m_{L,R}$ and $n_{L,R}$ label the descendants on the left and right boundaries and the indices $\mathfrak{i}_{L},\mathfrak{i}_{R}$ label Kac-Moody zero-modes belonging to the $p_{\pm}$ representation of $\SL(2,\mathbb{R})$. These zero-modes correspond to the $\SL(2,\mathbb{R})$ Kac-Moody generator fixed by the Drinfeld-Sokolov reduction (see \cite{Bershadsky:1989mf}), which reduces the symmetry algebra from $\widehat{\text{SL}(2,\mathbb{R})}$ to Virasoro.  Physical states in the 3d gravity Hilbert space are non-factorizable due to the matching of $p_{\pm}$ on the left and right boundary, corresponding to the shared zero mode $p_{\pm}$ in the bulk. The JT limit removes descendants, and gives the two-sided states of \eqref{kii}:
\begin{align}
 \ket{ p_{\pm} \,\mathfrak{i}_{L}\mathfrak{i}_{R}; m_{L}  m_{R}}  \to \ket{k_{\pm}\, \mathfrak{i}_{L}\mathfrak{i}_{R}}.
\end{align}

\begin{figure}[h]
\centering
\includegraphics[scale=.25]{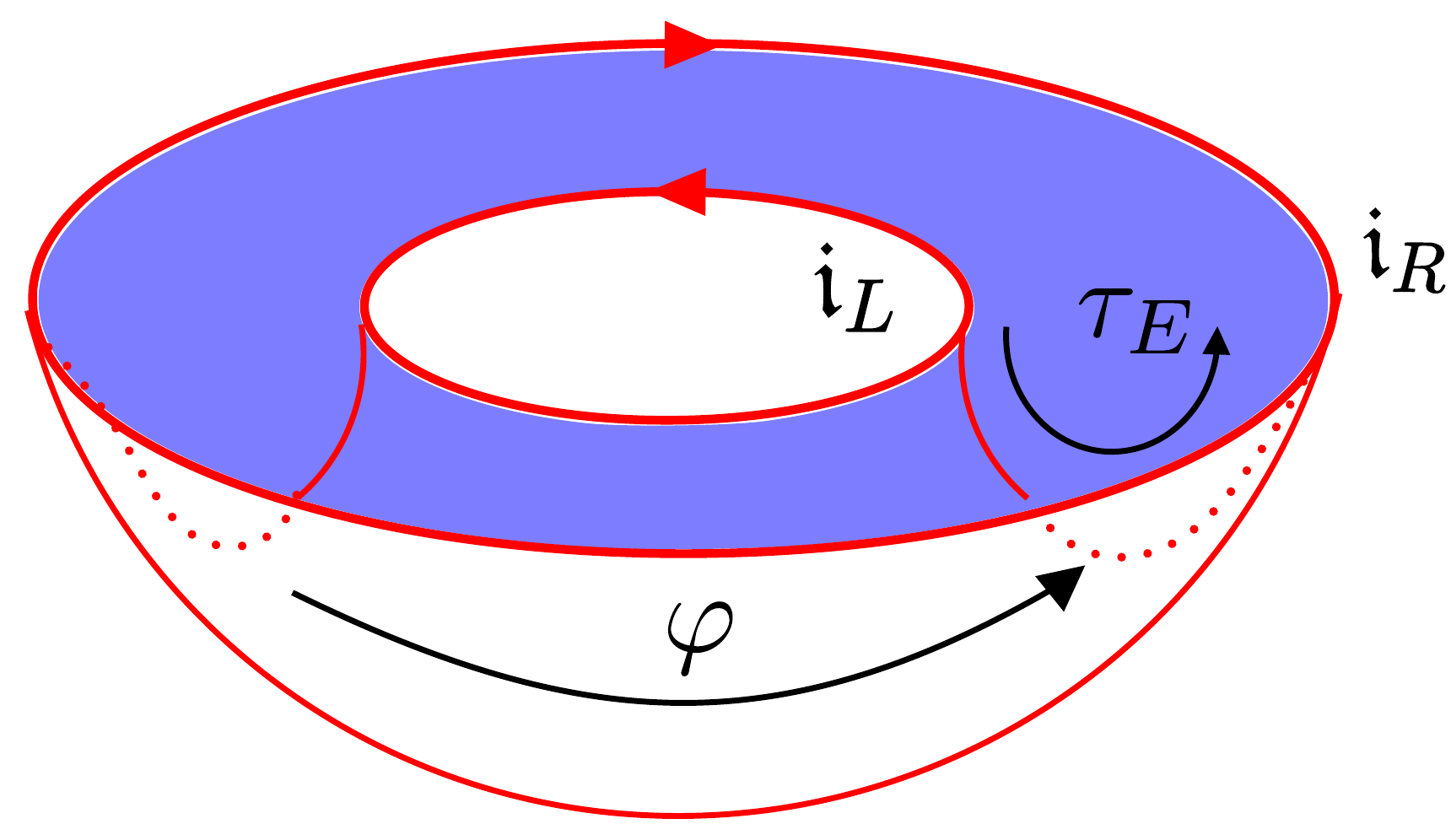}
\caption{The Hartle-Hawking state \eqref{HH} is prepared by the path integral on half of the Euclidean geometry. In the channel defined by eq. \eqref{gcpf}, we would insert Wilson lines in representations $p_{\pm}$ that link with the boundaries $\mathfrak{i}_{L,R}$ (these are not shown in the figure).}
\label{3DHH}
\end{figure}

The bulk Hartle-Hawking state $\ket{\text{HH}_{\beta,\mu}}$ whose norm squared produces the 3d gravity partition function \eqref{solidtorus} $Z(\beta,\mu) \equiv \braket{\text{HH}_{\beta,\mu} \vert \text{HH}_{\beta,\mu}}$ is
\begin{align}\label{HH}
  \ket{\text{HH}_{\beta,\mu}} &= \int_0^{+\infty} \hspace{-0.2cm} dp_{+} \int_0^{+\infty} \hspace{-0.2cm} dp_{-} \sqrt{\dim_{q}(p_{+})}\sqrt{ \dim_{q}(p_{-}) } e^{-\frac{\beta}{\ell} (p_{+}^{2} +p_{-}^{2})+ i \mu \frac{\beta}{\ell} (p_{+}^{2} -p_{-}^{2})} \nn
   &\times  \sum_{m_{L}=m_{R}} q^{N/2} \ket{ p_{+}\,\mathfrak{i}_{L}\mathfrak{i}_{R}; m_{L}m_{R} }\otimes \sum_{n_{L}=n_{R}} \bar{q}^{N/2}\ket{ p_{-}\,\mathfrak{i}_{L}\mathfrak{i}_{R}; n_{L}n_{R} },
\end{align}
where $q=e^{\frac{\beta}{\ell}(-1+i\mu)}$ and $\bar{q}=e^{\frac{\beta}{\ell}(-1-i\mu)}$, as before. Notice the descendants on the $L$ and $R$ side are matched. This is the standard TFD state of the boundary theory, obtained from the path integral over half of the Euclidean geometry as shown in Figure \ref{3DHH}. Below, we will derive the bulk factorization of this state and compute its entanglement entropy.

\subsubsection{Edge sector of 3d gravity from the shrinkable boundary condition}
\label{sec:edge_shrink}

\begin{figure}[h]
\centering
\includegraphics[scale=.3]{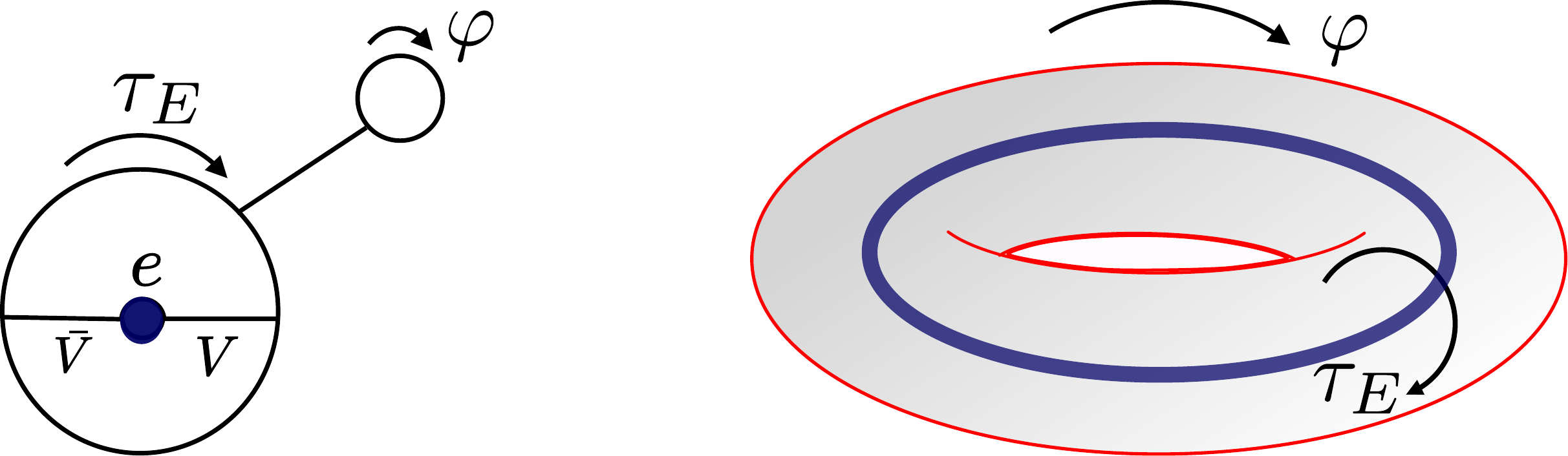}
\caption{In 3d gravity, introducing the stretched entangling surface involves excising a tubular neighborhood of a circle, corresponding to the darkblue region.}
\label{torusv2}
\end{figure}

To define factorization in 3d gravity, we need to identify its edge sector, i.e. the edge symmetry group, its relevant irreducible representations and its density of states.  A first route to answer this question is to demand that the shrinkable boundary condition is satisfied.  In this case the stretched entangling surface $S_{\epsilon}$ describes a thin torus inside the Euclidean spacetime, which is the boundary of the darkblue region in Figure \ref{torusv2}.

To find the bulk edge modes, we follow the procedure in JT, as outlined below eq \eqref{eq:gluingJT}. Consider one of the chiral sectors in our proposal for 3d gravity in \eqref{eq:3dfactor}. The addition of the stretched entangling surface corresponds to putting the Chern-Simons theory on $\mathbb{T}^2\times I \equiv \mathcal{A} \times S^1$, where $\mathcal{A}$ is an annulus.  The path integral on this spacetime can be viewed as an amplitude between the inner and outer boundary with modular parameter $\tau_{1},\tau_{2}$, where $\tau_n = \frac{\beta_n}{2\pi \ell}(\mu_n+i)$. Inserting a complete set $\mathbb{1} = \int d\lambda \ket{\lambda}\bra{\lambda}$ between the boundary states of this amplitude can then be interpreted as inserting a Wilson loop labeled by $\lambda$ in the interior of the two solid tori.  This corresponds to decomposing the partition function on $\mathbb{T}^2\times I$  as
\begin{equation}
\label{eq:CSglue}
 Z_{\mathbb{T}^2\times I}(\tau_1,\tau_2) = \sum_{\hat{\lambda}} \chi_{\hat{\lambda}}(\tau_1) \chi_{\hat{\lambda}}(\tau_2)\, , 
\end{equation}
which in the limit $\tau_1\to 0$ is a 3d Chern-Simons analog of equation \eqref{eq:gluingJT}:
\begin{equation}
Z(\beta_2,\mu_2)= \int_{-\infty}^{+\infty} d\lambda\, Z_{\text{inner}} (\tau_1 \to 0,\lambda)\, Z_{\text{outer}}(\tau_2,\lambda).
\end{equation}
However, for 3d gravity we will see that the ``inner'' piece is not quite correct in \eqref{eq:CSglue}.

\begin{figure}[h]
\centering
\includegraphics[width=0.75\textwidth]{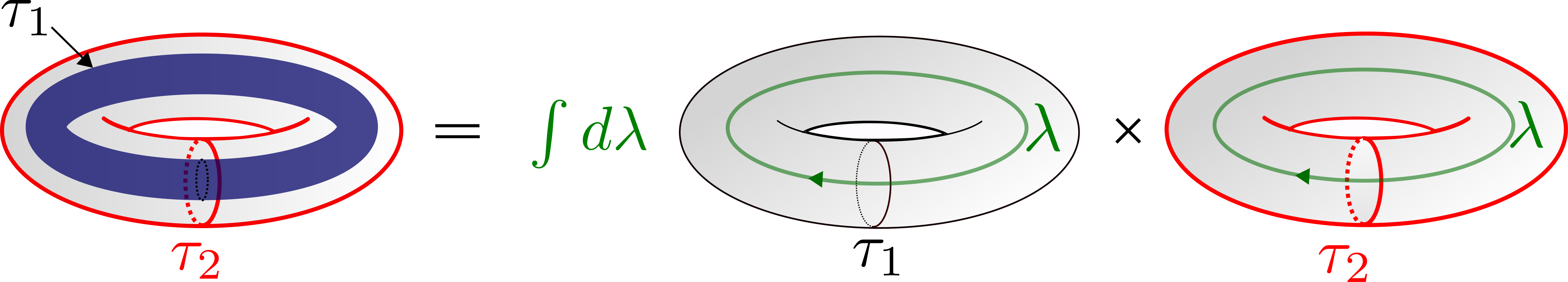}
\caption{Two-boundary amplitude (hollow torus) in 3d gravity obtained by gluing defect amplitudes together. The holographic boundary is drawn in red color.}
\label{DecompAnn3d}
\end{figure}

The gravitational Wilson loops in the representation $\lambda$ play the role of the defect insertions in Figure \ref{DecompAnn}, here depicted in Figure \ref{DecompAnn3d}.

 The outer boundary labeled by $\tau_2$ is at asymptotic infinity. As noted earlier, the gravitational boundary conditions fix a zero-mode $\mathfrak{i}_{L,R}$ of the Kac-Moody algebra, leading to a partition function given by Virasoro characters rather than Kac-Moody ones. As in JT, the fixing of these indices leads to a coset partition function (this is a chiral sector of \eqref{defecttr}):
\begin{equation}
Z_{\text{outer}}(\tau_2,\lambda) = \chi^{\text{Vir}}_{\lambda}\left(-\frac{1}{\tau_2} \right) = \frac{1}{\eta(\tau_2)} \int_0^{+\infty} dp\, \cos (2 \pi \lambda p)\, e^{2\pi i \tau_2 p^2}.
\label{eq:Virext}
\end{equation}
This is interpreted as a Wilson loop insertion in the interior of the solid torus as in equation \eqref{defecttr}.

The inner boundary labeled by $\tau_1$ corresponds to an entangling surface.  In ordinary Chern-Simons theory with compact gauge group $\text{G}$, we saw in section \ref{sec:CS-review} that the appropriate shrinkable boundary condition $A_{\tau}-A_{\varphi}\vert_{\partial\mathcal{M}}=0$ leads to Kac-Moody edge modes.  Thus we would naively be tempted to associate the Kac-Moody character of the loop group $\widehat{\SL(2,\mathbb{R})}$ to the inner torus:
\begin{equation}
  Z_{\text{inner}} (\tau_1,\lambda) \stackrel{?}{=} \chi_{\hat{\lambda}}(-1/\tau_1) = \int_0^{+\infty} dp\, \cos (2\pi \lambda p)\,\chi_{\hat{p}}(\tau_1)\,,
\label{eq:choice}
\end{equation}
where $\chi_{\hat{p}}(\tau_1) \sim 1/ \eta(\tau_1)^3$, as written in \eqref{eq:CSglue}. However, \eqref{eq:choice} is neither compatible with the gravitational shrinkable boundary condition nor finite as $\tau_1 \to 0$, since most noticeably, it describes an infinite degeneracy due to all the descendants as $\tau_1 \to 0$. The latter is interpretable as coming from all of the modes with non-trivial spatial profile along the black hole horizon.

\begin{figure}[h]
\centering
\includegraphics[width=0.55\textwidth]{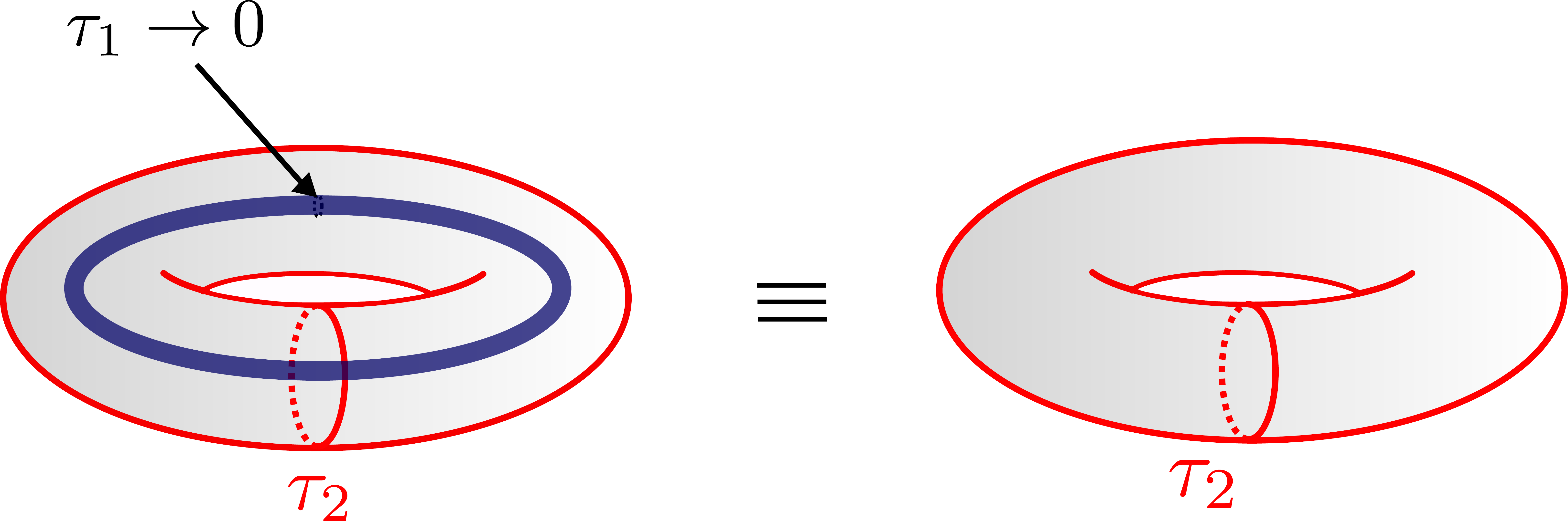}
\caption{Shrinkable boundary in 3d gravity, where we start with a hollow torus with interior modular parameter $\tau_1$. Taking $\tau_1 \to 0$ makes the torus solid, and should reproduce the solid torus amplitude \eqref{solidtorus}.}
\label{ShrinkableBC3d}
\end{figure}

We then set out to find $Z_{\text{inner}} (\tau_1 \to 0,\lambda)$ by demanding consistency with the shrinkable boundary condition 
\begin{equation}\label{3dshrink}
Z(\beta_2,\mu_2)= \int_{-\infty}^{+\infty} d\lambda\, Z_{\text{inner}} (\tau_1 \to 0,\lambda)\, Z_{\text{outer}}(\tau_2,\lambda) \stackrel{!}{=} \int_0^{+\infty} dp\, 4\sqrt{2} \sinh (2\pi b p) \sinh (2\pi b^{-1} p)\, \frac{e^{2 \pi i \tau_2 p^2}}{\eta(\tau_2)}\,,
\end{equation}
as illustrated in Figure \ref{ShrinkableBC3d}.

This shrinkability condition then requires\footnote{Fourier transforming \eqref{3dshrink} with respect to $\tau_2$, and then multiplying by $\cos (2 \pi \lambda p')$ directly leads to this result.}
\begin{equation}
\boxed{
  Z_{\text{inner}} (\tau_1,\lambda) \stackrel{!}{=} \int_0^{+\infty} \hspace{-0.2cm} dp\, 4\sqrt{2} \sinh (2\pi b p) \sinh (2\pi b^{-1} p) \cos (2\pi \lambda p)\, e^{2\pi i \tau_1 p^2}
} \,, 
\label{eq:edge-3d}
\end{equation}
which most noticeably contains no descendants! 

This inner partition function \eqref{eq:edge-3d} provides an interpretation for the density of states  
\begin{align}\label{dimqp}
   \dim_{q}(p) = 4\sqrt{2}\sinh ( 2 \pi b p) \sinh( 2 \pi b^{-1} p),
\end{align}
as counting edge modes living on the bulk entangling surface.  Crucially, \eqref{dimqp} coincides with the Plancherel measure on the quantum semi-group $\SL^+_{q}(2,\mathbb{R})$.  We can think of $\SL^+_{q}(2,\mathbb{R})$ as an object defined by the spectral decomposition \cite{Ponsot:1999uf,ip2012representation}
\begin{equation}
\label{L2}
\text{L}^{2}(\SL^+_{q}(2,\mathbb{R})) = \int_{\oplus_{p \geq 0} }   \,\dim_{q}(p)\, \mathcal{P}_{p} \otimes \mathcal{P}^{*}_{p} \quad \text{with} \quad q= e^{\pi i b^{2}}.
\end{equation}
This is a quantum group generalization of the Peter-Weyl theorem \eqref{PWJT} in which 
 the  $\mathcal{P}_{p}$ are representations of the \emph{modular double} of $\mathcal{U}_{q}(\SL(2,\mathbb{R}))$. To explain what this is, first recall that $\mathcal{U}_{q}(\SL(2,\mathbb{R}))$ denotes the $q$-deformation of the universal enveloping algebra $\mathcal{U}(\SL(2,\mathbb{R}))$; the latter being an algebra generated by the Lie algebra elements of $\SL(2,\mathbb{R})$, subject to the standard commutation relations. When $q=1$, the right hand side of \eqref{L2} corresponds to a basis of representation matrix elements $R_{ab}(g)$, which naturally transforms as $\mathcal{P}_{R} \otimes \mathcal{P}^{*}_{R}$ under left and right action of $g$, as in \eqref{lrregular}. By considering infinitesimal versions of left and right multiplication, we can interpret $\mathcal{P}_{R} \otimes \mathcal{P}^{*}_{R}$ as representations of  $\mathcal{U}_q(\SL(2,\mathbb{R}))$.
 
Modulo some subtleties which we will return to shortly, these representations can be $q$-deformed, giving representations of $\mathcal{U}_{q}(\SL(2,\mathbb{R}))$.
The modular double of $\mathcal{U}_{q}(\SL(2,\mathbb{R}))$ refers to a special subclass of continuous series representations of $\mathcal{U}_{q}(\SL(2,\mathbb{R}))$ appearing in Liouville theory: they are  simultanenous representations of $\mathcal{U}_{q}(\SL(2,\mathbb{R}))$ and its modular dual $\mathcal{U}_{\tilde{q}}(\SL(2,\mathbb{R}))$ with $\tilde{q}=e^{ i\frac{ \pi}{b^{2}}}$. This is the origin of the self-duality transformation mapping $b\to \frac{1}{b}$, which plays a prominent role in Liouville theory. As Hopf algebras, $\SL^+_{q}(2,\mathbb{R})$ and the modular double of $\mathcal{U}_{q}(\SL(2,\mathbb{R}))$ are dual spaces. We can interpret \eqref{L2} to mean that their representations are in one-to-one correspondence. 
 
Our conclusion is hence that, unlike $\PSL(2,\mathbb{R})\otimes \PSL(2,\mathbb{R})$ Chern-Simons theory, in 3d gravity the boundary category associated to bulk entangling surfaces is given by two copies of $\text{Rep}(\SL^+_{q}(2,\mathbb{R}))$.   
This distinction between gravity and gauge theory is forced on us by the shrinkable boundary condition.
We will provide a physical picture of the meaning of this difference in boundary category in section \ref{s:appgaugegra}. 

Note that from the perspective of the edge sector, the transition from JT to 3d gravity requires precisely a $q$-deformation of the underlying group theoretical structure of the edge degrees of freedom. We will see below that these degrees of freedom fully account for the black hole density of states, and carry the entire black hole entropy. There are descendants on the asymptotic boundary that also carry entropy, contributing to the total entropy in spacetime, but we do not associate these degrees of freedom to the black hole itself.\footnote{As a final comment, we also note that 2d Liouville gravity is governed by the same $q$-deformation of the JT structure, and can be formulated in the same language. In particular, its disk partition function, amplitudes and (presumably) its factorization can be similarly developed as we have done here. The amplitudes differ from 3d gravity in that there are no descendants anywhere, and the energy eigenvalues (in the exponentials) are also $q$-deformed. We refer the reader to \cite{Fan:2021bwt,Mertens:2020hbs,Mertens:2020pfe} for the expressions and discussions.}

\subsubsection{One-sided states and the factorization map}
\label{sec:grav_fact}

Next, let us introduce a boundary label $e$ on the entangling surface, which is a circle splitting the wormhole into two regions $V$ and $\bar{V}$. Each of these regions is an annulus with one boundary on the stretched horizon and one at asymptotic infinity.  We denote the associated one-sided Hilbert spaces as $\mathcal{H}_{\mathfrak{i}_L e} $ and $\mathcal{H}_{ e\mathfrak{i}_R}$.  

\begin{figure}[h]
\centering
\includegraphics[scale=.25]{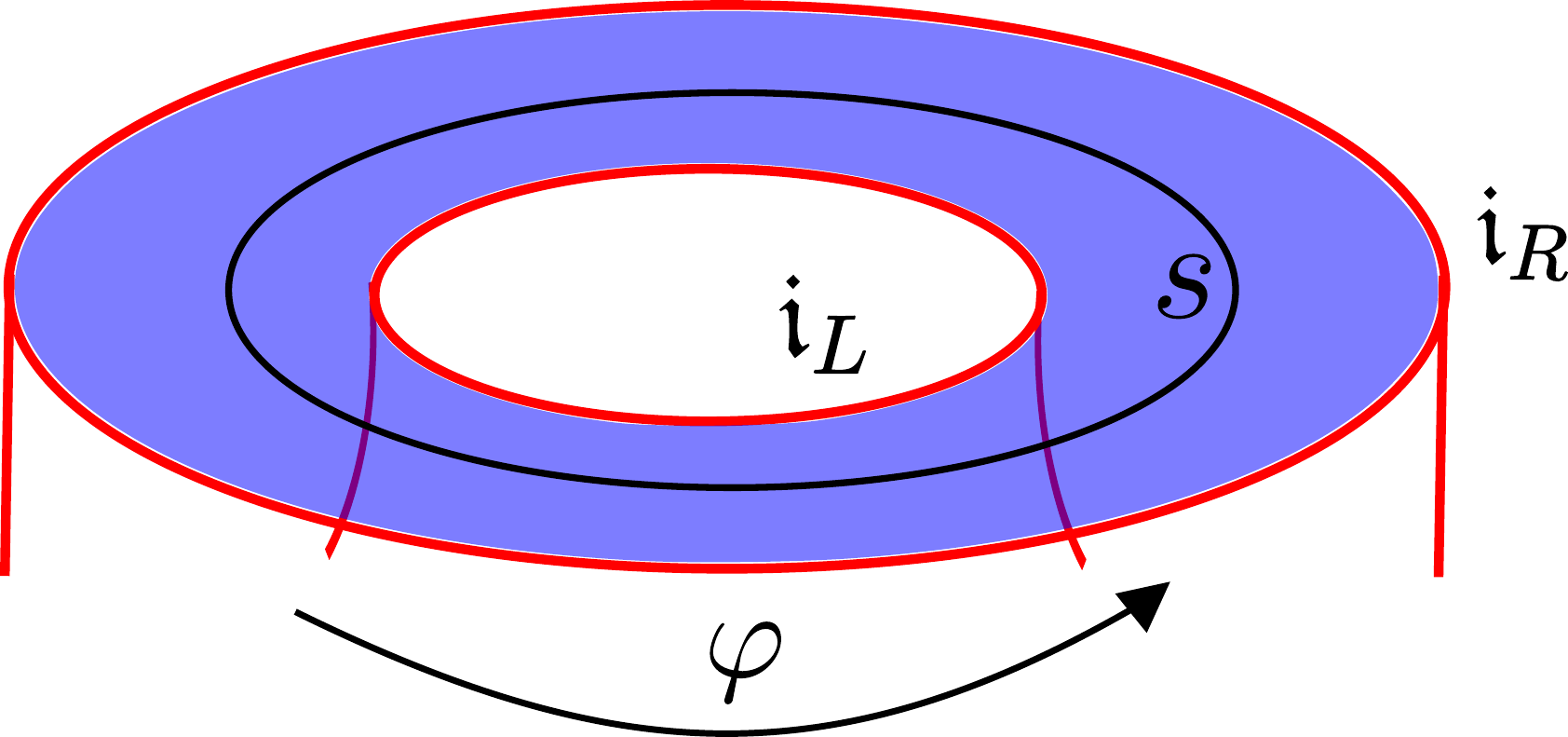}
\caption{Splitting a two-sided state using the factorization map, which introduces an entangling surface (the black circle) and an edge sector label $s$.}
\label{3DHHPRES}
\end{figure}

In 3d gravity, the shrinkable boundary condition $e$ labels a complete set of representations $\SL^+_{q}(2,\mathbb{R})$, corresponding to anyonic edge modes living at the black hole horizon.  The edge mode Hilbert spaces are $q$-deformations of those in JT. When neglecting descendants, one-sided wavefunctions are $\text{L}^{2}$-functions on a quotient of $\SL^+_{q}(2,\mathbb{R})$ induced by the AdS$_3$ asymptotic boundary conditions. In particular, the zero-mode subspace $\mathcal{H}_{e \mathfrak{i}}^0$ is given by
\begin{align} 
\mathcal{H}^{0}_{ e\mathfrak{i}_R}=
\text{L}^{2}(\SL^+_{q}(2,\mathbb{R})/\sim) &= \int_{\oplus_{p \geq 0} }   \,\dim_{q}(p)\, \mathcal{P}_{p} \otimes \mathcal{P}^{*}_{p,\mathfrak{i}_R}.
\end{align} 
Similarly, $\mathcal{H}_{\mathfrak{i}_{L} e}^{0} $ can be identified with functions on the left coset, while $\mathcal{H}_{\mathfrak{i}_{L} \mathfrak{i}_{R}}^{0} $ corresponds to functions on the double coset. The parallel structures exhibited in the interval Hilbert spaces of 3d gravity and JT gravity suggests one should define a factorization map by $q$-deforming \eqref{rfact}, i.e. by interpreting the edge mode indices $s$ as states transforming under  $\SL^+_{q}(2,\mathbb{R})$ with a density of states given by the Plancherel measure \eqref{dimqp}. Explicitly, such a factorization map is given by
\begin{align}\label{3dfact}
\boxed{
 i: \ket{ p_{\pm}\, \mathfrak{i}_{L} \mathfrak{i}_{R}; \, m_{L} \, m_{R}}  \to  \frac{1}{\sqrt{\dim_{q}(p)}}\int_{-\infty}^{+\infty}ds \,  \ket{p_{\pm}\, \mathfrak{i}_{L}s; m_{L}} \otimes \ket{p_{\pm}\, s\,\mathfrak{i}_{R}; m_{R}}}.
\end{align} 
Here $s$ labels a state in the representation $\mathcal{P}_{p_{\pm},\mathfrak{i}_{L}}$ and  $\mathcal{P}_{p_{\pm},\mathfrak{i}_{R}}^*$ respectively. Notice that the descendants $m_{L},m_{R}$ describing the boundary gravitons are spectators in this factorization map, which only acts non-trivially on the zero-mode subspace, and \emph{no descendants are introduced at the entangling surface}. We will not write down the analogous volume IR regulators that we wrote explicitly in the JT gravity case starting with equation \eqref{rfact}.  These should correspond to $q$-volume factors of both the quantum group $V_\text{G}^q$ and the subgroup of conjugacy class elements $V_C^q$, with well-defined limits $\lim_{q\to 1}V_\text{G}^q = V_\text{G}$ and $\lim_{q\to 1} V_C^q \to V_C$. Both would have to be suitably defined in the $q$-setting, which we leave as an important open problem.
We illustrate this factorization map in Figure \ref{3DHHPRES}.

To apply this factorization map \eqref{3dfact} to the Hartle-Hawking state \eqref{HH}, let us write the latter in terms of its chiral pieces, i.e. $\ket{\text{HH}_{\beta,\mu}} = \ket{\text{HH}_{\beta,\mu,+}} \otimes \ket{\text{HH}_{\beta,\mu,-}}$, where
\begin{align} 
\ket{\text{HH}_{\beta,\mu,+}} &=\int_0^{+\infty} \hspace{-0.2cm} dp_{+} \sqrt{\dim_{q}(p_{+})} e^{-\frac{\beta}{2\ell} (p_{+}^{2} + i \mu  p_{+}^{2} )} 
      \sum_{m_{L}=m_{R}} q^{N/2} \ket{ p_{+}\,\mathfrak{i}_{L}\mathfrak{i}_{R}; m_L,m_{R} },\nn
\ket{\text{HH}_{\beta,\mu,-}} &=\int_0^{+\infty} \hspace{-0.2cm} dp_{-} \sqrt{\dim_{q}(p_{-})} e^{-\frac{\beta}{2\ell} (p_{-}^{2} -i \mu  p_{-}^{2} )} 
   \sum_{n_{L}=n_{R}} \bar{q}^{N/2} \ket{ p_{-}\,\mathfrak{i}_{L}\mathfrak{i}_{R}, n_{L},n_{R} }.
\end{align} 
When restricted to a single chiral sector, the calculation of the reduced density matrix is, in essence, the same as in JT, except for keeping track of the boundary gravitons. The factorized state in the $+$ sector is
\begin{align}
    i\ket{\text{HH}_{\beta,\mu,+}} = \int_{0}^{+\infty}  \hspace{-0.2cm} d p_{+}\,  e^{-\frac{\beta}{2\ell} (p_{+}^{2} + i \mu p_{+}^{2} )} 
     \sum_{m_{L}=m_{R}} q^{N/2} \int_{-\infty}^{\infty}ds \, \ket{p_{+}\, \mathfrak{i}_{L}s; m_{L}} \otimes \ket{p_{+}\, s\, \mathfrak{i}_{R}; m_{R}}.
\end{align} 
The corresponding reduced density matrix after tracing over the left ($L$) degrees of freedom is
 \begin{align}
     \rho_{+}(\beta,\mu)=\int_{0}^{+\infty} \hspace{-0.2cm}  dp_{+}\,  e^{-\frac{\beta}{\ell} (p_{+}^{2} + i \mu p_{+}^{2} )} 
    \sum_{m_{R}} q^{N} \int_{-\infty}^{\infty} ds \ket{p_{+}\, s\,\mathfrak{i}_{R}; m_{R}} \otimes \bra{p_{+}\, s\,\mathfrak{i}_{R}; m_{R}}.
 \end{align}
 The shrinkable boundary condition guarantees that the partial trace of this reduced density matrix reproduces the 3d gravity partition function $Z(\mu, \beta)$ \eqref{solidtorus}, provided that we impose that the trace of the projector onto the $p_{+}$ representation matches the Plancherel measure: 
 \begin{align} 
\tr_{+} \int_{-\infty}^{+\infty} ds \ket{p_{+}\, s\,\mathfrak{i}_{R}; m_{R}} \otimes \bra{p_{+}\, s\,\mathfrak{i}_{R}; m_{R}} = \dim_{q}(p_{+}) .
\end{align} 

 The total reduced density matrix is obtained by combining both chiral sectors:
 \begin{align}
     \rho(\beta,\mu) = \rho_{+}(\beta,\mu) \otimes \rho_{-}(\beta,\mu).
 \end{align}
Its von Neumann entanglement entropy has contributions from both the entangled boundary gravitons labeled by $m_R$ and $n_R$, and the entangled ``bulk'' edge modes labeled by $s$.  By design, this entanglement entropy $S$ equals the thermal entropy of $Z(\beta,\mu)$ \eqref{eq:BHncl}, which we reproduce here for convenience:
\begin{align}
\label{eq:vNfinal}
&S =- \hspace{-0.15cm}\sum_{n,\bar{n}=0}^{+\infty}\int_0^{+\infty} \hspace{-0.35cm} dp_+ dp_-  \hspace{-0.1cm}\left[ p_{n,\bar{n}}(p_+,p_-\hspace{-0.07cm}) \log p_{n,\bar{n}}(p_+,p_-\hspace{-0.07cm}) - p_{n,\bar{n}}(p_+,p_-\hspace{-0.07cm}) \log \Big((\text{dim}_q p_+ \text{dim}_q p_-\hspace{-0.07cm}) p(n) p(\bar{n}) \Big) \right],
\end{align}
where
  \begin{align}
    p_{n,\bar{n}}(p_+,p_-) &= (\text{dim}_q p_+ \text{dim}_q p_-) p(n) p(\bar{n}) \frac{e^{-\frac{\beta}{\ell}(p_+^2 + p_-^2-\frac{1}{12}+n+\bar{n})}\,e^{i\mu \frac{\beta}{\ell}(p_+^2 - p_-^2+n-\bar{n})}}{Z(\beta,\mu)},
\end{align}
and $p(n)$ is the number of descendants at level $n$, where we have used the notation of section \ref{sec:2duniv} where the descendant \emph{level} is denoted by $n,\bar{n}$. As explained in section \ref{sec:3dgrav_def}, the contribution from the boundary gravitons is subleading in the semi-classical limit. Thus, to leading order in the semi-classical expansion, the bulk entanglement entropy gives
 \begin{align}
     S \, \approx \, \log \Big( \dim_{q}(p_{+})\dim_{q}(p_{-}) \Big) \, \to \, 2\pi b^{-1} (p_+ + p_-),
\label{eq:3d-ent-entropy}
 \end{align}
 consistent with the Bekenstein-Hawking entropy \eqref{eq:BHcl}.

\subsubsection{The coordinate algebra for $\SL^+_{q}(2,\mathbb{R})$ and its (co-)representations}
\label{sec:co-rep}

To understand why our factorization map \eqref{3dfact} is not an ad hoc construction, we must understand how it arises as part of a constrained algebraic structure that defines $\SL^+_{q}(2,\mathbb{R})$.  In particular, we will explain that the factorization map \eqref{3dfact} is a representation of the co-product for a Hopf algebra $\mathcal{F}(\SL^+_{q}(2,\mathbb{R}))$ associated with functions on $\SL^+_{q}(2,\mathbb{R})$.  $\mathcal{F}(\SL^+_{q}(2,\mathbb{R}))$ is referred to as the coordinate algebra of $\SL^+_{q}(2,\mathbb{R})$ and underlies the $\text{L}^2$-space appearing on the left hand side of the Peter-Weyl theorem \eqref{L2}. We also note that an additional set of constraints is imposed on the co-product factorization map, when we embed it within a putative extended TQFT for 3d gravity associatedto the boundary category $\text{Rep}(\SL^+_q(2,\mathbb{R}))$, see subsection \ref{sec:3dext}.\footnote{For example, in 2d the extended TQFT forms a ``knowledgeable" Frobenius algebra. So in addition to the Hopf algebra compatibility relations, the co-product has to satify further compatibility relations of the Frobenius algebra.}  There are indications that such a TQFT exists, based on work related to the quantization of Teichm\"uller space \cite{EllegaardAndersen:2011vps}: we will briefly comment on this in the conclusion.  For now, our goal is to identify the ingredients needed to define factorization within such a framework.  With this in mind, we turn to a more detailed explanation of the algebraic structure of SL$^+_{q}(2,\mathbb{R})$. 

\paragraph{Coordinate algebra for $\SL(2,\mathbb{R})$.} We begin by defining  $\text{L}^{2}(\SL(2,\mathbb{R}))$ in a manner that will be convenient for $q$-deformation. Consider a basis given by products of matrix elements $g_{ij}$ of $\SL(2,\mathbb{R})$ in the \emph{fundamental} representation.  

This basis of matrix elements generates a commutative  algebra $\mathcal{F}(\SL(2,\mathbb{R}))$ which consists of sums of products of $g_{ij}$, subject to the relation
\begin{align}
    ad-bc=1,\quad \text{where }\,\, g\equiv \begin{pmatrix}a&&b\\c&&d
    \end{pmatrix}.
\end{align}
 $\mathcal{F}(\SL(2,\mathbb{R}))$ is referred to as the \emph{coordinate algebra}.

As for compact gauge groups, $\mathcal{F}(\SL(2,\mathbb{R}))$ is a Hopf algebra that has a product defined by the ordinary multiplication of matrix elements, together with a unique co-product $\Delta$
\begin{align} \label{Delta}
    \Delta: \mathcal{A} \to \mathcal{A} \otimes \mathcal{A}, \qquad     g_{ij} \to \sum_{k} g_{ik} \otimes g_{kj} ,
\end{align}
that can be used to define factorization. A further property of a Hopf algebra is an operation called the antipode $S$, which can be used to define conjugate representations. In the case of an ordinary group, $S$ is just the inverse operation:
\begin{align}
    S:\mathcal{A}\to \mathcal{A}, \quad      g_{ij} \to g_{ij}^{-1} .
\end{align}

Now consider the quantum group $\SL_{q}(2,\mathbb{R})$. In the original formulation due to Fadeev, Reshetikhin and Takhtajan \cite{Faddeev:1987ih}, this quantum group was defined by its coordinate algebra $\mathcal{F}(\SL_{q}(2,\mathbb{R}))$, which is a non-commutative deformation of $\mathcal{F}(\SL(2,\mathbb{R}))$ satisfying the relations
\begin{align} \label{msl2commutators}
ab&= q^{1/2} ba, \quad  ac= q^{1/2} ca,\quad  bd= q^{1/2} db,\quad cd= q^{1/2} dc,  \nn
bc&=cb,\quad ad-da = (q^{1/2}-q^{-1/2}) bc.
\end{align} 
These commutation relations are captured by the $\mathcal{R}$-matrix of $\SL_{q}(2,\mathbb{R})$ (see Appendix \ref{app:hopf}), which specifies its braiding property. In addition, one imposes a $q$-deformed version of the condition $\det g = 1$:
\begin{align}\label{det}
    ad-q^{1/2}bc=1.
\end{align}

Since the ``matrix elements" of $\SL_{q}(2,\mathbb{R})$ are non-commutative, they should be represented as operators on a Hilbert space rather than numbers. Explicit ``integrable" representations of these operators can be constructed, see e.g. \cite{Klimyk:1997eb}.
It is important not to confuse the representation of the operator algebra defined by \eqref{msl2commutators} and \eqref{det},  with the representations appearing on the right hand side of the spectral decomposition \eqref{L2}. The former give a representation of the Hopf algebra of functions on $\SL_{q}(2,\mathbb{R})$ preserving the multiplication rule for the matrix elements. When $q=1$, this corresponds to the commutative algebra of four elements $a,b,c,d$ subject to the determinant condition, which is indeed different from the ${\cal P}_R$ representations appearing in \eqref{L2}. When $q\neq 1$, the antipode is given by
\begin{equation}
S(g) = \begin{pmatrix} d & -q^{-1/2} b \\ -q^{1/2} c & a \end{pmatrix}, \qquad S(g)g = gS(g) = \mathbb{1}.
\end{equation}
As in the undeformed case, this allows us to define conjugate representations $\bar{R}$, which can fuse with the representations $R$ to form singlets. This completes the definition of the Hopf algebra $\mathcal{F}(\SL_{q}(2,\mathbb{R}))$, which we can complete into the space $\text{L}^{2} (\SL_{q}(2,\mathbb{R}))$ with an appropriate norm.

We are now ready to define $\SL^+_{q}(2,\mathbb{R})$.  We simply take  $\mathcal{F}(\SL_{q}(2,\mathbb{R}))$, but  restrict to a representation of $a,b,c,d$ in terms of \emph{positive} self-adjoint operators (i.e. operators with a \emph{positive} spectrum). This is the $q$-deformed analog of the positivity condition for the matrix elements of  $\SL^+(2,\mathbb{R})$.  As in the case of $\SL^+(2,\mathbb{R})$, the ``inverse" $S(g)$ may not belong to $\SL^+_{q}(2,\mathbb{R})$ even if $g \in \SL^+(2,\mathbb{R})$; hence this symmetry is non-invertible. A norm on operators in $\mathcal{F}(\SL^+_{q}(2,\mathbb{R}))$ was defined in \cite{Ponsot:1999uf}, and the completion of $\mathcal{F}(\SL^+_{q}(2,\mathbb{R}))$ with respect to this norm gives $\text{L}^{2}(\SL^+_{q}(2,\mathbb{R}))$.

 \paragraph{Co-representation of $\SL^+_{q}(2,\mathbb{R})$.}  Having defined the configuration space $\SL^+_{q}(2,\mathbb{R})$ on the left-hand side of the Peter-Weyl theorem \eqref{L2}, we now turn to a more detailed understanding of the representations which appear on its right-hand side.  As we alluded to previously, in the original statement of \eqref{L2}, these are certain irreducible representations of the modular double of $\mathcal{U}_{q}(\SL(2,\mathbb{R}))$.  Intuitively, these act on ``representation matrix elements" $R_{ab}(g)$ via left and right multiplication of $g$.  

 However we have yet to give a proper definition of a representation matrix for $\SL^+_{q}(2,\mathbb{R})$.  This is not entirely straightforward; for example, the usual definition of a representation is a mapping $\rho_{R}$ that defines the action of a group  of $G$ on a vector space $V_{R}$:
 \begin{align}\label{rhorep}
     \rho_{R}: G \otimes V_{R} \to V_{R}, \qquad (g,v) \to R(g)v= \sum_{b} R_{ba}(g)v^{b}.
  \end{align} 
However, this does not make sense even for the ``fundamental" matrix elements of $\SL_{q}(2,\mathbb{R})$ described above, since $R_{ba}(g)$ is then given by the operators $a,b,c,d$ acting on $\text{L}^{2}(\mathbb{R})$ which has no apriori relation to the 2 dimensional vector space $V_{R}$. 

In the quantum group literature, these issues are addressed by the general philosophy of replacing a group $G$ with the algebra of functions on $G$. Thus one replaces the notion of a representation of $G$ with a co-representation of the coordinate algebra $\mathcal{F}(G)$, defined by 
\begin{align}
    \rho_{R} &: V_{R} \to V_{R} \otimes \mathcal{F}(G) , 
\end{align}
which satisfies a set of compatibility conditions involving the co-product $\Delta$ and the co-unit $\epsilon$ on $\mathcal{F}(G)$:
\begin{align}\label{coreprules}
    (\rho_R \otimes \text{id}) \circ \rho_{R} = (\text{id} \otimes \Delta)\circ \rho_{R}, \qquad    (\text{ id} \otimes \epsilon ) \circ \rho_{R} = \text{id}.
\end{align}
When $\mathcal{F}(G)$ is the coordinate algebra (possibly of a quantum group), a concrete characterization of its co-representations $\rho_{R}$ satisfying the compatibility relations \eqref{coreprules} is given by a map
\begin{align}\label{corep}
    \rho_{R} : V_{R} \to V_{R} \otimes \mathcal{F}(G), \qquad  v^{i} \to \sum_{j} v^{j} \otimes R_{ji}, 
\end{align}
where $R_{ji}$ is a \emph{matrix of elements} in $\mathcal{F}(G)$ satisfying   
\begin{align}
\label{cofact}
   \Delta ( R_{ij})  =  \sum_{k} R_{ik} \otimes R_{kj}, \qquad  \epsilon(R_{ij}) = \delta_{ij}.
\end{align}
We will refer to $R_{ij}$ as matrix elements of the co-representation $\rho_{R}$: in the case of an ordinary group $G$, one can show that these coincide with the usual representation matrix elements. Indeed, next to the above ``bootstrap'' argument, for $\mathcal{F}(G)$ one can derive the co-product and co-unit of representation matrix elements \eqref{cofact} directly from the Peter-Weyl theorem, as we show in Appendix \ref{app:hopf}.

The upshot is that the co-product \eqref{cofact} on these co-representations can be used to define the bulk factorization map \eqref{3dfact}.   To make this totally explicit, let's write down the action of the factorization map on the zero-mode sector, and restore the frozen boundary indices $\mathfrak{i}_{L},\mathfrak{i}_{L}$ as in JT:
\begin{align}
    i: \ket{p_{\pm} \mathfrak{i}_{L}, \mathfrak{i}_{R}} \to \frac{1}{\sqrt{\dim_{q}(p_{\pm})}} \int ds \ket{p_{\pm} \mathfrak{i}_{L}, s}\otimes \ket{p_{\pm} s, \mathfrak{i}_{R}}.
\end{align}
If we define the wavefunctions as co-representation matrices:
\begin{align} 
\braket{g|p_{\pm} \mathfrak{i}_{L} \mathfrak{i}_{R} } =R^{p_{\pm}}_{ \mathfrak{i}_{L} \mathfrak{i}_{R}}(g),\qquad \braket{g|p_{\pm} \mathfrak{i}_{L} s } =R^{p_{\pm}}_{ \mathfrak{i}_{L} s}(g),\qquad \braket{g|p_{\pm}s \mathfrak{i}_{R} } =R^{p_{\pm}}_{s  \mathfrak{i}_{R}}(g).
\end{align} 
Then the factorization map just corresponds to the co-product \eqref{cofact} for the continuous series co-representations of $\SL^+_{q}(2,\mathbb{R})$. An explicit formula for the highly nontrivial representation matrices $R^{p_{\pm}}_{ss}(g)$ has been computed: see equation (7.35) of  \cite{ip2012representation}.  Given their importance in 3d gravity, it would be useful to understand these kinds of expressions better, see \cite{Mertens:2022aou} for recent progress and explicit expressions relating these different types of gravitational wavefunctions.

\subsection{Summary}

Paralleling the gauge theory mathematical structure of section \ref{sec:review}, we showed that factorization in gravity can be implemented by introducing edge states at the entangling surface, tagged by $\SL^+(2,\mathbb{R})$ representation labels for 2d JT gravity, and by quantum semi-group $\SL^+_{q}(2,\mathbb{R})$ labels for 3d gravity. The Plancherel measure appearing in the Peter-Weyl like spectral decomposition of $\text{L}^2\left(\SL^+_q(2,\mathbb{R}\right)$ \eqref{L2} is determined by the Virasoro modular S-matrices $S_0^{p_\pm}$ in \eqref{eq:S-modular}. This spectral decomposition allows us to define a factorisation map \eqref{3dfact} that matches the co-product of the underlying Hopf algebra. Thus, mathematically, our gravity factorization is a rather natural construction given our BF review in section \ref{sec:review}.

Physically, the q-deformation parameter equals $q=e^{i\pi b^2}$ and is related to the original microscopic 2d CFT by $c=1 + 6(b+b^{-1})^2$. The group theory Plancherel measure can be interpreted as the density of edge states at the entangling surface when implementing the shrinkable boundary condition using euclidean path integral methods. Most notably, the gravitational shrinkable boundary condition in 3d gravity does not allow descendant labels for the edge states at the entangling surface, leading to a finite entanglement entropy.

\section{Gravity versus gauge theory}
\label{CSdefect}

The purpose of this section is to offer a physical interpretation of our results in section \ref{sec:facgrav} in a broader context. We will compare our gravitational entanglement entropy formula to more standard entanglement entropy formulas in QFT or semiclassical gravity, paying special attention to Chern-Simons theory where we will identify key differences between its standard gauge theory formulation and the gravitational interpretation described in this work. 

\subsection{Embedding in entropy formulas}
\label{s:anyon}

The finiteness of the bulk entanglement entropy \eqref{eq:3d-ent-entropy} for a two-sided black hole in 3d gravity stands in contrast to the standard entanglement entropy formulas in QFT in a fixed background, which includes Chern-Simons theory and semiclassical gravity. Here, we revisit the validity of these expressions and explain their differences with our quantum gravity result  \eqref{eq:3d-ent-entropy}.

\paragraph{QFT in curved spacetimes.} 
For any continuum quantum field theory in $d$ spacetime dimensions, the entanglement entropy in any state of the system across a cut has the form (see e.g. \cite{Solodukhin:2011gn}):\footnote{The sum of subleading divergences ends in a logarithmic term only for $d$ even.}
\begin{equation}
\label{eq:SQFT}
S_{\text{QFT}} = \# \frac{A}{\epsilon^{d-2}} + \frac{s_{d+4}}{\epsilon^{d-4}} + \hdots + s_0 \ln \epsilon + S_{\text{finite}},
\end{equation}
where $\epsilon$ is a UV-regulator with dimensions of length encoding the short-distance entanglement of the QFT across the entangling surface. The divergent terms are the same for any state, whereas the specific information of the state itself is encoded in $S_{\text{finite}}$ only.

The leading contribution $S_{\text{leading}} \equiv \# \frac{A}{\epsilon^{d-2}}$ is proportional to the area $A$ of the entangling surface, but has a cut-off dependent prefactor. Since the near-cut region is always locally Rindler, $S_{\text{leading}}$ can also be calculated as the thermal entropy of the gas described by the modular Hamiltonian associated to a half-line (the Rindler Hamiltonian). For example, for a scalar field in $d=4$ one obtains \cite{Dowker:1994fi}:
\begin{equation}
S_{\text{leading}} = \frac{A}{360 \pi \epsilon^2},
\end{equation}
where $\epsilon$ is the regulated proper length to the Rindler horizon. This leads to an infinite amount of correlation between quanta in both subregions. Given the locality of these arguments, they extend to arbitrary cuts (see e.g. \cite{Solodukhin:2011gn}), in particular to the black hole event horizon entangling surface considered in this work.

From the thermal entropy perspective, one can hence view this divergence as an IR volume divergence, originating from the existence of infinite space close to the Rindler horizon, something that can be made more explicit by using tortoise coordinates. We should stress here that the thermal entropy obtained in subsection \ref{sec:grand-can} did not contain an IR volume divergence. Hence, according to the current discussion, there is no infinite correlation across the entangling surface. It follows our gravitational entanglement entropy is of a different kind than the QFT entanglement entropies \eqref{eq:SQFT}. We will make this comment more precise below.

\paragraph{Anyon defect entropy.}  Focussing on the relevant dimension $d=3$ discussed in this work, where
\begin{equation}
\label{eq:3dent}
S_{\text{QFT}} = \# \frac{l}{\epsilon} + S_{\text{finite}},
\end{equation}
we now compare this generic QFT expectation with specific analytic results applicable to Chern-Simons theory. Chern-Simons theory is a topological gauge theory in which (part of) the finite part of \eqref{eq:3dent} can be interpreted in terms of the factorization of Wilson lines via edge modes that transform under a quantum group.
For compact gauge groups, it is well known this quantum group is a ``hidden symmetry" of Chern-Simons theory \cite{Slingerland:2001ea}; these are symmetries commuting with all observables of the theory. In general, the edge mode symmetry satisfies this definition of hidden symmetry for a subregion, since the symmetry generators are in the center of the subregion operator algebra.  This is why the edge state labels split the Hilbert space into superselection sectors. We make some further comments on this in the concluding section \ref{Sec:conclusion}.

\begin{figure}[h]
\centering
\includegraphics[scale=.4]{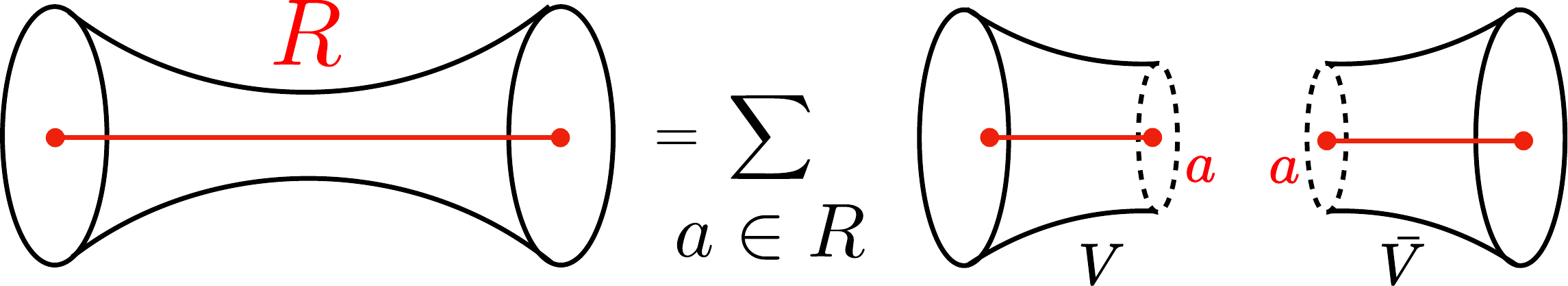}
\caption{Splitting a Wilson line in representation $R$ in two pieces requires introducing anyonic charges $a$ labeling states in the representation $R$.}
\label{WH}
\end{figure}

Physically, quantum groups are the symmetries of \emph{anyons}, i.e. the particles that couple to the Chern-Simons gauge field and provide the charges on which Wilson lines can end.  As a result, one expects the factorization of a state with a Wilson line inserted across an entangling surface would require the introduction of anyonic edge modes (see Figure \ref{WH}). 

Unlike ordinary charged particles,  anyons transforming in a representation $R$ of the quantum group are assigned a quantum dimension $\dim_{q}R$ which does not have to be an integer.  This is an \emph{effective} dimension for the Hilbert space associated to a single anyon, defined so that as $N \to \infty$ the fusion Hilbert space of $N$ anyons scales as $(\dim_{q}R)^N$.   Heuristically, we expect to obtain a contribution of  $\log \dim_{q}R$ to the entanglement entropy each time we cut a Wilson line in the representation $R$.  This expectation is almost correct.  

Using the standard shrinkable boundary condition for Chern-Simons theory \cite{Wong:2017pdm,Wen:2016snr}, the edge partition function is counting the states of a chiral WZW model at the entangling surface:
\begin{equation}
\label{eq:chaWZW}
\chi_R(e^{-\frac{2\pi \epsilon}{l}}) \to S_{R}{}^{0} \chi_0(e^{-\frac{2\pi l}{\epsilon}}).
\end{equation}

Hence the entanglement entropy of a state containing a Wilson line insertion crossing the entangling surface is
\begin{align} \label{Stop} 
S_{\text{CS}} ~=~  \frac{\pi^2 c}{3}\frac{l}{\epsilon}  + \log S_{R}{}^{0} ~=~  \frac{\pi^2 c}{3}\frac{l}{\epsilon}  + \log S_{0}{}^{0} + \log \dim_{q}(R),
\end{align}
where $l$ is the length of the entangling surface and $\epsilon$ a UV regulator separating the two subregions.

In the second equality, we used the relation between the quantum dimension and the regulated dimension for the representation $R$ of a chiral algebra \cite{Harvey:1999gq}
\begin{align}
    \dim_{q} R = \lim_{\epsilon \to 0} \frac{\chi_{R}(e^{-\frac{2 \pi \epsilon}{l}} ) }{\chi_{0}(e^{-\frac{2 \pi \epsilon}{l}})} = \frac{ S_{R}{}^{0}}{S_{0}{}^{0}},
\end{align}
which holds in a rational CFT. This is the analogous entanglement entropy formula to \eqref{eq:3dent} specific to Chern-Simons theory in an excited state that includes a Wilson line. Let us discuss the different terms in \eqref{Stop} in more detail.
\begin{itemize}
\item
The first two terms in \eqref{Stop} define the vacuum entanglement entropy $S_{\text{vac}}= \frac{\pi^2 c}{3}\frac{l}{\epsilon}  + \log S_{0}{}^{0}$, i.e. in the absence of any Wilson line excitation. This is an explicit realization of the general statement relating the divergent structure in \eqref{eq:SQFT} to the vacuum entanglement entropy.  Let us comment on the $l$ and $\epsilon$ dependences.

Since the Chern-Simons action does not make explicit reference to any bulk background metric, it is natural to ask what the origin of the length $l$ dependence in \eqref{Stop} is. This comes about because the gauge theory shrinkable boundary condition $A_{\tau_E}-A_\varphi\vert_{\partial \mathcal{M}}=0$ at the entangling surface requires the addition of a boundary term to the Chern-Simons action that introduces a metric dependence. For example, the standard boundary term $\oint_{\partial \mathcal{M}} \tr A\wedge *A$ depends on the background metric via the Hodge star $*$. 
This metric dependence is part of the edge CFT data and defines the length $l$ appearing in the ``area law" divergence. 

 Regarding the $1/\epsilon$ divergence, it can be traced to the fact that the number of descendants per primary is infinite in the edge CFT at the entangling surface. Indeed, the left hand side of \eqref{eq:chaWZW} is just $\chi_R(1) = \lim_{\mathfrak{q}\to 1} \text{Tr}_R\mathfrak{q}^{L_0}$.

\item
The $\log \dim_{q} R$ term is associated with the entanglement of anyons arising as a \emph{defect entropy}.  This is defined by subtracting the background entanglement entropy $S_{\text{vac}}$ from \eqref{Stop}
\begin{align} \label{sdef}
S_{\text{def}} \equiv  S_{\text{CS}} - S_{\text{vac}} = \log \dim_{q}(R).
\end{align}
\end{itemize}

\paragraph{Difference with 3d pure gravity.} Since 3d gravity can be formulated in terms of Chern-Simons theory, one may wonder if we can directly apply \eqref{Stop} to find the entanglement entropy of the gravitational degrees of freedom. 
This does not work. The failure to satisfy the gravitational shrinkability condition manifests itself in two ways:
\begin{itemize}
\item
In the gravitational Chern-Simons theory, the topological (or defect) entanglement entropy one would write down using the gauge theory arguments above is \cite{McGough:2013gka}:
\begin{equation}\label{sdef?}
S_{\text{def}} \overset{?}{=} \log \Big(\frac{S_{p_{+}}{}^{0}}{S_{0}{}^{0}}\frac{S_{p_{-}}{}^0}{S_{0}{}^{0}} \Big).
\end{equation}
This is not correct since both $S_{0}{}^{0}=0$ and $S_{p}{}^{0}=0$ in the gravitational case, where these are just the Virasoro modular $S$-matrices. Even stripping off the normalization factors $S_{0}{}^{0}=0$ does not work. As we stressed in footnote \ref{fn2}, for compact groups, the modular $S$-matrix is symmetric, but this is not so for the non-compact case at hand. 
\item
The divergent area term $\sim l/\epsilon$ itself is also problematic in the gravitational theory. As noted earlier,  the length dependence arises from the background metric on the entangling surface, which makes sense when describing quantum matter in a fixed background.  
 However, this does not make sense in gravity, where the  \textit{dynamical} spacetime metric is encoded into the Chern-Simons connection itself as $A^\pm = \frac{e}{\ell} \pm \omega$.   There is simply no consistent way to match the background metric in the boundary term $\oint_{\partial \mathcal{M}} \tr A\wedge *A$ to the dynamical one encoded in $A_{\mu}$. 
 For example,  if we allowed for a varying background that matches with the one determined by $A^\pm$, the classical variational principle would not be satisfied due to the hidden $A$ dependence in the Hodge star of the boundary term.  This differs from the situation at the asymptotic boundary of AdS$_3$, where the metric is fixed and this matching in principle makes sense.

Altogether, this argument says the ``area'' dependence found in the first term of \eqref{Stop} is \textit{not} the correct area of the black hole horizon, which should be measured with the metric information encoded in $A^\pm$ instead. Hence this term has no meaning at all for the gravitational model, and we have to focus on $S_{\text{def}}$ instead. 
\end{itemize}
 Our results from section \ref{sec:edge_shrink} show that the correct gravitational black hole entropy is instead
\begin{align}\label{Sbh}
    S_{\text{def}}= \log (\dim_{q}p_{+} \dim_{q}p_{-}) = \log (S_{0}{}^{p_{+}}S_{0}{}^{p_{-}}).
\end{align}
How do we interpret this formula?

\paragraph{Gravitational anyons.} Anyons allow excited states with Wilson lines to split, and the quantum state describing a two-sided black hole is one such Wilson line. Since we expect the entanglement entropy associated with the splitting of such Wilson line to be the logarithm of the effective dimension of the anyon Hilbert space, we are left to ask what is the appropriate quantum dimension of these anyons. This quantum dimension is effectively just the non-vanishing $S$-matrix element leading to the entropy \eqref{Sbh}.
This does appropriately account for the black hole entropy in terms of the entanglement of anyon edge modes. 

The defect entropy equation \eqref{Sbh} holds for a state with an anyon excitation $(p_+,p_-)$. To properly compare with our gravitational von Neumann entropy \eqref{eq:BHncl}, we should consider the mixed state that is a classical mixture of such anyon excitations,
\begin{equation}
\rho = \bigoplus_{p_{\pm},n,\bar{n}}p_{n,\bar{n}}(p_+,p_-) \rho_{p_{\pm},n,\bar{n}},
\end{equation}
with probability distribution $p_{n,\bar{n}}(p_+,p_-)$ \eqref{eq:probdis3d}, and with $\rho_{p_{\pm},n,\bar{n}}$ the reduced density matrix for a state with anyon label $p_\pm$ and asymptotic descendant labels $n,\bar{n}$.\footnote{Asymptotic boundary descendants are added in a somewhat ad hoc way from the perspective of this section.} \\

To what extent do these ``gravitational anyons'' behave the same as the well-studied (compact) non-abelian anyons? 
A first question is whether there exist anti-anyons whose fusion with the original anyon contains the identity. The answer is mixed. Firstly, our anyon degrees of freedom are labeled by $p_\pm$. Within the representation theory of SL$_q^+(2,\mathbb{R})$, the conjugate representation is found by letting $p_\pm \to -p_\pm$, and these are to be interpreted as the anti-anyon degrees of freedom. However, these representations do \emph{not} fuse into the identity module since the latter is not present in the Hilbert space.\footnote{ What is instead true is the (weaker) property that the DOZZ formula for the bulk Liouville three-point function $C(\alpha_1,\alpha_2,\alpha_3)$ with operator labels $\alpha_1 = Q/2 + iP_1$, $\alpha_3 = Q/2 + iP_2$ and $\alpha_2 = \epsilon \to 0$ yields precisely a Dirac delta function on $P_1+P_2=0$. The main difference with the compact case is that the identity representation is not normalizable.}
These are well-known features in the context of 2d Liouville CFT that also appear here.

A second question one might ask is whether these anyons and anti-anyon degrees of freedom correspond to positive and negative energy. This is not true due to two seperate reasons. Firstly, the energy for both anyon and anti-anyon, according to a fiducial observer, is the same and $\geq 0 $ (up to a zero-point energy), and is proportional to the quadratic Casimir $H = \frac{p_+^2+p_-^2}{\ell}$. Secondly, the edge degrees of freedom live at the horizon, and hence for any outside observer these energies are redshifted to zero.
The above discussion applies without modification in the JT limit.\footnote{An interesting feature that appears there is that for $\mathcal{N}=2$ JT supergravity, the shift between the vacuum and the continuum disappears (in technical terms, the Weyl vector for the supergroup OSp$(2|2,\mathbb{R})$ vanishes). It would be interesting to see whether in this case the anyon fusion rules are more in line with the compact case.}

\paragraph{Comparison with quantum bulk entanglement entropy.} 
Finally, we compare our result with those obtained in semi-classical gravity. This involves working in a bulk semi-classical expansion\footnote{ Such a bulk semi-classical expansion was performed around thermal AdS$_3$ in \cite{ Belin:2019mlt}.  Their computation of $S_{\text{vN}}(\Lambda)$ involves Virasoro edge modes assigned to the bulk entangling surface. As expected, the resulting entanglement entropy of bulk gravitons and photons is divergent, but with the correct area dependence to allow for an interpretation in terms of the renormalization of $G_{\mt{N},\text{bare}}$.} in $G_{\mt{N}}$ where the relevant notion of bulk black hole entropy is the \textit{generalized entropy} \cite{Bekenstein:1974ax,Jafferis:2015del,Engelhardt:2014gca}
\begin{equation}
  S_\text{gen} = \frac{\text{Area}}{4G_{\mt{N},\text{bare}}(\Lambda)} + \text{(corrections)} + S_\text{vN}(\Lambda)\,,
\label{eq:Sgen}
\end{equation}
containing a classical area piece, plus possible corrections of local integrals of combinations of the Ricci scalar and the extrinsic curvature at the entangling surface, combining into the Wald entropy \cite{Wald:1993nt}. The quantum loop effects $S_\text{vN}$ across the entangling surface are of the form discussed earlier \eqref{eq:SQFT} . 

Both types of contributions to \eqref{eq:Sgen} are cut-off $\Lambda$ dependent and divergent, but their sum is expected to be finite and typically interpreted as renormalizing $G_{\mt{N}}$ and the couplings of higher-derivative gravitational terms added to the effective action. This procedure underlies the slogan that gravity is expected to make sense of QFT entanglement entropy, by providing an effective physical regulator \cite{Susskind:1994sm}. This expectation has recently been supported by abstract arguments based on von Neumann operator algebras \cite{Witten:2021unn,Chandrasekaran:2022eqq}, building on \cite{Leutheusser:2021frk,Leutheusser:2021qhd}.

In the context of AdS/CFT, quantum error correction provided significant insights into the holographic nature of these kinds of formulas \cite{Harlow:2016vwg}. In particular, this perspective stresses the relevance of the \textit{code subspace}, as a subspace embedded in the full UV complete $\mathcal{H}_\text{physical}$ where an emergent bulk geometric description together with a bulk effective QFT describing small energy excitations is reliable. In usual discussions, the "dimension" of the code subspace is much smaller than the full Hilbert space. Thus, one does not have access to the UV degrees of freedom in the bulk. However, when the size of the code subspace equals the size of $\mathcal{H}_\text{physical}$, and we have reached the UV-scale $\Lambda_{\text{Planck}}$, the Wald entropy term in \eqref{eq:Sgen} disappears, and the entanglement entropy should be entirely quantum. In the above notation, this is the finite quantity $S_\text{vN}(\Lambda_{\text{Planck}})$. This is precisely the result in our gravity proposal \eqref{eq:3d-ent-entropy}. At the Planck scale, the gravitational degrees of freedom become visible and one just computes their von Neumann entropy, instead of the semi-classical area term. In our scenario, we have accounted for these gravitational degrees of freedom as edge states instead. 

We stress this does not mean our proposal has a full UV-complete description of \emph{all} gravitational physics. We do have an effective quantum description of the Hilbert space that can count the black hole microstates correctly, in a low-energy description \cite{Harlow:2015lma}.
To better appreciate this last point, it may be useful to observe an analogous phenomenon in the study of topological phases in condensed matter physics. There is a variety of UV-complete lattice models (Kitaev's Toric code \cite{Kitaev:1997wr} and the  Levin-Wen models \cite{Levin_2006}) exhibiting topological order in their ground state.   Since the Hilbert space manifestly factorizes, the entanglement entropy can be computed directly from the microscopic degrees of freedom.  However, one can also appeal to the description of the ground state as a network of anyon Wilson lines \cite{Levin_2006,Bonderson:2017osr},\footnote{These Wilson lines are identified with the lattice spin configurations for the Levin-Wen model, while for the toric code, they live on the dual lattice.} in which the entanglement entropy is determined by anyon fusion category data. Our 3d gravity calculation of the entanglement entropy is analogous to the one provided by the anyon worldline description. In particular, each anyon excitation $R$ that crosses the entangling surface contributes an entanglement entropy of $\log \dim_{q} R$.  Notice that in the lattice model, the quantum dimension does not count the dimension of the microscopic Hilbert space: instead it measures the dimension of a \emph{non-local} Hilbert space which arises as the fusion space of anyons with label $R$. This fits well with the bulk gravitational picture, where there are no local degrees of freedom: as a result the usual QFT entanglement due to UV-modes near the horizon is absent in the gravitational computation.  Instead, ``long-range" entanglement makes up the entire  contribution to the gravitational entanglement entropy.

\subsection{Wilson line factorization in Chern-Simons theory}
\label{s:linefactori}

Our earlier discussions showed that quantum group symmetry plays a crucial role in labeling the edge states required for factorizing 3d gravity. In the context of CS theory, we also saw it was natural to introduce the defect entropy $S_{\text{def}}$ in \eqref{sdef}. One can ask whether there exists a factorization map which only cuts the Wilson line degrees of freedom (and not the vacuum contribution), and provides a canonical definition of $S_{\text{def}}$ in terms of a reduced density matrix. Here, we briefly comment on this question and supplement this perspective with a more explicit form of the gravitational Chern-Simons wavefunctions and their factorization across an entangling surface.

The results of \cite{Donnelly:2020teo,Jiang:2020cqo} imply there exists a factorization map cutting only the Wilson line degrees of freedom provided we introduce anyon edge modes transforming under a $q$-deformation of the gauge group. A $q$-deformed entanglement entropy can then be defined which counts the quantum dimension of the anyon: 
\begin{align}
    S_{\text{def}} = -\tr_{q}\rho_{R} \log \rho_{R} = \log \dim_{q}(R) .
\end{align}
In this formula, $\rho_{R}$ is the reduced density matrix for a Wilson line in a fixed representation $R$, and $\tr_{q}$ is a quantum partial trace which is defined to be invariant under the adjoint action of the quantum group surface symmetry. This $q$-deformed notion of entanglement entropy has also been used to measure entanglement on spin chains with quantum group symmetries \cite{Pasquier:1989kd}.

 To illustrate these ideas, we will here use the results of reference \cite{Guadagnini:1989tj}, which give explicit Chern-Simons wavefunctions in the presence of punctures corresponding to the endpoints of Wilson line insertions.\footnote{Technically, the Chern-Simons description of bulk gravity goes beyond the scope of reference \cite{Guadagnini:1989tj}, which restricts to compact gauge groups, since gravity requires a theory of $\SL(2,\mathbb{R})$ Wilson lines. The proper TQFT-like framework underlying our gravitational Chern-Simons description was developed mainly by J. Teschner and collaborators, see e.g. \cite{Teschner:2000md,Teschner:2013tqy}, and generalizes the usual formulation in terms of modular tensor categories which is appropriate to the case of compact gauge groups.}
In the canonical quantization of Chern-Simons theory, the two spatial components  $A^{a}_{i}(x), \quad i=1,2=r,\varphi$ are conjugate variables.  In the Schr\"odinger representation, wavefunctions depend on the ``position" component, which we pick to be $A^{a}_{1}(x)$. Moreover, Gauss' law requires the wavefunctions to be gauge-invariant, which implies they only depend on $A^{a}_{1}(x)$ via the holonomy variable $U$: 
\begin{align} 
 U(x) \equiv P \exp \Big( - i \int^{x} L^R_{a} A^{a}_{1}(x)\Big), \qquad a=1\hdots \text{dim G}.
\end{align} 
Let  $\Psi_{0} \equiv \Psi_{0}[ U] $ be the vacuum wavefunctional of the Chern-Simons theory on a spatial cylinder with no Wilson line insertions. An explicit expression can be written down, but it will not be needed here. If the cylinder is heuristically identified with a Cauchy slice connecting the two asymptotic boundaries,\footnote{This is heuristic because the gravitational Wilson line is what produces the ER bridge connecting the two sides to begin with.} the insertion of a Wilson line in the irreducible representation $k$ connecting the boundary points $y_{L}$ and $y_{R}$ produces the wavefunction
\begin{align}\label{Wpsi}
    \hat{W}^{k}_{\mathfrak{i}_{L},\mathfrak{i}_{R}} (y_{L},y_{R}) \Psi_{0}[U]\,.
\end{align}
Here $ \hat{W}_{\mathfrak{i}_{L},\mathfrak{i}_{R}}$ is the matrix element of a  Wilson line operator acting as an operator on the above Hilbert space of wavefunctionals:
\begin{align}
    \hat{W}^{k}= P \exp \Big( i \int_{y_{L}}^{y_{R}} \hat{A}_{i}^{a}(x) L_{a}^{k} dx^{i} \Big), \qquad L_{a}^{k},\,\, a=1\hdots 3  \,= \,\left\{L_0,L_{+},L_-\right\}.
\end{align}
The $L_{a}^{k}$ are the generators of SL$(2,\mathbb{R})$ in the irreducible representation $k$, and the objects $\hat{A}_{i}^{a}(x) $ are quantum operators satisfying the canonical commutation relations for Chern-Simons theory. We view \eqref{Wpsi} as a ``position space'' realization of the two-sided states in our gravitational Hilbert space
\begin{align}
    \hat{W}^{k}_{\mathfrak{i}_{L},\mathfrak{i}_{R}} (y_{L},y_{R}) \Psi_{0}[U] \equiv \braket{U(x) | k,\mathfrak{i}_{L},\mathfrak{i}_{R} },
\end{align}
where the matrix indices $\mathfrak{i}_{L},\mathfrak{i}_{R}$ correspond to the asymptotic AdS$_3$ boundary conditions. Reference \cite{Guadagnini:1989tj} shows that this wavefunctional can be ``split'' as:
\begin{align} \label{qfact}
    \hat{W}^{k}_{\mathfrak{i}_{L},\mathfrak{i}_{R}} \Psi_{0}[U] =\hat{V}^{-1}_{\mathfrak{i}_{L} \, s }(y_{L} ). \hat{V}_{s\, \mathfrak{i}_{R}}(y_{R}) \Psi_{0} [U].
\end{align}
Here, for each $a$ and $s$, $\hat{V}^{-1}_{as}(y_L)$ is an operator carrying the quantum numbers of the puncture at $y_L$ associated with the representation $k$ of the gauge group $G$. It can also be interpreted as a Wilson line from some arbitrary (common) reference point to the point of interest $y_L$. The left index $a$ transforms as an ordinary representation of $G$, whereas the right index $s$ transforms under the \emph{$q$-deformation} of $G$. The latter can be shown by braiding the punctures around each other: this operation transforms the index $s$ by the monodromy matrix associated with the relevant quantum group \cite{Guadagnini:1989tj}. In the context of our gravitational theory, we have labeled the left index by $a=\mathfrak{i}_{L}$, as it corresponds to the particular Kac-Moody generator that is fixed by the asymptotic AdS$_3$ boundary conditions.  Similarly $\hat{V}_{sa}(y_R)$ carries an ordinary group index $a$ on the right, and a quantum group index $s$ on the left. The contraction of the $s$ indices is what makes the quantum group symmetry \emph{hidden}: the quantum group symmetry cannot be detected in the bulk theory because all states transform in its singlet representation.

Equation \eqref{qfact} should be viewed as a concrete realization of our factorization map \eqref{3dfact}: the gravitational Wilson line $\hat{W}^{k}_{\mathfrak{i}_{L},\mathfrak{i}_{R}}$ is factorized with quantum group edge modes labeled by $s$.  This is consistent with our discussion in section \ref{s:anyon}, where we pointed out that the correct gravitational bulk factorization should cut only the Wilson line degree of freedom and not the reference background represented here by the wavefunction $\Psi_{0}[U]$. 

\subsection{Gravitational shrinkability and conical singularities}
\label{s:appgaugegra}
The distinction between gauge theory and gravity, encoded in the gravitational shrinkability condition, can be given a clean physical interpretation in terms of partition functions with conical defects. Let us elaborate on this calculation.

Let us go back to the defect Wilson line insertions within the solid torus partition function. Analytically continuing \eqref{defecttr} $\lambda \to i \theta$, leads to the amplitude:
\begin{align}
\label{defectan}
Z_{\text{conical}} = \int_0^{+\infty} \hspace{-0.2cm} dp_+ dp_- 4\cosh(2\pi \theta p_+)\cosh(2\pi \theta' p_-) \frac{e^{-\beta(p_+^2 + p_-^2)} e^{i\mu\beta(p_+^2 - p_-^2)}}{\left|\eta(\tau)\right|^2}.
\end{align}
Geometrically, this corresponds to a conical defect inserted in the Euclidean time direction. We will write down the corresponding classical (saddle) metrics below. 

For the specific case when $\theta$ or $\theta'$ are integers, the expression \eqref{defectan} changes.
Indeed, inserting the $((1,n),(1,n'))$ degenerate Virasoro representation, corresponds to setting $\theta =n$ and $\theta'=n'$, and leads to a different measure:
\begin{align}
\label{eq:defect}
Z_{\text{exc}} &= \chi_{(1,n)}\left(-\frac{1}{\tau}\right)\chi_{(1,n')}\left(-\frac{1}{\bar{\tau}}\right) \\
&\hspace{-0.4cm} = \int_0^{+\infty} \hspace{-0.2cm} dp_+ dp_- 32 \sinh (2\pi n b^{-1} p_+) \sinh (2 \pi b p_+) \sinh (2\pi n' b^{-1} p_-) \sinh (2 \pi b p_-) \frac{e^{-\beta(p_+^2 + p_-^2)} e^{i\mu\beta(p_+^2 - p_-^2)}}{\left|\eta(\tau)\right|^2}. \nonumber
\end{align}

The classical metric description corresponding to the insertion of these defect Wilson lines for representations satisfying $n=n'$ is, setting $\beta = 2\pi$ here for simplicity:
\begin{equation}
ds^2 =\ell^2 \left[n^2\sinh^2\rho d\tau_E^2 + d\rho^2 + n^2 \cosh^2\rho d\varphi^2\right]\,.
\end{equation}
This has a conical periodicity in the $(\tau,\rho)$ plane of $2\pi n$. Similarly, defect Wilson lines in representations satisfying $n\neq n'$ correspond to spinning conical defects. The Euclidean section of this geometry has a complex metric ($dt = -i d\tau_E$) as usual:
\begin{equation}
ds^2 =\ell^2 \hspace{-0.1cm} \left[\left(nn'\sinh^2\rho -\frac{(n-n')^2}{4}\right)\hspace{-0.1cm}d\tau_E^2 + d\rho^2 + \left(nn'\cosh^2\rho +\frac{(n-n')^2}{4}\right)\hspace{-0.1cm}d\varphi^2 + i \frac{n^2-n'^2}{2}d\varphi d\tau_E\right],
\end{equation}
with a conical singularity of periodicity $2\pi \sqrt{nn'}$ in the plane ($\tilde{\tau_E},\rho$), where $\tilde{\tau_E}$ is the direction along which $d\varphi = -i \frac{n-n'}{n+n'}d\tau_E$ in the $(\tau_E,\varphi)$ plane, normalized such that $\tau_E = \tilde{\tau}_E$ along this direction. Note that this geometry becomes real in Lorentzian signature.

We need to make one important modification to the amplitude \eqref{eq:defect}. The actual evaluation of the Alekseev-Shatashvili path integral that would lead to \eqref{eq:defect} is formal: as soon as $\theta > 1$, (or $n \geq 2$), unstable modes appear in the one-loop fluctuations, and the extremum becomes a genuine saddle. The expression itself \eqref{eq:defect} is a bit formal from that perspective but can be motivated by other means. 

Dealing with negative modes in the Euclidean gravitational path integral is an old problem that dates back to \cite{Gibbons:1978ac,Page:1978zz}, where a simple proposal was made to complexify the unstable modes, making them stable again. Going through this process requires an extra factor of $i$ per such unstable mode, coming from the $\text{det}^{-1/2}$ fluctuation determinant.
Increasing $n$ by $1$ leads to 2 extra modes that become negative. This leads to a factor of $(-1)$ in the path integral coming from the inverse square root of the quadratic operator. This is the same type of procedure as in \cite{Gibbons:1978ac}.

Hence for a chiral sector with a defect $(1,n)$, the effective spectral density factor in the amplitude \eqref{eq:defect} becomes (up to a factor of $4\sqrt{2}$):
\begin{equation}
\label{repldos}
\rho_n(s) = (-1)^{n-1}\sinh (2\pi n b^{-1} s) \sinh (2 \pi b s).
\end{equation}

The above results let us stress the important differences between the first-order gauge theoretic formulation and the second order formulation of gravity. These can be made visible especially in lower dimensional models where we have sufficient control over the computations. Indeed, in JT gravity, an intuitive argument was given in \cite{Fan:2021wsb} on how to go back from the JT spectral measure to the correct Plancherel measure of the full group SL$(2,\mathbb{R})$. Armed with the above discussion on defects in 3d gravity, we can mirror this argument and use it to conjecture a new $q$-deformed measure for SL$_q(2,\mathbb{R})$. 

In the gauge theory formulation, Wilson line defects with unit monodromy matrix as one goes around the
Euclidean time cycle require $\theta, \theta' \in \mathbb{N}$ \cite{Raeymaekers:2014kea}. However, in gravity, only the case $n = n'= 1$ is a geometrically smooth configuration. Indeed, any other value of $n$ or $n'$ yields a (rotating) conical defect as illustrated above. 

What we learn from this discussion is that going back from gravity to gauge theory requires forgetting these winding numbers again by summing over $n, n' \in \mathbb{N}$. Since these two integers are independent, we can focus on a chiral sector to go through the argument. Summing the resulting spectral densities \eqref{repldos}, we regularize the expression by including an exponential dampening factor $e^{-2\pi r n}, \, r\geq 0$ and resum: 
\begin{equation}
\label{sumtric}
2\sum_{n=1}^{\infty} (-)^{n-1}  e^{-2\pi r n} \sinh (2\pi n b^{-1} s ) \sinh (2\pi b s) = \frac{\sinh (2\pi b s) \sinh (2 \pi b^{-1} s)}{\cosh (2 \pi r) + \cosh (2 \pi b^{-1} s)}.
\end{equation}
Letting $r \to 0$, we get the effective measure:
\begin{equation}
\label{qdim}
\rho(s) = \tanh (\pi b^{-1} s) \sinh (2 \pi b s).
\end{equation}
This object has the following properties:
\begin{itemize}
\item The double-scaling limit 
\begin{equation}
   \lim_{b\to 0} \frac{\rho(bk)}{2\pi b^2} =  k \tanh \pi k
\end{equation} 
reproduces the correct group-theoretical Plancherel measure for SL$(2,\mathbb{R})$. This makes us interpret \eqref{qdim} as the quantum dimension of the principal series representations of SL$_q(2,\mathbb{R})$, with standard deformation parameter $q=e^{\pi i b^2}$. Notice the asymmetry of \eqref{qdim} under $b\to 1/b$. This is in stark contrast to the $\sinh (2\pi b^{-1} s ) \sinh (2 \pi b s)$ measure of the modular double governing gravity, which by definition has this symmetry.
\item This quantum dimension makes an appearance as the modular $S$-matrix of the characters of the corresponding Kac-Moody algebra $\widehat{\mathfrak{sl}(2,\mathbb{R})}$ \cite{Fotopoulos:2004ut,Israel:2004jt,Jego:2006ta}. In particular, the right hand side of \eqref{sumtric} is a candidate formula for the quantum dimension of the same set of representations of the universal cover of SL$_q(2,\mathbb{R})$.
\end{itemize}

To sum up, the gravitational shrinkable boundary condition excludes bulk geometries with a conical defect at the entangling surface. On the other hand, these conical defects are naturally summed over in the gauge theory with gauge group $\PSL(2,\mathbb{R})\otimes \PSL(2,\mathbb{R})$. Thus, the measure for gauge theory and gravity differs in an essential way, leading to a different choice of bulk edge modes. 

\subsection{Summary}
Gravity and gauge theory differ in the implementation of the shrinkable boundary condition at the entangling surface. The gravitational entanglement entropy does not contain contributions from the fluctuations parallel to the entangling surface (descendants), because gravity does not contain the local degrees of freedom which produce the reference vacuum entanglement entropy across the horizon. We illustrated this last statement by the analogy with anyon defect entropy, and the splitting of a Wilson line on top of the vacuum. We also showed that the absence of conical singularities in the Euclidean 3d bulk is what distinguishes the gravitational description from the gauge theory formulation.

\section{Gravity and gauge theory as extended TQFT}
\label{s:exTQFT}

Our earlier discussions on factorization in gauge theory and 3d gravity relied on group theory and path integral considerations. Here, we retake the issue of factorization with an emphasis on the categorical formulation of symmetry and path integrals, in an attempt to explain and unify, in a relatively self-contained and abstract way, some of the statements we made in the previous sections. There are two reasons for introducing this level of abstraction.  

First, to understand the meaning of splitting Wilson lines in representations of $\SL^+(2,\mathbb{R})$ or $\SL^+_{q}(2,\mathbb{R})$, we must appeal to a more abstract notion of a symmetry: a generalized symmetry $\text{G}$ is a space defined by the (generally noncommutative) algebra $\mathcal{F}(\text{G})$ of functions on $\text{G}$.  As in the commutative case, $\mathcal{F}(\text{G})$ can be identified with a dual category of representations $\text{Rep}(\text{G})$.\footnote{A related idea appears in the Tannaka-Krein duality (see e.g. \cite{Chari:1994pz}), which says that a group $G$ can be reconstructed from the category $\text{Rep}(\text{G})$ of its representations.} In physical applications, one is often first presented with the latter, which contains data such as the Plancherel measure and fusion rules, without knowing exactly what $\text{G}$ is a priori.  Thus it is useful to view the representation category as a definition of $\text{G}$. 

Second, while identifying the factorization map \eqref{Wsplit} as a co-product of a generalized symmetry $\text{G}$ imposes algebraic constraints from the Hopf algebra structure of $\mathcal{F}(\text{G})$, these constraints have a priori no relation to spacetime locality.  On the other hand, the Euclidean path integral does impose local constraints on the factorization map.  To incorporate the latter, we appeal to a categorical formulation of the (Euclidean) path integral called extended TQFT, in which spacetime locality is captured by sewing relations. Geometrically, these sewing relations express the rules for the consistent cutting and gluing of spacetime subregions.  Crucially, extended TQFT translates subregions and their gluing rules into abstract objects and algebraic relations. It thus provides a fundamental substrate from which spacetime ``emerges".

\subsection{Extended TQFT and factorization}
\label{sec:eTQFT}

We now explain how the framework of (extended) topological quantum field theory (TQFT) unifies the group theory and path integral formulation of factorization described so far.\footnote{2d Yang-Mills and BF theory are not strictly topological theories. In particular, they both have infinite dimensional Hilbert spaces. However, they can still be formulated within the framework of extended TQFT, provided we make some small modifications to the standard axioms.}  

In a TQFT, the path integral is replaced by a set of algebraic data that satisfy constraint equations capturing spacetime locality.  A familiar example of this viewpoint is given by the modern formulation of a conformal field theory. Rather than appealing to path integration over local fields, a CFT is defined by its spectrum of primaries and the OPE coefficients $C_{ijk}$  appearing in their three-point function (see e.g. \cite{DiFrancesco:1997nk}):
\begin{align} 
\mathtikz{\muCijk{0}{0}}~=~ \left\langle O_i(0)O_j(z)O_k(\infty)\right\rangle_{S^2} ~=~ \frac{C_{ijk}}{z^{h_i+h_j-h_k}}
\label{Cijk}.
\end{align} 
The tensors $C_{ijk}$ are obtained by solving a set of sewing relations: this is the modular bootstrap program. When a Lagrangian presentation is available, the same OPE data can be obtained from the path integral on a sphere with three insertions, or equivalently, the ``pair of pants" diagram. Sewing relations arise from different slicings of the same manifold. For example, crossing symmetry corresponds to equating two different slicings of a sphere with 4 holes in the path integral perspective. This is equivalent to the associativity constraint satisfied by the algebra defined by the tensors $C_{ijk}$.

We want to take a similar perspective with the path integral \eqref{split} computing our factorization map. Compared to the pair of pants cobordism, our factorization map has additional labels determining the boundary conditions at the physical and entanglement boundaries. For example, in $d=2$, the most general factorization map is given by a 6-index tensor:
\begin{align}
\mathtikz{\deltaA{0}{0};\node at (.5,-1.3) {\small $k$};\node at (-.5,-1.2) {\small $j$};\node at (0,.2) {\small $i$};\node at (0,-1) {\small $e$};\node at (-.6,-.3) {\small $a$};\node at (.6,-.3) {\small $b$} }
\end{align}
 Instead of computing this path integral directly,  we shall obtain it by solving the sewing relations of an \emph{extended} TQFT, which include the shrinkable boundary condition. In 2d, the sewing relations imply the extended TQFT is a \emph{Frobenius algebra} \cite{Abrams:1996ty,Quinn:1991kq}, and the factorization map is its co-product.  A Frobenius algebra is an associative algebra equipped with a trace that we denote as $\tr_{\text{Fr}}$.  The aforementioned unification of the path integral and group theory factorization refers to the following statement: the same factorization map serves as a co-product of a Hopf algebra (as discussed above eq. \eqref{eq:Ifact-2}) and as a co-product of a Frobenius algebra. The former captures the group multiplication axioms whereas the latter captures the sewing relations of the extended TQFT.
 
 Besides providing a definition of local factorization,\footnote{As explained in \cite{Donnelly:2018ppr} together with the factorization map, the extended TQFT also defines the basic ingredients such as subregion Hilbert spaces, partial traces, and reduced density matrices which are needed to compute quantum information measures in a continuum theory.} the extended TQFT can be formulated in a higher category language that is useful for capturing the parallel structure between low-dimensional gravitational theories and their gauge theory counterparts.   
 In this framework, the boundary labels are viewed as objects in a category of boundary conditions. For example, in 2d gauge theories, this boundary category is the category of algebras, and the entanglement boundary condition $e$ corresponds to the group algebra $\mathbb{C}[\text{G}_\mt{S}]$ of the surface symmetry group.  Since in a 2d extended TQFT, all amplitudes can be reconstructed by the category assigned to a point, the TQFT itself can be defined by its boundary category.\footnote{The cobordism hypothesis, originally formulated in \cite{Baez:1995xq}, (see \cite{Lurie:2009keu,Freed:2012hx} for more recent discussions) states that an extended TQFT in any dimension can be reconstructed from what it assigns to a point. } 
 
 The boundary category thus provides a precise mathematical definition of the bulk gravity theory.  For example, in 2d, by replacing the ordinary surface symmetry group of 2d YM or BF by $\SL^+(2,\mathbb{R})$, we can define an $``\SL^+(2,\mathbb{R})$ gauge theory", which can be viewed as a formulation of JT gravity.  Similarly, in 3d, we propose an analogous construction in which $\SL^+(2,\mathbb{R})$ is replaced by its $q$-deformation $\SL_{q}^+(2,\mathbb{R})$. 
In section \ref{sec:3dext}, we will comment on hints that the resulting bulk theory corresponds to an extension of the Teichm\"uller TQFT, which was developed in \cite{Verlinde:1989ua, EllegaardAndersen:2011vps}. 

\subsubsection{Definition of a (closed) TQFT}

Heuristically, a topological field theory is a map:
\begin{align}
    Z: \text{geometry}\to \text{algebra}.
\end{align}
This idea is most naturally formulated in the language of category theory. In particular, a $d$-dimensional (closed) TQFT is a functor
\begin{align}\label{Zfunct}
    Z: \mathbb{Bord}^{\text{closed}}_{d,d-1} \to \mathbb{Vect}_{\mathbb{C}}
\end{align}
between the geometric category of $d$-dimensional cobordisms to the category of complex vector spaces. 
The objects of $\mathbb{Bord}^{\text{closed}}_{d,d-1} $ are $(d-1)$-manifolds and the morphisms are the $d$-dimensional cobordisms, while the objects of $\mathbb{Vect}_{\mathbb{C}}$ are vector spaces and the morphisms are linear maps between them.

More explicitly, $Z$ is a rule that maps \cite{Atiyah:1989vu}
\begin{itemize}
\item every closed, oriented $(d-1)$-manifold $\Sigma$ to a vector space $Z(\Sigma)$ over $\mathbb{C}$. The empty set is mapped to $\mathbb{C}$ and disjoint unions of manifolds are mapped to the tensor product of vector spaces.
\item every $d$-dimensional cobordism $B$ between two closed ($d-1$)-manifolds $\Sigma_{\text{in}}$ and $\Sigma_{\text{out}}$ to a complex linear map $Z(B): Z(\Sigma_{\text{in}}) \to  Z(\Sigma_{\text{out}})$.  
\end{itemize}  
These ``two tiers" are related by the fact that a $d$-manifold $B$ with only an outgoing boundary $\Sigma_{\text{out}}$ is mapped to $Z(B) \in Z(\Sigma_{\text{out}})$.  This rule follows from the fact that $B$  is a cobordism from the empty set into $\Sigma_{\text{out}}$, and the corresponding linear map $Z(B): \mathbb{C} \to Z(\Sigma_{\text{out}})$  is just a choice of an element in $Z(\Sigma_{\text{out}})$. The functorial property of $Z$ implies the gluing of cobordisms along the in/out boundaries is mapped to composition of linear maps.  Intuitively, this means a TQFT provides a linear representation of manifolds.  

To illustrate these ideas, consider the case of $d=2$.  In this case the $d-1$ manifolds
are disjoint unions of circles, and the cobordisms are ``closed string" worldsheets with circular ingoing and outgoing boundaries. All 2d closed, oriented manifolds can be generated by gluing a finite set of cobordisms (read from top to bottom):
\begin{align} 
\mathtikz{\muC{0}{0}},\quad \mathtikz{\etaC{0}{0}},\quad \mathtikz{\epsilonC{0}{0}},\quad  \mathtikz{\deltaC{0}{0}}
\label{eq:gen}.
\end{align} 
The vector space $Z(S^1)\equiv \mathcal{H}_{S1}$ is endowed with a multiplication rule due to the pair of pants cobordism, which gives the analog of the OPE coefficients:
\begin{align}
\mathtikz{\muC{0}{0}}:\mathcal{H}_{S1}\otimes \mathcal{H}_{S1} &\to \mathcal{H}_{S1},\nn
\ket{i}\otimes \ket{j} &\to \sum_{k} c_{ijk} \ket{k} .
\end{align}
This makes $\mathcal{H}_{S1}$ an algebra, with a unit, co-unit, and co-product given by the 3 rightmost diagrams of \eqref{eq:gen}.  The properties of this algebra are given by the sewing relations of $\mathbb{Bord}^{\text{closed}}_{d,d-1} $. These ensure that different ways of cutting up a manifold into the generators \eqref{eq:gen} should give the same partition function when gluing them back together. These TQFT sewing relations imply that $\mathcal{H}_{S^1}$  is a \emph{commutative Frobenius algebra} \cite{Abrams:1996ty,Quinn:1991kq}. 

A simple example is given by a ``classical limit" of a 1+1d rational CFT, in which we take all primary dimensions $h_{i,j,k}$ to go to zero. Then the OPE coefficients $C_{ijk}$ go to the structure constants $c_{ijk}$ of a TQFT (see e.g. \cite{Dijkgraaf}).

\subsubsection{Extended open-closed TQFT in 2d}

For applications to the two-sided black holes in JT and 3d gravity, we need to introduce $(d-1)$-manifolds with boundaries.  These codimension-2 boundaries are decorated by a label denoting an abstract boundary condition. This is a  familiar aspect of boundary conformal field theory in 2D, where a spatial boundary is labeled by a conformal boundary condition that satisfies a set of sewing constraints \cite{Lewellen:1991tb}.  An \emph{extended} TQFT incorporates the extra structure associated with   codimension-2 or higher boundaries into the TQFT framework.  Different formulations of extended TQFT exist in the mathematical literature.   The approach we follow in this work, based on a version called ``open-closed" TQFT \cite{Moore:2006dw,Lazaroiu:2000rk}, is tailored for computations of entanglement entropy.  

As before, we begin by introducing these ideas in $d=2$. An open TQFT is a functor $Z$
\begin{align}
Z: \mathbb{Bord}^{\text{open}}_{2,1} \to \mathbb{Vect}_{\mathbb{C}},
\end{align} 
where $\mathbb{Bord}^{\text{open}}_{2,1}$ is a geometric category whose objects are intervals 
$\hspace{-0.25cm}\mathtikz{\Int{0}{0}; \node at (-.35,.22) {\small $a$};\node at (.35,.25)  {\small $b$}}\hspace{-0.25cm}$ with labeled endpoints, and the morphisms are cobordisms between these intervals.  A vector space $\mathcal{H}_{ab}$ is assigned to a labeled interval, and a cobordism maps ingoing intervals into outgoing intervals.  The open TQFT makes a distinction between the ``gluing" boundary corresponding to the initial and final slice, and the ``free boundary" which describes the time evolution of the interval endpoint where a boundary condition is assigned.
A generating set of cobordisms for the open TQFT is given by
\begin{align}\label{opencob}
    \mathtikz{\epsilonA{0}{0};\node at (0,-0.45) {\small $a$}},\quad \mathtikz{\muA{0}{0};\node at (0,0) {\small $b$};\node at (-.6,-.6) {\small $a$};\node at (.6,-.6) {\small $c$}},  \quad \mathtikz{\etaA{0}{0};\node at (0,0.45) {\small $a$}},  \quad  \mathtikz{\deltaA{0}{0};\node at (0,-0.85) {\small $b$};\node at (-.65,-.35) {\small $a$};\node at (.65,-.35) {\small $c$}  }.
\end{align}
Notice that open cobordisms satisfy a superselection rule for the boundary labels : these are always unchanged along a free boundary. As in the closed TQFT, the total  Hilbert space $\oplus_{a,b} \mathcal{H}_{ab}$ of the open TQFT forms a Frobenius algebra, with the ``open string fusion" (second diagram of \eqref{opencob}) as the multiplication rule.  However in this case the Frobenius algebra need not be commutative.   

We have now defined two Frobenius algebras describing the closed and open sector of a TQFT ``path integral".  The combined, \emph{open-closed} TQFT relates these two sectors via  algebra homomorphisms which split the circle into an interval and vice versa:
\begin{align} \label{zip}
\mathtikz{\zipper{0}{0};\node at (0,-.7) {\small $a$}},\quad \mathtikz{ \cozipper{0}{0};\node at (0,-.3) {\small $a$}}
\end{align} 
The resulting combined open-closed TQFT satisfies sewing relations making the total Hilbert space $\oplus_{a,b} \mathcal{H}_{ab}$ a \emph{knowledgeable} Frobenius algebra \cite{Lauda:2005wn}.  

\subsubsection{The higher category viewpoint and the boundary category}
 The gluing of manifolds along codimension-2 boundaries introduces a higher categorical structure that we now explain. The idea is to treat the labeled codimension-2 surfaces as objects in a geometric 2-category $\mathbb{Bord}_{d,d-1,d-2}$, and iterate the same structure that was previously associated with $(d-1)$-dimensional objects.  In particular, the 1-morphisms of this category are $(d-1)$-dimensional cobordisms between these $(d-2)$-dimensional objects. Gluing along $(d-2)$-manifolds corresponds to composition of these 1-morphisms.    For $d\geq 3$, we can have a $(d-1)$-manifold $M$ with a single outgoing boundary $\pd M$.  In this case the  TQFT assigns to $M$ an object $Z(\pd M)$ of the boundary category $\mathcal{B}_{s}$.  The $d$-dimensional cobordisms are now viewed as 2-morphisms (morphisms between 1-morphisms), and the composition of these 2-morphisms corresponds to gluing along $(d-1)$-manifolds. These two types of gluing satisfy natural compatibility conditions.

 Let's illustrate these ideas in $d=2$.  The objects of $\mathbb{Bord}_{2,1,0}$ are labeled points and the 1-morphisms are labeled intervals $\hspace{-0.25cm}\mathtikz{\Int{0}{0}; \node at (-.35,.22) {\small $a$};\node at (.35,.25)  {\small $b$}}\hspace{-0.25cm}$.  The composition of these 1-morphisms is determined by the open string fusion, which satisfies the requirement of associativity. The 2-morphisms are strips swept out by an interval:
 \begin{align}
    \mathtikz{\idA{0}{0};\node at (-.4,-.5) {\small $a$};\node at (.4,-.5) {\small $b$}   }
\end{align}
The compatibility condition says we can interchange the order of vertical versus horizontal gluing.

In higher categorical language, a $d=2$ extended TQFT is defined to be a functor  
 \begin{align}
  Z:  \mathbb{Bord}_{2,1,0} \to \mathcal{B}_{S},
\label{eq:2d-category}
\end{align}
where $\mathcal{B}_{S}$ is a 2-category of boundary conditions.  For the $d=2$ examples considered in this paper, $\mathcal{B}_{S}$ is the category of algebras, with bimodules as the 1-morphisms and bimodule homomorphisms as the 2-morphisms \cite{Freed:2012hx}.\footnote{In 2d BCFT, one would view the objects of $\mathcal{B}_{S} $ as conformally invariant boundary conditions, and the 1-morphisms as boundary condition changing operators. This defines a vector space $\mathcal{H}_{ab}= \text{Hom}_{\mathcal{B}_{S}}(a,b)$, where $a,b$ labels Cardy boundary states.} The composition of 1-morphisms is the relative tensor product of modules, which coincides with the entangling product \eqref{eq:fusion}. 

What do we gain by this higher categorical description?  Note that in any dimension the extension can be iterated all the way down to a point, allowing us to consistently glue along manifolds of all codimension.  This increases the computational power of the TQFT when $d \geq 3$.
The enhanced computational power arises because the mathematical structure becomes more refined as we go to higher codimensions,  capturing more information about the total theory. For the particular $d=2$ and $d=3$ theories we consider, the partition functions, Hilbert spaces, and quantum information measures can be fully determined by the boundary category assigned to a codimension-2 surface, i.e. an entangling surface.\footnote{In fact, the cobordism hypothesis states that a fully extended TQFT is completely determined by what it assigns to a point.} 

\subsection{The shrinkable boundary condition in extended TQFT}
\label{sec:sh-eTQFT}

To discuss entangling surfaces, we introduce the shrinkable boundary condition $e$ into the open-closed TQFT formalism.  For simplicity, consider first the open-closed TQFT with only this boundary condition. Thus there is a single interval Hilbert space $\mathcal{H}_{ee}$ with the multiplication rule
\begin{align}
    \mathtikz{\muA{0}{0};\node at (0,0) {\small $e$};\node at (-.6,-.6) {\small $e$};\node at (.6,-.6) {\small $e$}} :\mathcal{H}_{ee} \otimes \mathcal{H}_{ee} \to \mathcal{H}_{ee},
\end{align}
required to satisfy all sewing relations of a knowledgeable Frobenius algebra. In addition, we impose the shrinkable boundary condition given by \cite{Donnelly:2018ppr}:
\begin{align}\label{emult}
    \mathtikz{\etaA{0}{0}  \cozipper{0}{0}; \node at (0,.46) {\small $e$}} =\mathtikz{\etaC{0}{0}}
\end{align}
This implies all holes created by the $e$ boundary can be closed. The factorization map on the interval is then given by the co-product of the Frobenius algebra $\mathcal{H}_{ee}$:
\begin{align}\label{cop}
\mathtikz{\deltaA{0}{0}\node at (0,-1) {\small $e$};\node at (-.5,-.3) {\small $e$};\node at (.5,-.3) {\small $e$}}
\end{align} 
For 2d gauge theories, we give an explicit realization of the Frobenius algebra $\mathcal{H}_{ee}$ in appendix \ref{app:2dgauge}.

In the open-closed TQFT formalism,
the $e$ label essentially turns a ``free" boundary into a gluing boundary along which we can fuse or split an interval using \eqref{cop} and \eqref{emult}.  As we alluded to in section \ref{sec:path-integral}, this should be viewed as cutting and gluing along a codimension-2 entangling surface.  This is a useful way to view codimension-2 gluings because it gives an open TQFT interpretation to any replica calculation of entanglement entropy on a closed 2d manifold: one simply  introduces holes with the shrinkable boundary condition at each connected component of the entangling surface. In particular, the sphere can be interpreted either as a ``closed string'' amplitude, or an ``open string'' trace  as shown in \eqref{ZS2}.   The extended TQFT language provides a precise algebraic description of this type of open-closed duality.  As illustrated in section \ref{sec:facgrav}, such a duality provides a mechanism by which the Euclidean gravity path integral can ``know" about black hole microstates.

Since this is an important point, let us spell out the details.  First, consider computing  $Z(S^2)$ in the unextended TQFT by making a codimension-1 cut along the equatorial circle. $Z(S^2)$ is then obtained by composing 
\begin{equation}
    \text{the unit}\,\,=\mathtikz{\etaC{0}{0}} \,\,\text{with the co-unit} \,\,\tr_{\text{Fr}}(\cdot)=\mathtikz{\epsilonC{0}{0}}\,,
\end{equation}
which is the trace function for the closed Frobenius algebra $\mathcal{H}_{S1}$.  This gives
\begin{align} \label{TrFr1-m}
    Z(S^2) = \tr_{\text{Fr}}(\mathbb{1}).
\end{align}
These operations are fixed by the Frobenius algebra corresponding to the closed TQFT.
However the trace \eqref{TrFr1-m} corresponds to the evaluation of an amplitude in the ``closed string channel'' and is not relevant to counting quantum states.

On the other hand,  a state-counting interpretation of $Z(S^2)$ can be obtained by introducing codimension-2 cuts along the equator. This is achieved by applying the factorization map to the unit, which gives a cobordism describing the ``thermofield double state":
\begin{align}
\mathtikz{\etaC{0}{0} } \to  \mathtikz{\etaC{0}{0} \mzipper{0}{0} \deltaA{0cm}{-1cm}} 
\end{align}
A similar procedure acting on the co-unit gives a second algebraic characterization of $Z(S^2)$: 
\begin{align}
    Z(S^{2}) = \mathtikz{\etaC{0}{0} \mzipper{0}{0} \deltaA{0cm}{-1cm} \epsilonC{0}{-4cm}  \cozipper{0}{-3cm}\muA{0cm}{-2cm} } =\mathtikz{\copairA{0}{0} \pairA{0}{0}}  = \tr_{\mathcal{H}_{ee}}(\mathbf{id}_{\mathcal{H}_{ee}}) 
\end{align}
In the second equality, an open-closed sewing relation was used. The final expression for $Z(S^2)$ is a quantum mechanical trace on the Hilbert space assigned to a subregion.  In other words, unlike \eqref{TrFr1-m}, this trace counts the dimension of an  ``open string" Hilbert space  $\mathcal{H}_{ee}$.     In appendix \ref{sec:CS-review}, we give a $d=3$ TQFT analog of this calculation in the context of Chern-Simons theory with compact gauge group.

\paragraph{Coset and shrinkable boundary conditions.}
In the fully extended higher category description of 2d gauge theory, a point is assigned to the category of algebras.   In particular, the shrinkable boundary label $e$ corresponds to the object $\mathbb{C}[\text{G}]$, viewed as a Frobenius algebra: we refer the reader to \cite{Freed:2012hx} for the details of this construction.
Here, we observe that the Peter-Weyl theorem implies that the interval Hilbert space $\mathcal{H}_{ee}=\text{L}^{2}(\text{G})$ is indeed a bimodule of $\mathbb{C}[\text{G}]$, consistent with the definition of 1-morphisms in the category of algebras, as stated below \eqref{eq:2d-category}.  The relative tensor product which describes the composition of 1-morphisms just corresponds to the entangling product defined by the surface symmetry group $G$.   

Consider now a subgroup $H$. In the extended TQFT language, two objects $e$, $H$ are introduced in the boundary category $\mathcal{B}_{S}$ of algebras.  These correspond to the group algebras $\mathbb{C}[\text{G}]$, $\mathbb{C}[\text{G/H}]$. Following the TQFT rules, a bimodule of these algebras is assigned to the intervals $\hspace{-0.25cm} \mathtikz{\Int{0}{0};\node at (-.35,.25) {\small $e$};\node at (.37,.28)  {\small $H$}}\hspace{-0.25cm}$ and $\hspace{-0.25cm} \mathtikz{\Int{0}{0};\node at (-.37,.28) {\small $H$};\node at (.35,.25)  {\small $e$}}\hspace{-0.25cm}$.  Using the relative tensor product, these two intervals can be glued along the endpoint labeled by $e$.  This is the entangling product producing the interval $\hspace{-0.25cm} \mathtikz{\Int{0}{0};\node at (-.37,.28) {\small $H$};\node at (.37,.28)  {\small $H$}}\hspace{-0.25cm}$ and the Hilbert space 
\begin{align}
    \mathcal{H}_{HH}= \bigoplus_{R} \mathcal{P}_{R,0}\otimes \mathcal{P}_{R,0}.
\end{align}
The coset boundary condition $H$ is relevant to the one-sided black hole states in JT and 3d gravity, where the gluing process just described provides a categorical description of how entanglement generates spacetime, or ER=EPR.

\subsection{JT gravity as an extended TQFT}
\label{sec:JTtqft}
Let us now apply the  extended TQFT paradigm to JT gravity.   As alluded to earlier, we interpret $e$ as the semi-group algebra 
 $\mathbb{C}[\text{SL}^{+}(2,\mathbb{R})]$.  The interval Hilbert space $\mathcal{H}_{ee}$ is a bimodule of $\mathbb{C}[\text{SL}^{+}(2,\mathbb{R})]$, which we identify as the space of functions on the semi-group
\begin{align}
    \mathcal{H}_{ee}=L^{2}(\text{SL}^{+}(2,\mathbb{R})).
\end{align}
As we saw previously, this space supports the left-right regular representation \eqref{lrregular} of the semigroup, and has a basis given by representation matrix elements of $\text{SL}^{+}(2,\mathbb{R})$ \eqref{L2JT}.  In this basis,  we can define the Frobenius algebra structure of $\mathcal{H}_{ee}$ by
\begin{align}
&\mathtikz{ \etaA{0cm}{0cm};\draw (0cm,0.5cm) node {\footnotesize $e$ } }= \int_{0}^{+\infty}  d k\, \int_{-\infty}^{\infty} ds  \sqrt{k \sinh 2 \pi k}\,e^{-\frac{\beta C(k)}{2}} \ket{k\, s\,s}, \nn
&\mathtikz{ \muA{0cm}{0cm};\draw (0cm,-.1 cm) node {\footnotesize $e$ } ;\draw (-.5cm,-.7cm) node {\footnotesize $e$ };\draw (.5cm,-.7cm) node {\footnotesize $e$ }  } :  \ket{k\,s_{1} s} \otimes \ket{k\, s' s_{2} } \to   \frac{1}{\sqrt{k \sinh 2 \pi k}} \delta(s-s')  \ket{k\,s_{1} s_{2}}, \nn 
&\mathtikz{\epsilonA{0}{0};\draw (0cm,-0.5cm) node {\footnotesize $e$ } }= \int_{0}^{+\infty}  d k\, \int_{-\infty}^{\infty} ds  \sqrt{k \sinh 2 \pi k}\,e^{-\frac{\beta C(k)}{2}} \bra{k\, s\,s}, \nn 
&\mathtikz{\deltaA{0}{0}\node at (0,-1) {\small $e$};\node at (-.5,-.3) {\small $e$};\node at (.5,-.3) {\small $e$}}: \ket{k s_{1} s_{2} } \to \frac{1}{\sqrt{k \sinh 2 \pi k} } \int_{-\infty}^{\infty} ds \ket{k\,s_{1} \,s} \ket {k\, s s_{2} }.
\end{align} 
These are formally identical to the cobordism generators of two-dimensional gauge theory, except that the representation basis have continuous labels.  From a formal point of view, this modification does not alter the sewing relations provided that we account for the associated volume factors. For this reason, this set of cobordisms define a Frobenius algebra just as in the compact case.

One aspect of the JT TQFT that is a bit more subtle than the compact case involves the 
identification of the unit element in the Frobenius algebra with the unit element of the group:
\begin{align} \label{unit}
\mathtikz{ \etaA{0cm}{0cm};\draw (0cm,0.5cm) node {\footnotesize $e$}}= \,\,\, \ket{g=\mathbb{1}}.
\end{align}
For compact groups, this follows directly from the delta function identity: 
\begin{align}
    \frac{1}{V_C}\sum_{R} \sum_{a} \dim R \,R_{aa}(g) = \delta(g-\mathbb{1}),
\end{align}
which in turn follows from the completeness relation satisfied by the characters 
\begin{align}\label{complete}
    \frac{1}{V_C}\sum_{R} \chi_{R}(g) \chi_{R} (g') = \delta(g-g'),
\end{align}
with $g'$ set to the identity, and the delta-functions on the right hand side only contain the contribution from the maximal torus.  
For the Frobenius algebra associated to JT gravity,  \eqref{unit}  is still true provided we define the characters in \eqref{complete} with respect to the appropriate measure. Notice also that the ``closed string'' cap state $\mathtikz{\etaC{0}{0} } $ corresponds to the identity element  \eqref{unit} of the Frobenius algebra.  This is made explicit by mapping the full disk to the half disk using one of the sewing relations of the extended TQFT:
\begin{equation} \label{axiom1}
\mathtikz{ \zipper{0cm}{1cm} \etaC{0cm}{1cm};\draw (0cm,0.2cm) node {\footnotesize $e$ } } = \mathtikz{ \etaA{0cm}{0cm};\draw (0cm,0.5cm) node {\footnotesize $e$ } }= \,\, \ket{g=\mathbb{1}}. 
\end{equation}
See appendix \ref{app:unitgroup} for further details.

\paragraph{Coset boundary conditions in JT}
To capture the two-sided geometries in figure \ref{BTZ}, we introduce the asymptotic boundary labels $\mathfrak{i}_{R}$ , $\mathfrak{i}_{L}$, which are the gravitational analog of the coset boundary labels $H$ described at the end of section \ref{sec:sh-eTQFT}.   The extended TQFT assigns the right quotient algebra $\mathbb{C}[\text{SL}^{+}(2,\mathbb{R})/\sim]$ to the boundary  label $\mathfrak{i}_{R}$  and  the analogous left quotient to $\mathfrak{i}_{L}$.  The two-sided Hilbert space $\mathcal{H}_{\mathfrak{i}_{L}\mathfrak{i}_{R}}$ is then the 1-morphism corresponding to the bi-module of these algebras given by
\begin{align}
    \mathcal{H}_{\mathfrak{i}_{L}\mathfrak{i}_{R}} \equiv \text{L}^2( \sim \backslash \text{SL}^{+}(2,\mathbb{R})/\sim).
\end{align}
Similarly, the one-sided Hilbert spaces $\mathcal{H}_{e \mathfrak{i}_{R} }=\text{L}^{2}(\SL^+(2,\mathbb{R})/\sim)$ are bimodules on which $e$ and $\mathfrak{i}_{R}$ act by left and right multiplication.  We can think of each element $g \in \SL^+(2,\mathbb{R})/\sim$ as a Wilson line connecting the bifurcation surface with the asymptotic boundary.  The composition of one-morphisms that glue  $\mathcal{H}_{e \mathfrak{i}_{R} }$ to $\mathcal{H}_{\mathfrak{i}_{L} e}$  into $\mathcal{H}_{\mathfrak{i}_{L}\mathfrak{i}_{R}}$  can be viewed as a categorical description of how entanglement generates spacetime, or ER=EPR.

 This more abstract perspective is useful in this situation, since defining an explicit path integral over $\SL^+(2,\mathbb{R})$ gauge fields is difficult due to the  non-local constraints that one has to impose on the gauge field.
For 3d gravity, we will have to suitably $q$-deform these objects.

\subsection{3d gravity as an extended TQFT}
\label{sec:3dext}
In three dimensions, the boundary category which defines an extended TQFT is a modular tensor category.  For Chern-Simons theory with compact gauge group $G$, this is usually taken to be the representation category of the loop group $\text{LG}$.
As reviewed in appendix \ref{sec:CS-review}, the Hilbert space factorization, shrinkable boundary condition,  and computation of entanglement entropy can be defined within the extended TQFT framework: in particular the entanglement edge modes are identified with elements of the representation category  $ \text{Rep}(\text{LG})$.

Since we have identified  $\SL^{+}_{q}(2,\mathbb{R})\times \SL^{+}_{q}(2,\mathbb{R})$ as the edge mode symmetry of bulk 3d gravity, it is natural to ask whether there exists a corresponding extended TQFT with a boundary category given by $\text{Rep}(\SL^{+}_{q}(2,\mathbb{R})\times \SL^{+}_{q}(2,\mathbb{R})) $.   A natural candidate for such a TQFT is given by the Teichm\"uller TQFT \cite{EllegaardAndersen:2011vps}.  This is the TQFT defined via the quantization of Teichm\"uller space, which is the classical  phase space of 3d gravity \cite{Verlinde:1989ua, Kim:2015qoa}.\footnote{ This is not the phase space of PSL$(2,\mathbb{R})$ Chern-Simons theory:  Teichm\"uller space on a Riemann surface $\Sigma$ is a particular component of the space of flat connections on $\Sigma$.}  More precisely, canonical quantization of Teichm\"uller space on a Riemann surface $\Sigma$ introduces a representation of the mapping class group (large diffeomorphisms), and it is well known that such a ``modular functor" defines a 3d TQFT \cite{Teschner:2005bz}.  Concretely, the quantization of Teichm\"uller space on $\Sigma$ gives rise to the Hilbert space of Virasoro conformal blocks\footnote{In the literature, these are often referred to Liouville conformal blocks, since they arise in the quantization of Liouville theory on $\Sigma$.  However we avoid the Liouville terminology, since this is sometimes used to refer to the flat spectrum \eqref{eq:lioupf}, which does not give the Cardy density of states. } on $\Sigma$, and the fact that these conformal blocks define a modular functor means that they satisfy a continuum version of the modular bootstrap equations \cite{Ponsot:1999uf}.
There is a wealth of evidence suggesting that 
$\text{Rep}(\SL^{+}_{q}(2,\mathbb{R})\times \SL^{+}_{q}(2,\mathbb{R})) $ defines an analog of a modular tensor category which underlies the  Teichm\"uller TQFT. In particular, the fusion rules and $F$-matrices dictated by the Virasoro modular bootstrap are given by the Clebsch-Gordon decomposition and $6j$-symbols of $\text{Rep}(\SL^{+}_{q}(2,\mathbb{R})\times \SL^{+}_{q}(2,\mathbb{R})) $.\footnote{ This is a direct analog of the Kazhdan-Lusztig equivalence \cite{10.2307/2152763} for compact groups $G$, where $\text{Rep}(\mathcal{U}_{q}(G))$ data provides the solution to the bootstrap equations for the Kac-Moody conformal blocks associated to the loop group LG. } 

\paragraph{The identity Wilson line and the entanglement boundary state}
Finally, we address one well known feature of the Teichm\"uller TQFT that might raise concerns about its viability as our gravitational TQFT.   Unlike the compact case, the  Teichm\"uller TQFT does not contain the identity Wilson line.  This is related to the fact that the spectrum does not contain the vacuum Virasoro block, defined by a path integral on a solid torus with an identity Wilson loop inserted.  This seems like a serious defect in the context of entanglement calculations; the entanglement brane boundary state associated to our factorization of the BTZ black hole is given by such a solid torus path integral, in the limit that it shrinks to zero size.    However, we circumvented this problem by applying  a modular transform to the vacuum block, leading to a superposition of heavy Virasoro blocks with a density of states given by the $\SL^{+}_{q}(2,\mathbb{R})$ Plancherel measure.   This gives the entanglement boundary state
\begin{align}
\ket{e} \equiv \int dp \,  4\sqrt{2}\sinh ( 2 \pi b p) \sinh( 2 \pi b^{-1} p)  \ket{p} ,
\end{align} 
which dimensionally reduces to the entanglement boundary state in JT gravity corresponding to a bulk disk path integral.

\subsection{Summary}

\begin{figure}[h]
\centering
\includegraphics[width=0.8\textwidth]{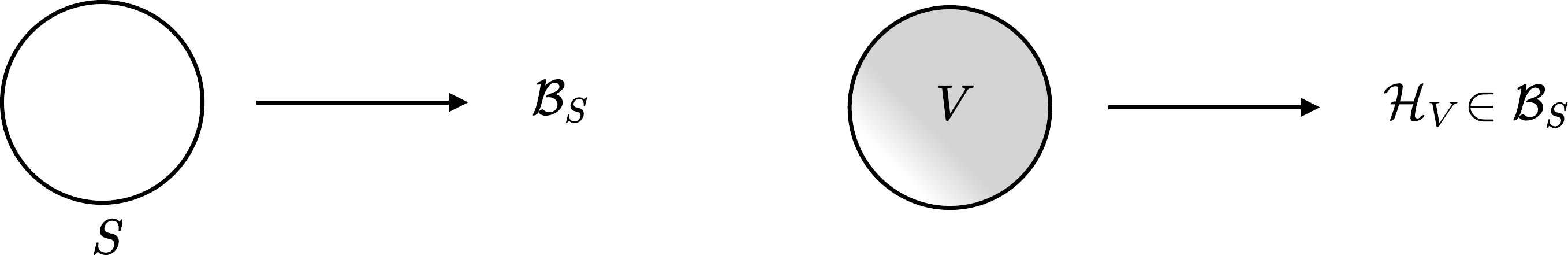}\\
\caption{The extended TQFT assigns a boundary category $\mathcal{B}_{S}$ to a circle $S$, and an object $\mathcal{H}_{\mt{V}} \in \mathcal{B}_{S}$ to a region $V$ bounded by $S$. }
\label{category}
\end{figure}

We reviewed how Hilbert space factorization can be embedded within the framework of extended TQFT, emphasizing a more abstract categorical perspective that we argue to be applicable to both gauge theories and gravity in d=2 and 3. In its standard formulation, the \emph{extended} TQFT framework assigns a \emph{boundary category} $\mathcal{B}_{S}$ to any closed oriented codimension-2 manifold $S$. In $d=3$,  $\mathcal{B}_{S}$ corresponds to a category of representations.  Furthermore, a codimension-1 manifold $V$ ending on $S$ is assigned to an object of the category $\mathcal{B}_{\mt{S}}$.  For example, for Chern-Simons theory with a compact gauge group, this is a representation of $\text{G}_{\mt{S}}$. In physical terms, this is a Hilbert space $\mathcal{H}_{\mt{V}}$ of the bounded region $V$. These facts are summarized in Figure \ref{category}.

To provide a precise framework for the emergence of spacetime from entanglement, 
we embedded the Euclidean path integral calculations provided in section \ref{sec:facgrav} into the extended TQFT formalism.  In particular, 
we introduced codimension-2 entangling surfaces for which the associated representation category is determined by the shrinkability constraint.  This constraint determines the gravitational edge modes necessary for the consistent cutting and gluing of spacetime subregions according to the rules of the extended TQFT. In this language, the ER=EPR paradigm reduces to the composition of morphisms. More generally, we proposed that the extended TQFT formulation of JT and 3d gravity provides the abstract algebraic structure from which classical geometry emerges.


\section{Concluding remarks} 
\label{Sec:conclusion}

We proposed an effective quantum mechanical model of 3d gravity with $\Lambda < 0$ based on the universal features of 2d holographic CFTs at high temperature. From the perspective of a microscopic AdS/CFT, our model is a theory of vacuum Virasoro blocks in the dual channel. While the bulk theory is 3d pure gravity, the boundary theory is given by the geometric action of Alekseev-Shatashvili \cite{Alekseev:1988ce,Alekseev:1990mp} for a Diff$(S^1)$ reparametrization field, with $S^1$ being the time coordinate. Instead of the usual Chern-Simons formulation, we argued the bulk Euclidean path integral should be defined as an extended TQFT with a boundary category given by $ \text{Rep}(\SL^{+}_{q}(2,\mathbb{R})\times \SL^{+}_{q}(2,\mathbb{R}))$. In particular, this viewpoint reproduces the Bekenstein-Hawking entropy as the bulk entanglement entropy of gravity edge modes transforming under the quantum semi-group $\SL^{+}_{q}(2,\mathbb{R})\times \SL^{+}_{q}(2,\mathbb{R})$. As known from gauge theory, once one introduces surface edge states, the Hilbert space can be split. This means that the gravitational Hilbert space can be split at the expense of introducing gravitational anyons. The entanglement of these anyons then allows us to glue spacetime back together again, providing a concrete picture of how gravitational entanglement generates connected spacetime. Forgetting the exterior of a subregion, the gravitational entanglement entropy associated to it is the black hole entropy. In the TQFT framework, the magic that allows the Euclidean path integral to give the correct counting of black hole microstates is attributed to the TQFT sewing relations which constrains the cutting and gluing of the path integral along codimension-2 surfaces.

Our proposal is incomplete. First, we did not give a full description of the putative extended TQFT describing the 3d bulk theory.  A natural candidate is given by the Teichm\"uller TQFT formulated in \cite{EllegaardAndersen:2011vps}, since this is the unitary TQFT associated with the representation category $ \text{Rep}(\SL^{+}_{q}(2,\mathbb{R})\times \SL^{+}_{q}(2,\mathbb{R}))$.\footnote{A physically intuitive way to understand this TQFT was provided in \cite{Mikhaylov:2017ngi}, which relates Teichm\"uller theory to analytically continued Chern-Simons theory.} Perhaps the most direct way to see the connection between our formulation of 3d gravity and Teichm\"uller TQFT is to consider the state sum model, in which the Teichm\"uller TQFT on a hyperbolic 3 manifold is defined by gluing together tetrahedra that gives a triangulation of the 3-manifold.   Such a state sum model defines an extension of the TQFT, since it must involve gluing along co-dimension 2 and higher surfaces (like the edges of the tetrahedra). The $6j$-symbols for $\text{Rep}(\SL^{+}_{q}(2,\mathbb{R})\times \SL^{+}_{q}(2,\mathbb{R})) $ are naturally identified with the tetrahedra in the state sum model, with each edge labelled by a quantum group representation.  Indeed, it has been shown that the semi-classical limit of the  $6j$-symbols for $\text{Rep}(\SL^{+}_{q}(2,\mathbb{R})\times \SL^{+}_{q}(2,\mathbb{R})) $  give the quantum volume of these  hyperbolic tetrahedra, which is given by the exponential of the Einstein-Hilbert action \cite{Teschner:2012em}. This relation suggests that the gluing of cauchy slices along co-dimension 2 entangling surfaces correspond to the fusion of quantum group representations along the edges of the tetrahedra.  We leave an explicit verification of this idea to future work.  One way to test our proposal would be to see if it gives a consistent set of rules to compute bulk entanglement entropy on different states and with different bi-partitions. In particular, one might wonder if the same gravitational edge modes can reproduce the Ryu-Takayanagi formula via a bulk entanglement entropy calculation: the answer is affirmative and reported on in \cite{Wong:2022eiu}.

We have given a description of gravitational edge modes as anyons: we showed that the bulk entanglement entropy measures their quantum dimension in a manner that is consistent with the black hole entropy.  However, our proposal does not provide a microscopic description of the edge modes. Similarly, our description of the subregion Hilbert was abstract: in particular, we did not give a realization of these states in terms of one-sided geometries. To understand the true implications of our proposal, we would need to make a connection to the microstates of the bulk string theory on AdS$_3 \times$ M$_{7}$.  As a first step in this direction, one might add the appropriate matter content due to reduction of string theory on M$_{7}$, and ask how the $\SL^{+}_{q}(2,\mathbb{R})\times \SL^{+}_{q}(2,\mathbb{R})$ symmetry is modified.   

Finally, the TQFT language of local cutting and gluing can only take us so far in gravity. The complication has to do with the group of large diffeomorphisms that would spoil locality. In 2d, this is the non-trivial mapping class group on Riemann surfaces of more complicated topology, leading to non-trivial global considerations. For the application to 3d chiral gravity, see \cite{Eberhardt:2022wlc}.

\paragraph{Edge states, hidden symmetries and superselection sectors.} 
The Lorentzian interpretation of the inner boundary in the shrinkable boundary condition is the horizon of the black hole. Hence, the resulting density of states should correspond to the gravitational edge modes localized at the black hole event horizon.

The ``hidden'' property of the edge mode symmetry structure is consistent with the idea that edge modes are not accessible from an outside observer perspective, since it would require an infinite amount of time to measure them.\footnote{However, if we were to introduce a cut-off, which is typically the most sensible thing to do in physics when an infinity arises, the argument may become less "sharp".} In a sense, edge modes can only be probed, when the system is cut or split into two pieces. 

This picture is also consistent with the physical picture of edge modes in Maxwell gauge theory in Rindler space \cite{Donnelly:2014fua,Donnelly:2015hxa,Blommaert:2018rsf}. One expects that due to infinite redshift, all excitations localized at the black hole event horizon should have vanishing energy for the outside observer, in terms of which the temperature and energies are measured. An explicit mode analysis shows the existence of vanishing $\omega_\mt{Rindler}=0$ at the Rindler horizon $\rho_\mt{Rindler}=0$ \cite{Blommaert:2018rsf}. Such bulk localization is consistent with these modes not being accessible for these observers in finite time. Since they should commute with any other exterior bulk operator, they should introduce superselection sectors decomposing the bulk Hilbert space
\begin{equation}
    \mathcal{H}_\mt{bulk} = \bigoplus_\alpha \mathcal{H}_\alpha \otimes \mathcal{H}_{\bar{\alpha}}.
\end{equation}
In our 3d gravitational discussion, $\alpha$ labels irreducible representations of $\SL^+_\text{q}(2,\mathbb{R})$: the $p_{\pm}$ labels describe the macroscopic properties of the black hole states, i.e. the energy $H$ and angular momentum $J$. Since the gravitational edge sector is based on a non-abelian group structure, there are additional edge labels: the $s$ and $\bar{s}$ are continuous microscopic labels counting the black hole degeneracy within a fixed ``macroscopic'' $(p_+,p_-)$ sector. Indeed, we already saw explicitly in JT gravity, see \eqref{eq:diagrho}, that the density matrix is block-diagonal in all of these edge labels. An analogous property holds in 3d gravity.

These edge quantum numbers $(p_+,p_- s,\bar{s})$ cannot be changed by any observer who only has access to the exterior of the black hole, or to the right half of the Hilbert space. This is obvious for the $s,\bar{s}$ quantum numbers since they are localized at the entangling surface. The macroscopic quantum numbers $p_\pm$ can also not be changed at the entangling surface in a finite amount of time, since attempting to inject some energy and/or angular momentum into the system only succeeds in creating multiple regions with different $p_\pm$ quantum numbers, without changing these quantum numbers at the black hole horizon.\footnote{This can be made more explicit in the 2d JT dimensional reduction, where injecting energy locally can be done by studying boundary correlation functions, which have multiple sectors in the spacetime characterized by different values of the energy \cite{Mertens:2017mtv}.}

Note that quantum group edge modes have been studied previously in the classical analysis of pure gravity edge modes, based on the covariant phase space formalism \cite{Dupuis:2020ndx}. However, the quantum group that arose in \cite{Dupuis:2020ndx} is the ordinary $\mathcal{U}_{q}(\mathfrak{sl}(2,\mathbb{R}))$ and not the modular double associated to $\SL^+_\text{q}(2,\mathbb{R})$.  It would be interesting to understand the relation between the two approaches. 

\paragraph{Generalization to other models and higher dimensions.} Since our work specifically applies to $d=3$, it is a natural question to ask whether the extracted lessons may carry over to higher dimensions. In such systems, gravity includes dynamical degrees of freedom. However, in a perturbative regime in $G_\mt{N}$, these can be carefully described as a further contribution to the matter sector propagating in a given background. Thus, the origin of the gluing of spacetime is still expected to be carried by gravitational edge modes, with symmetry transformation properties which may appear as "hidden", to outside observers, as in our 3d context.

An important outcome of our work uncovered the lack of descendants at the entangling surface in 3d gravity. An equivalent statement is that the edge sector of 3d gravity has no quantum numbers describing fluctuations along the horizon. This rephrasing makes the potential generalization of this statement to other models and higher dimensions straightforward. To be concrete, let us compare this description of the edge sector to that of dynamical Maxwell theory, as studied in this language by \cite{Donnelly:2014fua,Donnelly:2015hxa,Blommaert:2018rsf}. In spacetime dimensions $d \geq 3$, there are edge modes on the black hole horizon, in one-to-one correspondence with a surface electric flux perpendicular to the horizon:
\begin{equation}
\label{eq:edgeprofile}
E_\perp(\mathbf{x}) = \sum_\mathbf{k} \epsilon_\mathbf{k} e^{i\mathbf{k} \cdot \mathbf{x}},
\end{equation}
where $\mathbf{x}$ is the transverse coordinate along the black hole horizon and $\mathbf{k}$ is a (continuous) momentum label. Regularizing the black hole horizon with a brick wall boundary condition at $\rho_{\text{Rindler}} =\epsilon$, each edge mode labeled by $\mathbf{k}$ carries an energy
\begin{equation}
E_\mathbf{k}=\frac{\vert\epsilon_\mathbf{k} \vert^2}{2k^2\ln \frac{2}{k\epsilon}} \,\, \underset{\epsilon \to 0}{\to} \,\, 0,
\end{equation}
leading to the total energy in the coherent edge state $\ket{E_\perp}$, defined to be an eigenstate of the perpendicular electric field with eigenvalue \eqref{eq:edgeprofile}:
\begin{equation}
H_{\text{edge}}\ket{E_\perp}=\sum_{\mathbf{k}}E_\mathbf{k}\ket{E_\perp}.
\end{equation}
The fact that edge states are labeled by a transverse momentum label $\mathbf{k}$ immediately leads to a contribution to the entropy as $S \sim A/\epsilon^{d-2}$.
These tangential modes play the same role as the descendant labels in the 3d gravity story. Given the apparent similarity between the origin of edges states in 3d Chern-Simons models (reviewed in section \ref{sec:CS-review}) and in Maxwell's theory in $d\geq 3$, together with the absence of such edge sector in 3d gravity, we are lead to conjecture that no gravitational edge modes with non-trivial profile tangential to the black hole horizon exist
in higher-dimensional gravity. It would be interesting to get more clues on how this could be proven more generally.\footnote{There exist four dimensional perturbative calculations computing the entanglement entropy of linearized gravitons on a sphere starting with \cite{Benedetti:2019uej}, which was reproduced using the hyperbolic cylinder method in \cite{David:2020mls}. More recently, these were extended into higher dimensions in \cite{David:2022jfd}. }

\paragraph{Towards more general formulations of the shrinkable boundary condition.} It is not apparent how to formulate our approach for implementing the shrinkable boundary condition to higher dimensions. However, as reviewed in section \ref{sec:sh_gravity}, an alternative approach that generalizes more naturally to higher dimensions was given in \cite{Jafferis:2019wkd}. Here one combines a \emph{local} cutting map $\mathcal{J}$ and a defect operator insertion $\sqrt{D}$ to provide the factorization map in \eqref{sdj}. Even though the statistical mechanical interpretation of the bulk entanglement entropy seems to be gone from this perspective, it has the  advantage that such a defect operator is related to the area of the black hole.  This relation can be understood via Carlip and Teitelboim's \cite{Carlip:1994gc} off-shell formulation for the BTZ black hole.  They showed that the phase space associated with the Euclidean cigar topology is enlarged to include the conical angle and horizon area as a conjugate pair. This implies that the horizon area operator $\hat{A}$ generates translations in the conical angle.  Therefore the defect operator which imposes a $2\pi$ cone angle can be identified with the operator $D=\exp (\frac{ \hat{A}}{4G})$. A similar statement holds for higher dimensional black holes when we allow for a conical deficit at the Euclidean horizon \cite{banados1994black}.  It should be possible to obtain an explicit realization of $\hat{A}$ in a subregion Hilbert space along the lines of \cite{Jafferis:2019wkd}. Finally, it would be interesting to connect this Euclidean picture to the Lorentzian description of gravitational edge modes in \cite{Donnelly:2016auv}.  Here, the conical angle becomes a boost, and the area operator originates from a gauge-fixed version of an $\SL(2,\mathbb{R})$ gravitational surface symmetry associated with the normal geometry of the entangling surface. We hope to report on these ideas in future work.


\section*{Acknowledgments}
We would like to thank Nezhla Aghaei, Alexandre Belin, Andreas Blommaert, Andreas Brauer, Daniel Jafferis, Daniel Kapec, David Kolchmeyer, Alex Maloney, Samir Mathur, Du Pei,  Ingo Runkel, and Shinsei Ryu for discussions related to this work. GW especially thanks David Kolchmeyer and Daniel Jafferis for extended discussions on related topics. TM was supported by Research Foundation Flanders (FWO Vlaanderen), and acknowledges financial support from the European Research Council (grant
BHHQG-101040024). Funded by the European Union. Views and opinions expressed are however those of the author(s) only and do not necessarily reflect those of the European Union or the European Research Council. Neither the European Union nor the granting authority can be held responsible for them. JS is supported by the Science and Technology Facilities Council [grant number ST/T000600/1]. GW would like to thank the Aspen center for physics for hospitality while this work was being completed.

\appendix

\section{Boundary correlators in doubled JT limit}
\label{app:bcor}

In subsection \ref{sec:double-JT}, we identified a double scaling limit of our 3d gravity partition function leading to the squared JT gravity partition functions \eqref{doubleJT}. As mentioned in the main text, this is an interesting observation on its own right. In particular, it is natural to describe the insertion of boundary matter operators, extending the known techniques in the JT limit. We briefly discuss this below.

\begin{figure}[h]
\centering
\includegraphics[width=0.25\textwidth]{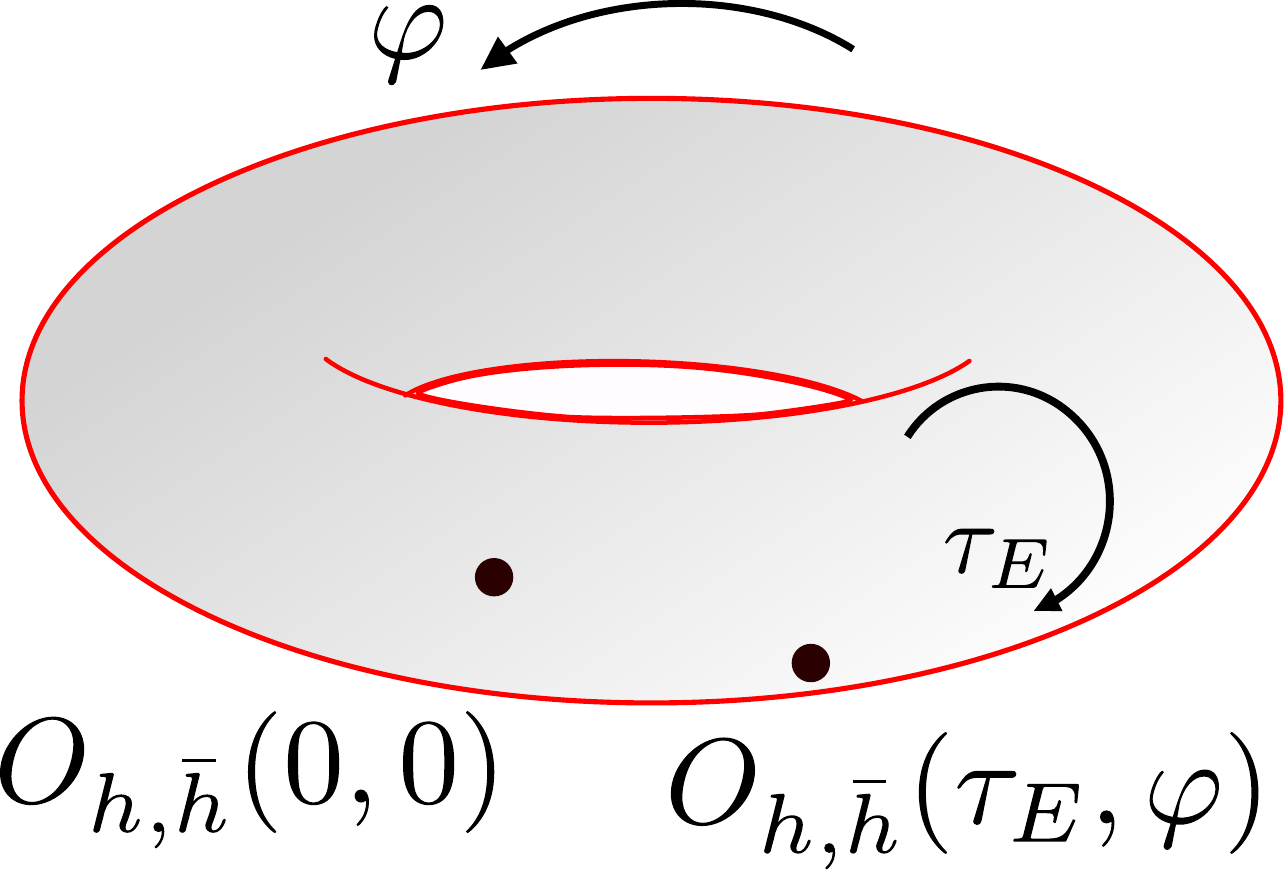}
\caption{Two boundary operators inserted on the boundary of the solid 2-torus.}
\label{2pttorus}
\end{figure}

Consider two identical operators $O_{h,\bar{h}}$, with conformal weights $h$ and $\bar{h}$,  inserted on the boundary of the solid torus and consider the Euclidean grand canonical correlator:
\begin{equation}
\text{Tr}[O_{h,\bar{h}}(\tau_E,\varphi)O_{h,\bar{h}}(0,0) e^{-\beta H + i\mu \frac{\beta}{\ell} J}].
\end{equation}
The set-up is illustrated in Figure \ref{2pttorus}.

The regime of interest to relate to known JT expressions is the following:
\begin{equation}
c \gg 1, \qquad \frac{\tau_E}{\ell} ,\frac{\beta-\tau_E}{\ell} \sim c\,, \qquad h, \bar{h} \sim 1,
\label{eq:double-scaleop}
\end{equation}
combined with a condition on the spectral gap being not too small: $\Delta_{\text{gap}} \gtrapprox \frac{\tau_E}{\ell} ,\frac{\beta-\tau_E}{\ell}$.
The above correlator can be written more explicitly as
\begin{align}
\label{eq:cor}
\text{Tr}[O_{h,\bar{h}}(\tau_E,\varphi)O_{h,\bar{h}}(0,0) e^{-\beta H + i\mu \frac{\beta}{\ell} J}] = \sum_{\text{primaries } O_1 O_2} \vert C_{O O_1 O_2}\vert^2 \mathcal{F}_{h_1,h_2}(\tau_E,\varphi) \bar{\mathcal{F}}_{\bar{h}_1,\bar{h}_2}(\tau_E,\varphi)),
\end{align}
in terms of primary sphere three-point functions $C_{O O_1 O_2}$, and Virasoro 2-point torus conformal blocks $\mathcal{F}_{h_1,h_2}$ and $\bar{\mathcal{F}}_{\bar{h}_1,\bar{h}_2}$. The latter can be obtained by inserting complete sets of states between the operators, when evolving in the Euclidean time direction:
\begin{align}
\label{eq:blocks}
\mathcal{F}_{h_1,h_2}(\tau_E,\varphi) &\equiv \hspace{-0.2cm}\sum_{N_1,N_2} \hspace{-0.1cm} \langle h_2,N_2\vert O\vert h_1,N_1\rangle \langle h_1,N_1 \vert O \vert h_2,N_2 \rangle e^{-\frac{\tau_E-i\varphi \ell}{\ell} (h_1 + \vert N_1 \vert -\frac{c}{24}) - \frac{\beta-\tau_E-i(\mu\beta-\varphi \ell)}{\ell}(h_2 + \vert N_2 \vert -\frac{c}{24})}, \nonumber \\
\bar{\mathcal{F}}_{\bar{h}_1,\bar{h}_2}(\tau_E,\varphi) &\equiv \hspace{-0.15cm} \sum_{\bar{N}_1,\bar{N}_2} \hspace{-0.1cm} \langle \bar{h}_2,\bar{N}_2\vert O\vert \bar{h}_1,\bar{N}_1\rangle\langle \bar{h}_1,\bar{N}_1 \vert O \vert \bar{h}_2,\bar{N}_2 \rangle e^{-\frac{\tau_E+i\varphi \ell}{\ell} (\bar{h}_1 + \vert \bar{N}_1 \vert -\frac{c}{24}) - \frac{\beta-\tau_E+i(\mu\beta-\varphi \ell)}{\ell}(\bar{h}_2 + \vert \bar{N}_2 \vert -\frac{c}{24})},
\end{align}
where $N_{1,2}$ is the descendant label, $\vert N_{1,2} \vert$ its level, and where we normalized the primary matrix element $\left\langle h_2\vert O\vert h_1\right\rangle=1$ (since the OPE coefficient was explicitly extracted). We largely follow the notation of \cite{Ghosh:2019rcj}. The diagrammatic representation of one of such blocks is drawn as:
\begin{equation}
\mathcal{F}_{h_1,h_2}(\tau_E,\varphi) \equiv
\begin{tikzpicture}[scale=0.25, baseline={([yshift=-0.1cm]current bounding box.center)}]
 \draw[thick] (-0.77,1.77) to (-2.5,3.5);
  \draw[thick] (-0.77,-1.77) to (-2.5,-3.5);
  \node at (-2.5,0) {\small $h_1$};
  \draw[thick](1,0) ellipse (2.5 and 2.5);
  \node at (4.5,0) {\small $h_2$};
  \node at (-3.5,-3.5) {\small $O$};
  \node at (-3.5,3.5) {\small $O$};
    \draw[->,thick, bend left=30] (5.5,3) to (5.5,-3);
  \node at (7.5,0) {\small $\tau_E$};
\end{tikzpicture}
\end{equation}

To proceed, we need to distinguish between two situations. One where our model is a proposal for 3d pure gravity as in section \ref{s:propgrav}, and a second one as containing universal dynamics of 2d CFT in the regime sketched in section \ref{sec:2duniv}. Let's start with the first case. In the regime of interest $\frac{\tau_E}{\ell} ,\frac{\beta-\tau_E}{\ell} \sim c \gg 1$, descendants scale out and the boundary correlator reduces to:
\begin{align}
&\text{Tr}[O_{h,\bar{h}}(\tau_E,\varphi)O_{h,\bar{h}}(0,0) e^{-\beta H + i\mu \frac{\beta}{\ell} J}] = \\
&\sum_{\text{primaries } O_1 O_2} \hspace{-0.4cm}\vert C_{O O_1 O_2}\vert^2 e^{-\frac{\tau_E-i\varphi \ell}{\ell} (h_1-c/24) +  \frac{\beta-\tau_E-i(\mu\beta-\varphi \ell)}{\ell} (h_2 - c/24)} e^{-\frac{\tau_E+i\varphi\ell}{\ell} (\bar{h}_1-c/24) +  \frac{\beta-\tau_E+i(\mu\beta-\varphi \ell)}{\ell} (\bar{h}_2 - c/24)}. \nonumber
\end{align}
For our 3d gravity proposal, the intermediate Hilbert space, and hence the precise range of primaries $O_{1,2}$ (including the measure) is known \eqref{solidtorus}. Thus, the intermediate operators have conformal weights $h_i = Q^2/4 + (p_i^+)^{2}$. Furthermore, for irrational Virasoro 2d CFTs with such weights, the sphere three-point function is determined by the conformal bootstrap and explicitly given by a DOZZ-like formula \cite{Dorn:1994xn,Zamolodchikov:1995aa}, derived in \cite{Collier:2019weq}. In principle, we should compute the Euclidean path integral of the Alekseev-Shatashvili model of subsection \ref{s:propgrav} with bilocal operator insertions to prove this statement explicitly purely from the gravity perspective. We leave this to future work. This DOZZ-formula simplifies in the scaling limit \cite{Mertens:2017mtv} as \cite{Ghosh:2019rcj} (where we set $p^\pm_i \to b p^\pm_i$):
\begin{equation}
\vert C_{O O_1 O_2} \vert^2\, \to \,  \frac{\Gamma(h\pm ip^+_1 \pm i p^+_2)}{\Gamma(2h)} \frac{\Gamma(\bar{h}\pm ip^-_1 \pm i p^-_2)}{\Gamma(2\bar{h})}\,.
\end{equation}
The resulting sum over primaries $O_{1,2}$ decouples into four integrals over $p^{\pm}_{1,2}$:
\begin{align}
\label{eq:twoptres}
&\text{Tr}[O_{h,\bar{h}}(\tau_E,\varphi)O_{h,\bar{h}}(0,0) e^{-\beta H + i\mu \frac{\beta}{\ell} J}] = \text{Tr}\left[ O_{h,\bar{h}} e^{-\tau_E H +i\varphi J} O_{h,\bar{h}} e^{-(\beta-\tau_E) H +i(\mu \frac{\beta}{\ell}-\varphi) J}\right]  \\
&\to \, \# b^{12} e^{\frac{\beta}{12\ell}}\int_0^{+\infty} \hspace{-0.2cm} dp^+_1 dp^+_2 p^+_1 \sinh(2\pi p^+_1) p^+_2 \sinh(2\pi p^+_2) e^{-\frac{b^2 \tau_E }{\ell}{p^+_1}^2} e^{-\frac{b^2 (\beta-\tau_E-i\mu\beta)}{\ell}{p^+_2}^2} \frac{\Gamma(h\pm ip^+_1 \pm i p^+_2)}{\Gamma(2h)} \nonumber \\
&\qquad \qquad \times \int_0^{+\infty} \hspace{-0.2cm} dp^-_1dp^-_2 p^-_1 \sinh(2\pi p^-_1) p^-_2 \sinh(2\pi p^-_2) e^{-\frac{b^2 \tau_E}{\ell}{p^-_1}^2} e^{-\frac{b^2 (\beta-\tau_E+i\mu\beta)}{\ell}{p^-_2}^2} \frac{\Gamma(\bar{h}\pm ip^-_1 \pm i p^-_2)}{\Gamma(2\bar{h})}, \nonumber
\end{align}
where we have not written some numerical prefactors similar as in \eqref{doubleJT}.
This expression is invariant under $\tau_E \to \beta- \tau_E$. Notice that even though all dependence on $\varphi$ gets scaled away in the double-scaled limit of interest discussed here, there is still an imprint of rotation on this doubled 1+1d JT system: the chemical potential $\mu$ is present and able to describe correlators probing a system with a rotating BTZ black hole as its saddle.

Let us next consider the situation describing the universal dynamics of an irrational 2d CFT in the regime \eqref{eq:double-scaleop}. Whenever $\Delta_{\text{gap}} \gtrapprox \frac{\tau_E}{\ell} ,\frac{\beta-\tau_E}{\ell}$, no additional primaries beyond the vacuum contribute in the \emph{dual} channel. This is the same argument as in subsection \ref{sec:2duniv}. In this channel, one inserts complete sets of states in the spatial $\varphi$-cycle instead, leading to an expansion in dual conformal blocks. In our case, the latter is dominated by identity blocks:\footnote{It is incompatible with the OPE to have both intermediate channels be the identity. There is a kinematic regime where this configuration dominates, and one where the complement (with swapped $\mathbb{1}$ and $O$ channels) dominates. See \cite{Ghosh:2019rcj} for the argument.}
\begin{equation}
\tilde{\mathcal{F}}_{O,\mathbb{1}}(\tau_E,\varphi) \equiv
\begin{tikzpicture}[scale=0.25, baseline={([yshift=-0.1cm]current bounding box.center)}]
 \draw[thick] (-0.77,1.77) to (-2.5,3.5);
  \draw[thick] (-0.77,-1.77) to (-2.5,-3.5);
  \node at (-2.5,0) {\small $O$};
  \draw[thick](1,0) ellipse (2.5 and 2.5);
  \node at (4.5,0) {\small $\mathbb{1}$};
  \node at (-3.5,-3.5) {\small $O$};
  \node at (-3.5,3.5) {\small $O$};
  \draw[->,thick, bend left=30] (5.5,3) to (5.5,-3);
  \node at (7.5,0) {\small $\varphi$};
\end{tikzpicture}
\end{equation}

To proceed, we need to transfer this information by transforming the above torus two-point conformal blocks \eqref{eq:blocks} between $S$-dual channels. Luckily this analysis has already been done for a chiral sector in \cite{Ghosh:2019rcj}, by applying a three-step sequence of fusion, modular $S$, and fusion transformations. Here we just apply it for both chiral sectors. The result for the boundary two-point function on the torus in terms of the conformal blocks \eqref{eq:blocks} is:
\begin{align}
\label{eq:res3d}
&\text{Tr}[O_{h,\bar{h}}(\tau_E,\varphi)O_{h,\bar{h}}(0,0) e^{-\beta H + i\mu \frac{\beta}{\ell} J}] = \\
&\int_0^{+\infty} \hspace{-0.2cm} dp_1^+dp_2^+ \rho(p_1^+) \rho(p_2^+) C_{h,p_1^+,p_2^+} \mathcal{F}_{p_1^+,p_2^+}(\tau_E,\varphi)
\int_0^{+\infty} \hspace{-0.2cm} dp_1^-dp_2^- \rho(p_1^-) \rho(p_2^-) C_{\bar{h},p_1^-,p_2^-} \mathcal{F}_{p_1^-,p_2^-}(\tau_E,\varphi), \nonumber
\end{align}
where
\begin{align}
\rho(p^\pm) &= 4\sqrt{2} \sinh ( 2 \pi b p^{\pm} ) \sinh ( 2 \pi b^{-1} p^{\pm} ), \\
C_{h,p_1^\pm,p_2^\pm} &= \text{DOZZ-like formula of \cite{Collier:2019weq}}.
\end{align}
In the double-scaling limit \eqref{eq:double-scaleop}, these blocks are further dominated by the primaries once again, and this expression reduces to \eqref{eq:twoptres}. However, the expression \eqref{eq:res3d} is fully valid in 3d, and is the boundary correlator on the solid torus for sufficiently small spectral gap. Note that this is the unnormalized correlator; the normalized one can be obtained by dividing by the partition function \eqref{solidtorus}.

Generalizations to higher-point functions and out-of-time ordered configurations can similarly be worked out, giving rise to results compatible with the doubled JT expressions in the double-scaling limit discussed in this appendix.

\section{Regularization of volume factors for non-compact groups}
\label{app:volume-reg}

Following Appendix C of \cite{Blommaert:2018iqz}, we review and improve here a natural strategy to regularize the different volume factors appearing in section \ref{sec:JT}. Our discussion should hold for non-compact groups with a continuous set of irreducible representations.

\subsection{Regularization of the representation dimensions}
The main idea in Appendix C of \cite{Blommaert:2018iqz} is to relate the Schur orthogonality relation
\begin{equation}
  \int_G dg\,R_{ab}(g) R^{\prime}_{cd}(g^{-1}) = V_\text{G}\,\frac{\delta_{RR^\prime}}{\text{dim}\,R}\delta_{ad}\delta_{bc}\,,
\label{eq:disc-rep}
\end{equation}
holding for discrete representations $R$ and $R'$, with the delta-regularized orthogonality relation
\begin{equation}
  \int_G dg\,R^k_{ab}(g) R^{k^\prime}_{cd}(g^{-1}) = \frac{\delta(k-k^\prime)}{\rho(k)}\delta_{ad}\delta_{bc}\,,
\label{eq:cont-rep}
\end{equation}
leading to the mathematically more accurate notion of Plancherel measure $\rho(k) \equiv \dim k$ for continuous representations. Here we associate $R \leftrightarrow k$. A formal comparison leads to
\begin{equation}
  \frac{\text{dim}\, R}{V_\text{G}} \delta (k-k^\prime) = \delta_{RR^\prime}\,\rho(k)\,.
\label{eq:formal-id}
\end{equation}

Taking the trace in \eqref{eq:disc-rep}, and using \eqref{eq:formal-id}, allows us to derive the orthogonality relation for characters
\begin{equation}
\label{eq:B4}
  \int_G dg\,\chi^k(g) \chi^{k^\prime}(g^{-1}) = \delta(k-k^\prime)\frac{\text{dim}\,R}{\rho(k)}\,.
\end{equation}
This integral over $G$ can be simplified using Weyl's integration formula:\footnote{Note that this is the compact version of Weyl's integration formula, which turns out to be correct as we explain in the next paragraph.}
\begin{equation}
\label{eq:Weylint}
\int_G dg \, f(g) = \frac{1}{\vert W\vert} \int_C d\alpha \, \vert \Delta(\alpha) \vert^2 \int_{G/C} dh\, f(h\alpha h^{-1}),
\end{equation}
where $W$ is the Weyl group, and the Weyl denominator $\vert \Delta(\alpha) \vert^2$ is the Jacobian in the change of variables on the group $g \to (\alpha,h)$. The subgroup of conjugacy class elements $C$ is a maximal torus in the group.
For the specific case of a class function $f$, such as the characters in our set-up, we can simplify this formula into:
\begin{equation}
\int_G dg \, f(g) = \frac{1}{\vert W\vert} \int_C d\alpha \, \vert \Delta(\alpha) \vert^2 f(\alpha)\left(\int_{G/C} dh\right).
\end{equation}
Taking the unit function $\mathbf{1}(g)$, we get the formal equality: $V_\text{G}=V_C \left(\int_{G/C} dh\right)$, where $V_\text{G}$ is the volume of $G$ and $V_C$ is the volume of a maximal torus (or a Cartan subgroup, or the subgroup $C$ of conjugacy classes).\footnote{The Jacobian factor $\vert \Delta(\alpha) \vert^2$ and size of the Weyl group $\vert W\vert$ are part of the measure $d\alpha$ of the Cartan subgroup.} The last factor represents the volume of any fixed conjugacy class, which we write as the ratio $V_\text{G}/V_C$. We can hence rewrite \eqref{eq:B4} into:
\begin{equation}
\label{eq:charcoj}
  \frac{1}{\vert W\vert} \int_C d\alpha\, \vert \Delta(\alpha) \vert^2\chi^k(\alpha) \chi^{k^\prime}(\alpha) = V_C\,\delta_{kk^\prime} = 2\pi \delta(k-k^\prime)\,.
\end{equation}
This leads to the formal relation $\delta(k-k) = V_C/2\pi$, which combined with \eqref{eq:formal-id} allows us to infer
\begin{equation}
  \text{dim}\, R = 2\pi\,\frac{V_\text{G}}{V_C}\rho(k)\,.
\end{equation}
This equation explicitly extracts the ``infinity'' from the dimension of the representation, and makes manifest its $k$-dependence.\footnote{As a simple check, for the abelian group $\mathbb{R}$, we have irreducible representation matrix elements $e^{ikx}$ and hence by \eqref{eq:cont-rep} $\rho(k)=1/2\pi$, and $V_\text{G}=V_C$, leading indeed to $\dim R =1$.}

The existence of several \emph{types} of inequivalent conjugacy classes for non-compact groups requires an additional sum over class types $C_i$ when applying Weyl's integration formula \eqref{eq:Weylint}, as developed in a general theorem for reductive Lie groups by Harish-Chandra. This sum over class types $C_i$ carries over into the left hand side of \eqref{eq:charcoj}. E.g. for the specific case of SL$(2,\mathbb{R})$, these class types are elliptic, parabolic and hyperbolic. However, the elliptic conjugacy classes do not contribute to \eqref{eq:charcoj} since the elliptic characters of the principal series irreps $\chi^k(\alpha)$ vanish. The parabolic classes are of measure zero, and hence also do not contribute to the left hand side of \eqref{eq:charcoj}.
For the positive semi-group SL$^+(2,\mathbb{R})$, elliptic conjugacy classes do not even exist \cite{Blommaert:2018iqz}.
We conclude that for the group-like structures of interest in the main text, equation \eqref{eq:charcoj} still holds but is restricted to the hyperbolic conjugacy class elements. These conjugacy classes are generated by the Cartan subalgebra:
\begin{equation}
\mathfrak{h} = \left\{\left(\begin{array}{cc}
s & 0 \\
0 & -s 
\end{array}\right), \quad s \in \mathbb{R} \right\}.
\end{equation}
Hence \emph{$V_C$ is the volume of the maximal torus for the hyperbolic conjugacy classes. } For the principal series irreps $k$ of these group-like structures, we use a continuous basis of quantum numbers $s$ to label states within the representation. Hence we formally have $\text{dim}\, R = \int_\mathbb{R} ds$. These equations are written in equation \eqref{eq:formal} in the main text.

\subsection{The $e$-brane boundary state and the unit group element}
\label{app:unitgroup}

For a compact group, we have the character completeness relations:
\begin{align}
\sum_R \chi_R(U) \, \chi_R(U') = V_C \delta (U- U'), \\
\label{eq:id1}
\sum_R \text{dim }R \, \chi_R(U) = V_C \delta (U- \mathbb{1}),
\end{align}
where $U$ is a group element in the maximal torus (or a conjugacy class). We can expand the (non-normalizable) unit state $|U = \mathbb{1}\rangle$ in the irrep basis as:
\begin{equation}
\label{eq:ecom}
|U = \mathbb{1}\rangle = \frac{1}{V_C}\sum_R \, \text{dim }R \, |R \rangle.
\end{equation}
As an explicit example, for the compact group SU(2) we have the character completeness relation:
\begin{equation}
\label{eq:su2}
\sum_{j\in \frac{\mathbb{N}}{2}} \frac{\sin (2j+1) \phi }{\sin \phi} \frac{\sin (2j+1) \phi' }{\sin \phi'} = 2\pi \frac{\delta(\phi-\phi')}{4\sin^2 \phi},
\end{equation}
where $\phi$ labels the different conjugacy classes.\footnote{Note that $\phi$ and $-\phi$ are in the same conjugacy class: this is simply a swap of the two eigenvalues of the SU(2) or SL$(2,\mathbb{R})$ matrix. We hence restrict to $\phi > 0$ here without loss of generality.} Taking the limit $\phi' \to 0$, we obtain the formal identity:
\begin{equation}
\sum_{j\in \frac{\mathbb{N}}{2}} \frac{\sin (2j+1) \phi }{\sin \phi} (2j+1) = 2\pi\frac{\delta(\phi)}{4\sin^2 \phi}.
\end{equation} 
We can view \eqref{eq:su2} as the natural regularization of the above limiting identity.

Let us now look at the same argument for the positive semigroup SL$^+(2,\mathbb{R})$. We start with a brief review on the principal series characters themselves, starting with the full group SL$(2,\mathbb{R})$. The characters of the principal series irrep of SL$(2,\mathbb{R})$ are computed by the following integral \cite{VK}
\begin{equation}
\label{eq:chara}
\chi_j(g) = \int_{\mathbb{R}} dx |bx+d|^{2j}\delta\left(\frac{ax+c}{bx+d}-x\right),
\end{equation}
where $j=-1/2+ik$. Using that the character is a class function, we write for an element in the hyperbolic conjugacy class:
\begin{equation}
g = \left[\begin{array}{cc} 
e^{\phi} & \epsilon \\
0 & e^{-\phi}
\end{array} \right],
\end{equation}
where $\epsilon$ can be viewed as a regulator since the delta-function generically has two solutions for $x$ unless $\epsilon=0$.\footnote{See \cite{Fan:2021wsb} for an analogous recent application of this calculation procedure to the supergroup OSp$(1|2,\mathbb{R})$.} A better way to think about this is we should instead of $\mathbb{R}$ attach the point at infinity as $\mathbb{R} \cup \infty$, and allow for a ``solution at infinity''. Hence
\begin{equation}
\chi_j(g) = \int_{\mathbb{R}} dx |\epsilon x+e^{-\phi}|^{2j}\delta\left(\frac{e^\phi x}{\epsilon x+e^{-\phi}}-x\right).
\end{equation}
The integral over $x$ boils down to the fixed point of the group action. We have two fixed points, and the delta function evaluates to
\begin{equation}
\delta\left(\frac{e^{\phi}x}{\epsilon x + e^{-\phi}}-x\right) = \frac{\delta(x)}{|e^{2\phi}-1|} + \frac{\delta(x-\frac{e^{\phi}- e^{-\phi}}{\epsilon})}{|1-e^{-2\phi}|},
\end{equation}
Hence 
\begin{equation}
\label{eq:charh}
\chi_j(g) = \frac{e^{-2j\phi}}{|e^{2\phi}-1|} + \frac{e^{2j\phi}}{|1-e^{-2\phi}|} = \frac{\cos 2k \phi}{ |\sinh \phi|}.
\end{equation}
If we specify to $g=\mathbb{1}$, then we would get from \eqref{eq:chara} the formal expression $\chi_j(\mathbb{1}) = \delta(x-x) \int_{\mathbb{R}} dx$, which is divergent both due to the proliferation of fixed points and the range of the $x$-integral. Instead letting $\phi \to 0$ in \eqref{eq:charh}, the result diverges due to $\sinh \phi \to 0$, as we approach the identity group element from the hyperbolic side.

The characters of SL$^+(2,\mathbb{R})$ are
\begin{equation}
\label{eq:charpl}
\chi_j(g) = \frac{\cos 2k \phi}{2|\sinh \phi|}
\end{equation}
only including a factor $1/2$ compared to the full SL$(2,\mathbb{R})$ manifold. This can be proven directly using the argument around footnote 21 of \cite{Blommaert:2018iqz}. 
Alternatively, from the above computation, we restrict to $x \in \mathbb{R}^+$. This causes both fixed point locations $x=0$ and $x=\infty$ to be at the integration endpoints, picking up an additional factor of $1/2$ in the process.\footnote{One might worry about the case $\phi < 0$ picking up only one of the solutions of the delta-function. This is misleading however, since in the case the second fixed point is at $x\to -\infty$, which in the one-point compactification of $\mathbb{R}$ is identified with $x\to +\infty$.}

We write again a completeness relation of these characters \eqref{eq:charpl}. By explicit calculation we have:
\begin{equation}
\label{eq:charcop}
\int dk \, \frac{\cos 2k \phi}{ 2|\sinh \phi|} \frac{\cos 2k \phi'}{ 2|\sinh \phi'|} = 2\pi\frac{\delta(\phi-\phi')}{16\sinh^2 \phi}.
\end{equation}
Letting $\phi' \to 0$, we obtain the limiting identity:
\begin{equation}
\int dk \, \frac{\cos 2k \phi}{ 2|\sinh \phi|} \frac{2\pi V_G}{V_C} k \sinh (2\pi k) = 2\pi \frac{\delta(\phi)}{16\sinh^2 \phi},
\end{equation}
or
\begin{equation}
\int dk \, \frac{\cos 2k \phi}{ 2|\sinh \phi|} k \sinh (2\pi k) =  \frac{V_C}{V_G} \delta(U-\mathbb{1}).
\end{equation}
This rather formal expression is given meaning as a limit of the well-defined relation \eqref{eq:charcop}.
This expression is as expected: using our finite volume regularization, we replace $\sum_R \to \frac{V_C}{2\pi} \int dk$ and $\text{dim }R \to \frac{2\pi V_G}{V_C} \rho(k)$ in \eqref{eq:id1}, and we find the non-compact version:
\begin{equation}
\int_k \rho(k) \, \chi_k(U) = \frac{V_C}{V_G} \delta (U- \mathbb{1}),
\end{equation}
where the delta-function on the maximal torus contains the Weyl denominator contragrediently to the measure factor: $\delta(U-\mathbb{1}) \sim \frac{\delta(\phi)}{\sinh^2 \phi}$. Hence this indeed matches the above expression found from the character completeness relation.

The unit element state $|U= \mathbb{1}\rangle$ is then expanded in the representation basis as
\begin{equation}
\label{eq:ebrane}
\boxed{
|U= \mathbb{1}\rangle = \frac{V_G}{V_C}\int dk \, k\sinh 2\pi k \, |k \rangle}.
\end{equation}
Indeed, using $\sum_R \to \frac{V_C}{2\pi} \int dk$ and $\text{dim }R \to \frac{2\pi V_G}{V_C} \rho(k)$ in \eqref{eq:ecom}, we reproduce \eqref{eq:ebrane}, and the overlap $\braket{k|U= \mathbb{1}}= \frac{V_G}{V_C} k \sinh 2 \pi k$ if we normalize the states such that $\braket{k|k'}=\delta(k-k')$.

We observe that the ``closed string'' $e$-brane boundary state, defined by the shrinkability constraint \eqref{eq:eoverlap} in the main text, is precisely this unit group element state (up to a $1/V_G$ normalization factor):
\begin{equation}
|e \rangle = \frac{1}{V_G}|U= \mathbb{1}\rangle = \frac{1}{V_C}\int dk \, k\sinh 2\pi k \, |k \rangle.
\end{equation}

\section{Hopf algebra of functions on a (quantum) group}
\label{app:hopf}
 The properties of a group $G$ can be captured by the algebra of functions on $G$.  The latter is a Hopf algebra $\mathcal{F} (G)$, which has both a co-product and a product.  The axioms satisfied by $G$ can be translated into compatibility relations for the algebraic structure of  $\mathcal{F}(G)$.  A quantum group can be defined via a non-commutative deformation of  $\mathcal{F} (G)$. The non-commutativity is encoded by the $R$-matrix.
Below we review these algebraic structures. 

\paragraph{Hopf algebra structure.} 
The quantum group $\mathcal{F}_q(G)$ is a quasi-triangular Hopf algebra.  To explain what this is, we start with the simpler structure of a bi-algebra $\mathcal{A}$, which is an algebra endowed with 4 operations
\begin{align}
     \text{product}& \quad\nabla: \mathcal{A}\otimes \mathcal{A} \rightarrow \mathcal{A},\nn
     \text{unit}& \quad\eta: \mathbb{C} \rightarrow \mathcal{A},\nn
     \text{co-product}&\quad  \Delta : \mathcal{A}\rightarrow \mathcal{A} \otimes \mathcal{A},\nn
     \text{co-unit}& \quad   \epsilon: \mathcal{A} \rightarrow \mathbb{C} .  
\end{align}
These operations are required to satisfy compatibility relations \cite{Klimyk:1997eb}. In particular the product and co-product are associative and co-associative respectively.  

Our main example is the set $\mathcal{F}(G)$ of $\mathbb{C}$-valued functions on a group $G$, where the operations are the following:
\begin{alignat}{3}
\nabla(f_1,f_2)(g) &= f_1(g)f_2(g), \quad &&f_1,f_2 \in \mathcal{F}(G), \\
\eta &= 1 \text{  (the identity function)}, \\
\label{DE}
\Delta(f) (g_1,g_2) &= f(g_1g_2 ),\quad &&g_1,g_2 \in G, \\
\epsilon (f) &= f(\mathbb{1}_{G}).
\end{alignat}
The first line is pointwise multiplication. Here $g_1g_2$ denotes the group multiplication of $g_1$ and $g_2$, and $\mathbb{1}_{G}$ is the identity element of $G$. 
The definition \eqref{DE} makes $\Delta: \mathcal{F}(G) \to \mathcal{F}(G \times G)$, which is not what we want. By the Peter-Weyl theorem, providing a dense basis of $\mathcal{F}(G)$ in terms of the representation matrix elements $R_{ab}(g)$, we have the isomorphism $\mathcal{F}(G \times G) \simeq \mathcal{F}(G) \times \mathcal{F}(G)$ since its action \eqref{DE} on a single basis element is:
\begin{equation}
\label{copr}
\Delta(R_{ab})(g_1,g_2) = R_{ab}(g_1g_2) = \sum_c R_{ac}(g_1)R_{cb}(g_2) = \sum_c R_{ac} \otimes R_{cb}(g_1,g_2),
\end{equation}
making it an element in $\mathcal{F}(G) \times \mathcal{F}(G)$. Linearly extending then proves the isomorphism. We can hence summarize the above natural co-product as:
\begin{equation}
\boxed{
\Delta(R_{ab}) = \sum_c R_{ac} \otimes R_{cb}}.
\end{equation}
The co-unit acting on representation matrix elements is:
\begin{equation}
\epsilon(R_{ab}) = R_{ab}(\mathbb{1}_G) = \delta_{ab}.
\end{equation}
A bi-algebra structure is upgraded into a Hopf algebra by the introduction of an anti-homomorphism:
\begin{align}
S: \mathcal{A}\rightarrow \mathcal{A}, \quad S(f_1f_2) &= S(f_2)S(f_1) ,\quad f_1,f_2 \in \mathcal{A} \label{anti} ,
\end{align}
called the antipode. For $\mathcal{F}(G)$, the natural definition is to act as the pullback of the inverse action:
\begin{align}
S(f)(g) = f(g^{-1}), \qquad
S(g_{ij} ) = g^{-1}_{ij} .
\end{align}
On a single basis element of $\mathcal{F}(G)$, we write
\begin{equation}
S(R_{ab})(g) = R_{ab}(g^{-1}) = R_{ab}^{-1}(g),
\end{equation}
mapping a representation matrix element to its inverse. Depicting the group element as an oriented interval, the inverse map changes this orientation by diagrammatically twisting:
\begin{align}
 \mathtikz{\SA{0cm}{0cm}} : f(g)\to f( S(g) ),\qquad
    \mathtikz{\SA{0cm}{0cm}, \SA{0cm}{-1cm} } f(g) \to f( S^{2}(g) )
    \end{align} 
In the underformed algebra, $S^2 =1$, so the double twist is just the identity.   Crucially, this will change under $q$-deformation.

\paragraph{$\mathcal{R}$-matrix.} 
Adding an $\mathcal{R}$-matrix to a Hopf algebra, makes it a quasi-triangular Hopf algebra. Given a vector space $V$ carrying the ``fundamental rep" of the group, the $\mathcal{R}$-matrix can be viewed as a linear operator on the tensor product $V\otimes V$:
\begin{align}
\label{eq:R}
     \mathcal{R} & \in \text{End } V (\otimes V).
\end{align}
We should view $V$ as the Hilbert space of an anyon, and the $\mathcal{R}$-matrix applies a braiding operation to a pair of anyons. Given the co-unit $\epsilon$ and antipode $S$, it is heavily constrained to satisfy several relations among itself, $\epsilon$ and $S$, including famously the Yang-Baxter equation.

\paragraph{$q$-deformation of $ \mathcal{F}(G) $.}

As a vector space, $\mathcal{F}(G)$ is defined over the complex numbers and spanned by the basis 
\begin{align}
     g_{i_1 j_1} g_{i_2 j_2} \cdots g_{i_n j_n}, 
     \quad  n=1,\cdots \infty.
\end{align}
where $g_{ij}$ are the matrix elements in the fundamental representation of $G$. The co-product \eqref{copr} in the fundamental representation can be written as:
\begin{equation}
\Delta(g_{ij}) = \sum_k g_{ik} \otimes g_{kj}.
\end{equation}
In the undeformed algebra, the matrix elements themselves commute:
\begin{align}
    g_{ij}  g_{kl} = g_{kl}g_{ij}.
\end{align}
However, in the quantum group $\mathcal{F}_q(G)$ this multiplication law (distinct from the matrix multiplication rule) becomes non-commutative.\footnote{It is customary to abuse language and refer to both the ``quantum space"  $G_{q}$ and the deformed algebra of functions $\mathcal{F}_q(G)$ as the quantum group.}

The precise nature of the non-commutative product in $\mathcal{F}_q(G)$ is determined by the $\mathcal{R}$-matrix of the quantum group. To express the product rule it is useful to consider an element $g$ as a matrix acting in the fundamental representation. Thus it acts on a vector space $V$ according to
\begin{align} \label{V}
g:V \rightarrow G_q \times V,\qquad 
    v_{i}\mapsto  \sum_{i} g_{ij} \otimes v_{j}.
\end{align}
If we analogously define matrices acting in $V \otimes V$ as
\begin{align}
    g_{(1)} = g_1 \otimes \mathbb{1},\qquad 
    g_{(2)} = \mathbb{1} \otimes g_2,
\end{align}
then the multiplication rule for the coordinate algebra is defined now using the $\mathcal{R}$-matrix \eqref{eq:R} as:
\begin{align}\label{Ru}
    \mathcal{R} g_{(1)} g_{(2)}= g_{(2)}g_{(1)} \mathcal{R},
\end{align}
where the composition of the operators above is defined with ordinary matrix multiplication. If $\mathcal{R}$ is not trivial, the group elements $g_{ij}$ no longer commute.   

The non-commutative product defined by the $\mathcal{R}$-matrix also implies that the antipode is no longer the usual inverse. However it still satisfies
\begin{align}
    \sum_{j} g_{ij} S(g_{jk})= \sum_j S(g_{ij})g_{jk}= \delta_{ik}.
\end{align}

\paragraph{Example: $\SL_{q}(2)$.}
To illustrate these definition, consider the quantum group $\SL_{q}(2)$. Its  coordinate algebra is generated by 4 elements $(a,b,c,d)$ of a matrix 
\begin{align}
g= \begin{pmatrix} a & b\\ c & d\end{pmatrix}.
\end{align} 
The commutation relations of the matrix elements are encoded in the $R$-matrix,
\begin{align}
\mathcal{R} = \begin{pmatrix} 
    q && 0       && 0       && 0 \\
    0 && q^{1/2} && 0       && 0 \\
    0 && q-1     && q^{1/2} && 0 \\
    0 && 0       && 0       && q
    \end{pmatrix}.
\end{align}
Then the multiplication rule \eqref{Ru} is equivalent to the commutation relations
\begin{align} \label{sl2commutators}
ab&= q^{1/2} ba, \quad  ac= q^{1/2} ca,\quad  bd= q^{1/2} db,\quad cd= q^{1/2} dc,\quad  \nn
bc&=cb,\quad ad-da = (q^{1/2}-q^{-1/2}) bc.
\end{align}
Additionally we impose the condition
\begin{equation} \label{sl2determinant}
ad-q^{1/2} bc = 1,
\end{equation}
which is the $q$-deformed version of the condition $\det g = 1$.
The antipode is given by
\begin{equation}
S(g) = \begin{pmatrix} d & -q^{-1/2} b \\ -q^{1/2} c & a \end{pmatrix}.
\end{equation}
Using the relations \eqref{sl2commutators} and \eqref{sl2determinant} we see that the antipode satisfies $S(g) g = g S(g) = \mathbb{1}$.

\section{Examples of extended TQFT}
\label{sec:ex-eTQFT}

We review the extended TQFT formulation of 2d gauge theories and Chern-Simons theories.

\subsection{2d gauge theory}
\label{app:2dgauge}
Two-dimensional YM or BF theories with compact gauge group $G$ can be formulated as an extended TQFT. In the open-closed TQFT description, the Frobenius algebra $\mathcal{H}_{ee}$ associated to the interval is the \emph{group algebra} $\mathbb{C}[\text{G}]$. This is the algebra with basis states $\ket{g}$ labeled by group elements, sum defined as $\sum_i \alpha_i \ket{g_i}, \, \alpha_i \in \mathbb{C}$, and the product $\ket{g_{1}}\cdot \ket{g_{2}} = \ket{g_{1}g_{2}} $ and linearly extending. The Frobenius form $\tr_{\text{Fr}}(g_1 \cdot g_2) \equiv \langle g_1,g_2 \rangle$ is defined by taking the coefficient of the identity group element in the expansion of $g_1 \cdot g_2$.

To relate the group algebra to the conventional description of the gauge theory Hilbert space, it is natural to identify $\mathbb{C}[\text{G}]$ with the function space $\text{L}^{2}(\text{G})$, whose elements are wavefunctions $\psi(g)$
\begin{align}
\label{eq:groups}
    \ket{\psi} = \int dg\, \psi(g) \ket{g}.
\end{align}
However, note the group product on $\mathbb{C}[\text{G}]$ corresponds to the \emph{convolution product} on $\text{L}^{2}(\text{G})$
\begin{align}
    \psi_{1} * \psi_{2} (g) \equiv \int d h \,\, \psi_{1}(gh) \psi_{2}(h^{-1} )  ,
\end{align}
rather than pointwise multiplication. This is due to the property: 
\begin{align}
\ket{\psi_1}\cdot \ket{\psi_2} = \int dg \left( \int d h \,\, \psi_{1}(gh) \psi_{2}(h^{-1} ) \right) \ket{g},
\end{align}
where we twice inserted \eqref{eq:groups} and used the right-invariance of the Haar measure. This is an important distinction: in the TQFT language, spacetime locality is captured by the convolution product on $\text{L}^{2}(\text{G})$, rather than the pointwise multiplication associated to the Hopf algebra. 

This distinction between the Hopf and the Frobenius algebras can be more explicitly seen by considering the group algebra in the representation basis. Here, the Frobenius algebra multiplication corresponds to a path integral process fusing the endpoints of a pair of intervals according to
\begin{align}
    \mathtikz{\muA{0}{0}}   (R_{ab}(g_{1}) , R_{cd}(g_{2})) &\to R_{ab} * R_{cd}(g)=\int dh\, R_{ab}(gh) R_{cd}(h^{-1}) \nn
    &= \sum_{e} R_{ae}(g) \int dh\,  R_{eb}(h) R_{cd}(h^{-1}) =\frac{1}{\dim R} R_{ad}(g) ( \delta_{bc}),
\end{align}
where we used the orthogonality relation \eqref{ortho}. On the other hand, the Hopf algebra product is simply pointwise multiplication
\begin{align}
    (R_{ab}(g_{1}) , R_{cd}(g_{2})) &\to R_{ab}(g)R_{cd}(g)\,,
\end{align}
and has \textit{no} path integral interpretation. 

The closed TQFT describing gauge theory states on the circle corresponds to the center $Z[\mathbb{C}[\text{G}]]$, which forms a \emph{commutative Frobenius algebra}. This can be identified with the gauge-invariant Hilbert space of  \emph{class functions} on the circle.\footnote{This is because such center has a basis consisting of averages within a conjugacy class C:
\begin{align}
    \ket{C}&\equiv \frac{1}{|C|}\sum_{g \in C} \ket{g}, \end{align} 
so a state $\ket{\psi} \in Z(\mathbb{C}[G])$ in the center has an expansion 
    \begin{align} 
     \ket{\psi} =\sum_{C} \psi(C) \ket{C}\,.
\end{align}} 

More details about the open-closed TQFT formulation of 2d gauge theory are given in \cite{Donnelly:2018ppr}.

\subsection{Chern-Simons theory}
\label{sec:CS-review}

Consider Chern-Simons theory with compact gauge group. In this case, the definition \eqref{Zfunct} of an unextended TQFT for $d=3$ is much less powerful because cutting 3-manifolds along a codimension-1 slice does not produce a finite, generating set of cobordisms as in the $d=2$ case. However, the extension of the TQFT down to codimension two, i.e. a circle, does provide the computational power needed to evaluate path integrals on 3-manifolds.  Whereas a $d=2$  extended TQFT is equivalent to a Frobenius algebra, a $d=3$ once-extended TQFT is equivalent to a modular tensor category (MTC),\footnote{Once extended means extending down to circles. A fully extended TQFT would extend down to a point.} which is the boundary category assigned to a codimension-2 surface.  
For gapless boundaries such as the entangling surface, the latter is given by the representation category of a loop group or quantum group.  When we extend down to a circle, it is possible to obtain a finite set of generating 1-morphisms analogous to \eqref{eq:gen}, along with a set of 2-morphisms satisfying  sewing relations \cite{Bartlett:2020cdc}.  

\begin{figure}[h]
\centering
\includegraphics[width=0.5\textwidth]{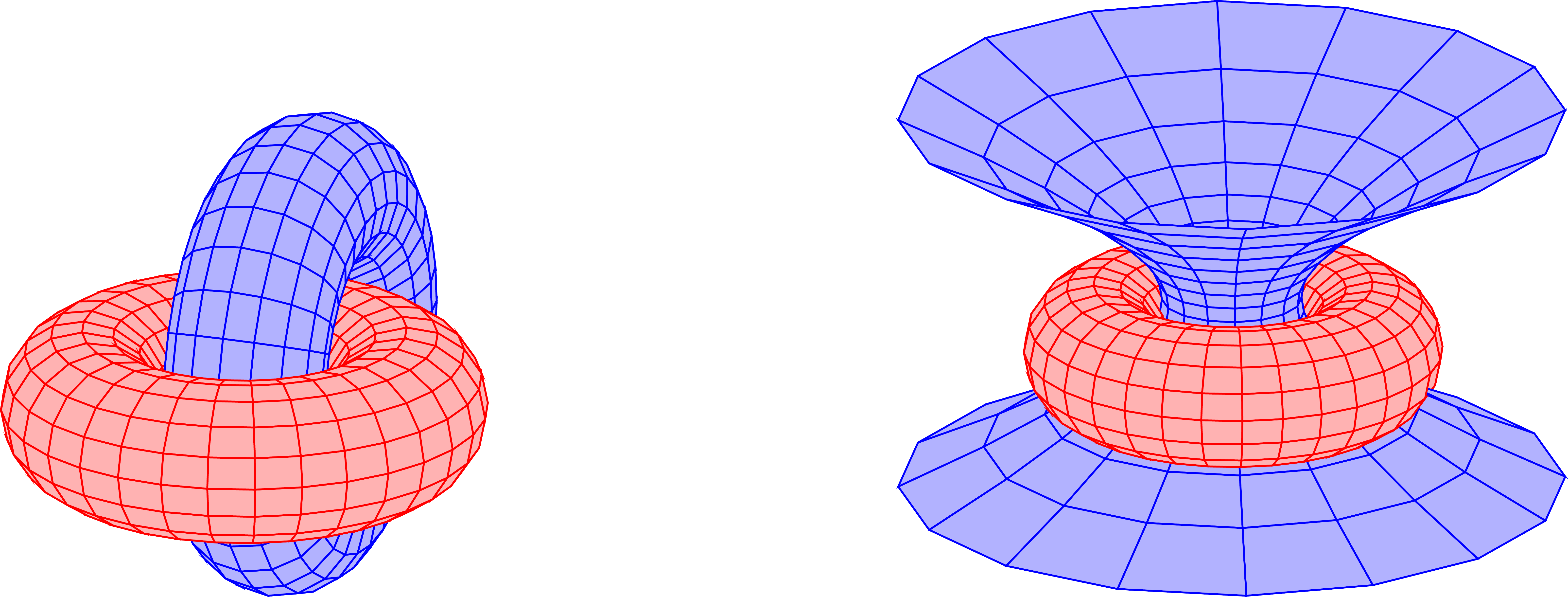}
\caption{The Heegaard splitting of the three-sphere.}
\label{Hee}
\end{figure}

An important part of the  MTC data is the modular S-matrix. Equipped with the S-matrix,  we can perform a ``closed channel" calculation of the $3$-sphere partition function $Z(S^3)$. Following the approach of \cite{Atiyah:1989vu,Witten:1988hf}, cut open $S^3$ along a codimension-1 surface $\Sigma=T^2$ with the topology of a torus. Using the Heegaard Splitting (Figure \ref{Hee}), this torus separates $S^3$ into two solid tori, to which the TQFT assigns a quantum state $\ket{0}\in \mathcal{H}_{T^{2}}$.
The Heegaard splitting means that we glue these solid tori with a large diffeomorphism, represented by an operator $\hat{S}$ on $\mathcal{H}_{T^{2}}$ defined by the S-matrix. We can thus represent the 3-sphere partition function as an S-matrix element:
\begin{align}
    Z(S^3) = \braket{0|\hat{S}|0}=S_{00}.
\label{eq:S00}    
\end{align}
Notice \eqref{eq:S00} expresses the ``path integral on $Z(S^3)$'' in terms of algebraic data associated with the boundary category. However, as in equation \eqref{TrFr1-m}, it does not yield a state counting interpretation\footnote{Moreover, when applying this to the irrational Virasoro CFT case relevant to 3d gravity, we encounter a modular S-matrix for which $S_{00}$ vanishes, so equation \eqref{TrFr1-m} is no longer valid.}.

\begin{figure}[h]
\centering
\includegraphics[width=0.42\textwidth]{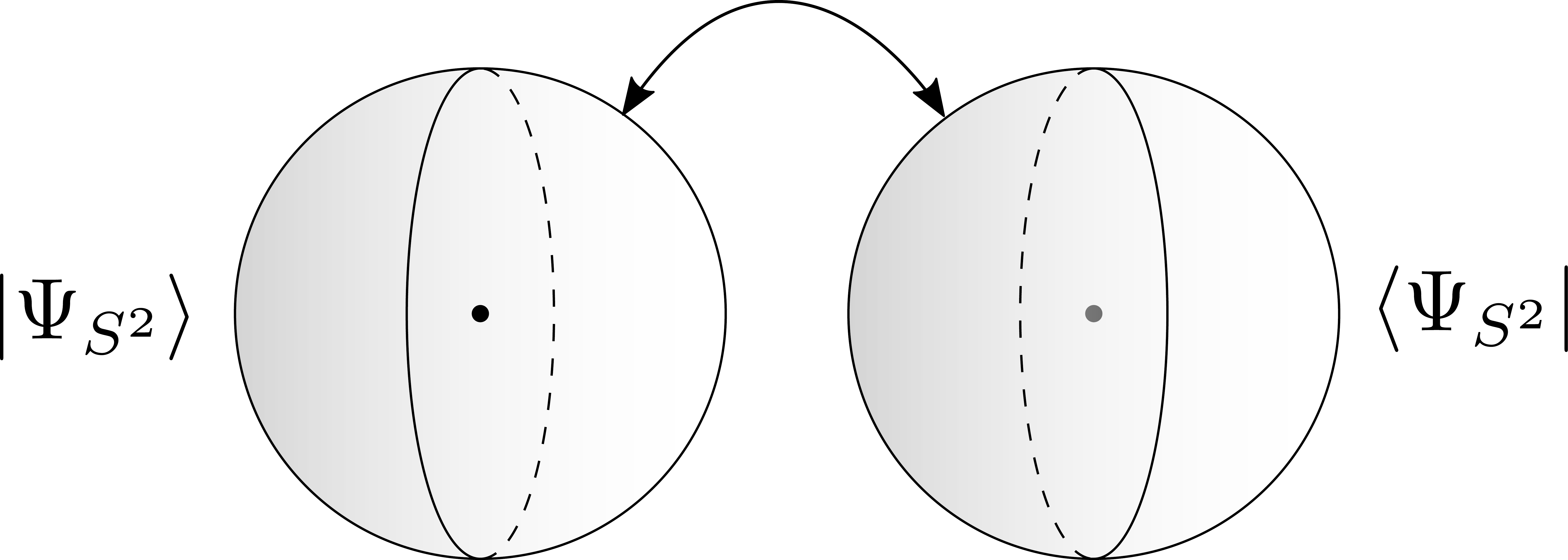}
\caption{Cutting the three-sphere symmetrically into two three-dimensional balls.}
\label{2balls}
\end{figure}

\paragraph{Open-slicing and ``microstate counting".} 
To get a state counting intepretation of $Z(S^3)$, consider an ``open slicing" analogous to the annulus interpretation of $Z(S^2)$ in two-dimensional gauge theory. This involves cutting along a codimension-2 surface, requiring us to perform a more intricate type of surgery. First, notice that a 3-sphere $S^3$ can be identified as two solid $3d$-balls glued along their 2-sphere boundary. The latter forms the equator of $S^3$ (see Figure \ref{2balls}) and can be viewed as our Cauchy slice. As a codimension-1 cut of $S^3$, it defines a Hilbert space $\mathcal{H}_{S^2}$. From the gluing of the two 3d-balls, we can just view $Z(S^3)=\braket{\Psi_{S^2}|\Psi_{S^2} }$, where $\ket{\Psi_{S^2}} \in \mathcal{H}_{S^2}$, is the ``Hartle-Hawking" state.

\begin{figure}[h]
\centering
\includegraphics[width=0.37\textwidth]{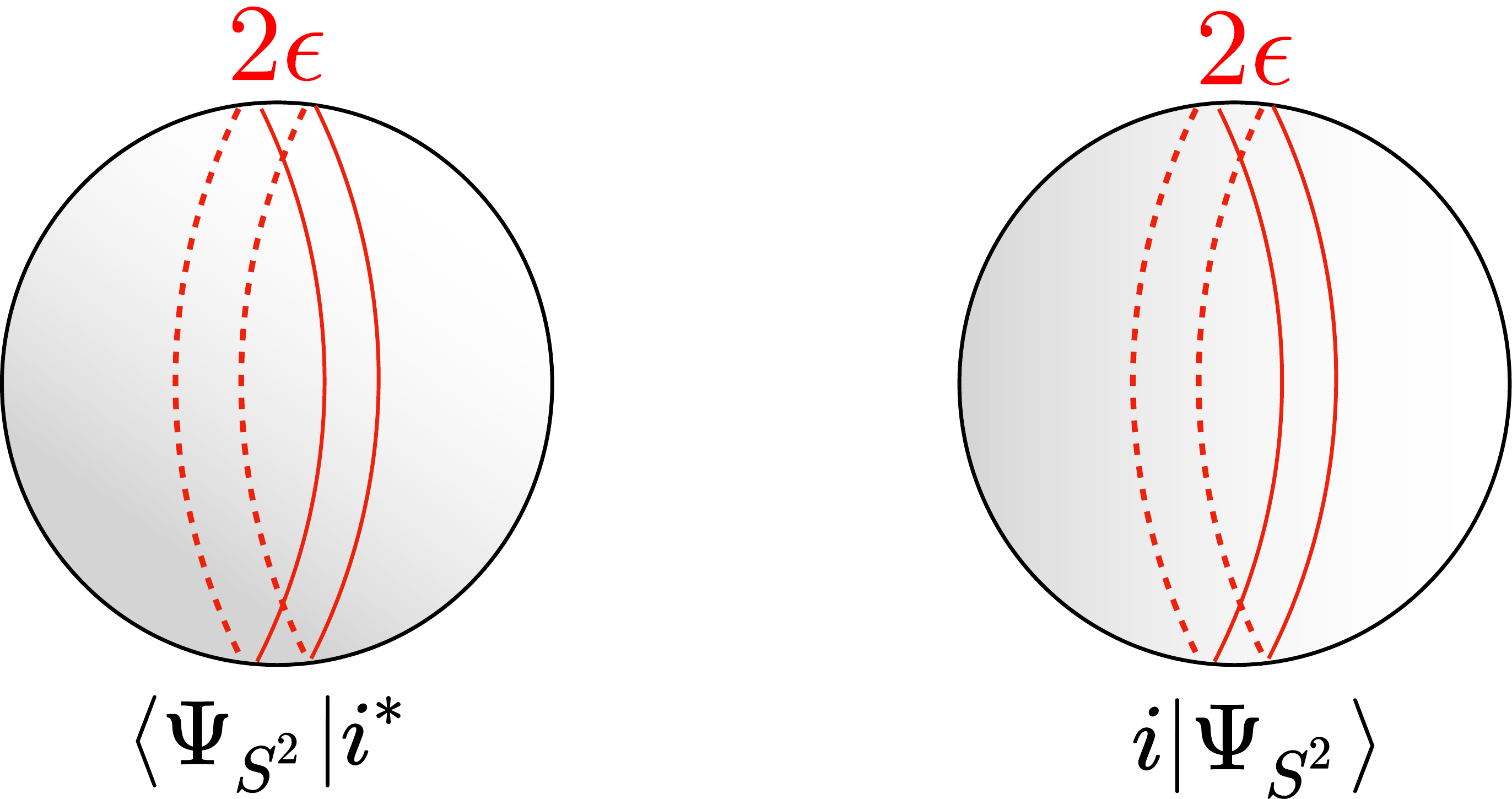}
\caption{The codimension-2 cut on the 2-sphere is implemented by a path integral with a stretched entangling surface.  Shown in red, this surface digs a trough along the equator of the 2-sphere. The resulting path integral defines the factorized states  $i \ket{\Psi_{S^2}}$ and $\bra{\Psi_{S^2}}i^*$.}
\label{trough}
\end{figure}

Second, let us cut the $S^2$ Cauchy surface along a codimension-2 surface with the topology of a circle, separating the 2-sphere into two disk like subregions $V$ and $\bar{V}$. As in the extended Hilbert space procedure described in section \ref{sec:algebra}, separate the subregions $V$ and $\bar{V}$ with a regulator $\epsilon$. This construction gives a quantum mechanical interpretation to the codimension-2 cut as the factorization of $\ket{\Psi_{S^2}}$ (and $\bra{\Psi_{S^2}}$) implemented by the path integral on a three ball with a trough dug out by a stretched entangling surface $S_{\epsilon}$. This is shown in Figure \ref{trough}.

The extended TQFT assigns a mathematical structure to the codimension-2 closed oriented manifold emerging from the $\epsilon\to 0$ limit of this construction: a linear category, with objects, the collection of Hilbert spaces assigned to codimension-1 manifolds that can end on it, and morphisms, the maps between these Hilbert spaces. To compute $Z(S^3)$, we reverse the cutting process and glue back together the basic building blocks. We observe that the cutting process has left us with two factorized states  $i \ket{\Psi_{S^2}}$ and $\bra{\Psi_{S^2}} i^*$, which when glued together give the partial trace of a reduced density matrix.  A successful gluing then implies
\begin{align}\label{S3shrink}
    Z(S^3)= \lim_{\epsilon \to 0} \tr_{\mt{V}} e^{-2\pi \epsilon H_{\mt{V}}},
\end{align}
where Hamiltonian $H_{V}$ is the generator of modular flow in region $V$.  Thus, the criteria for the codimension-2 cutting process to produce $Z(S^3)$ is precisely the shrinkability condition \eqref{shrink}.

\paragraph{The shrinkable boundary condition, the boundary category $\mathcal{B}_{S}$.} 
Unlike \eqref{shrink}, the shrinkability condition \eqref{S3shrink} can only be satisfied up to an infinite subtraction.  This is because the introduction of the stretched entangling surface $S_{\epsilon}$ breaks the topological invariance of the TQFT and introduces a UV divergence on the right hand side of \eqref{S3shrink}.  In particular, the standard holographic boundary condition setting $A_{\tau_E}-A_{\varphi}\vert_{\partial \mathcal{M}}=0$ on  $S_{\epsilon}$ introduces CFT edge modes which depend on the complex structure of $S_{\epsilon}$.   Under modular evolution by $2\pi$, $S_{\epsilon}$ becomes a thin torus which pinches as $\epsilon \to 0$, giving a divergence due to the infinite tower of descendants contributing to the CFT partition function at infinite temperature. 

Technically, one can subtract this divergence via zeta-function regularization or a counterterm subtraction.  However, when Chern-Simons theory is viewed as an emergent, low energy description of an underlying microscopic model, this divergence is a feature rather than a bug. In particular, this divergence gives the leading area law term in the entanglement entropy  $S_{V}$ of a subregion which is needed to ensure the positivity of $S_{V}$.   In subsequent sections, we will see that gravity regulates this divergence such that shrinkability can be satisfied exactly. 
The upshot is that for Chern-Simons theory with a compact gauge group $G$, we relax the definition of shrinkability to allow for this divergence.  The corresponding shrinkable boundary condition $A_{\tau_E}-A_{\varphi}\vert_{\partial \mathcal{M}}=0$ leads to the usual holographic duality, where the Hilbert space $\mathcal{H}_{V}$ on a disk is given by the representations of the loop group of $G$ associated with boundary edge modes on $S=\pd V$.  The boundary category $\mathcal{B}_{S}$ on the codimension-2 entangling surface is thus identified with the representation category of the loop group. Expression \eqref{S3shrink}  then implies that $Z(S^3)$ can by computed from the thermal partition function of the associated WZW model, with the CFT Hamiltonian $H_{V}= \frac{2\pi}{l} \left(L_{0} + \bar{L}_{0}-\frac{c}{24}\right)$ as the modular Hamiltonian, where $l$ is the length of $S$.\footnote{To relate this to the 3d coordinates $(t,r,\varphi)$ of section \ref{sec:3dgrav_def}, where the entangling surface is a $\varphi$-circle, we have $l=2\pi \ell$.} 

\paragraph{Factorization Map.} 
The factorization map $i$ on $S^2$ is represented by a cobordism from $S^{2}$ to a disjoint union $V \sqcup \bar{V}$ of two disks, which according to Atiyah's axioms, is assigned to the tensor product $\mathcal{H}_{V} \otimes \mathcal{H}_{\bar{V}}$. Rather than computing the path integral on such a topologically nontrivial space, we simply treat the factorization as a linear map
\begin{align}
    i: \mathcal{H}_{S^2} \to \mathcal{H}_{V}\otimes \mathcal{H}_{\bar{V}},
\end{align}
subject to the shrinkable boundary condition. In \cite{Wong:2017pdm}, it was shown that a solution is given by embedding $\ket{\Psi_{S^2}}$ into a regulated Ishibashi state: 
\begin{align}
        i \ket{ \Psi_{S^2}} =\frac{1}{\sqrt{Z(2\pi\epsilon)}}
        \sum_{m} e^{\frac{-2 \pi^2 \epsilon}{l} (L_{0}+\bar{L_{0}})} \ket{ m}_{V} \otimes \ket{ \bar{m}}_{\bar{V}} , 
\end{align}
where $m$ is a schematic label for the descendants, $l$ is the length of $S$, and $2\pi\epsilon$ is the effective temperature at the entangling surface. We can deduce the mapping $i$ from the fact that the right hand side is the thermofield double state for the thermal CFT on the stretched horizon. The norm of the resulting state is $1$, which is the partition function on $S^2 \times S^1$.

In the presence of a Wilson line in the representation $R$, which adds a puncture on $S^2$, the state is modified to $\ket{ \Psi_{S^2}(R)}$, and the factorization map is given by
\begin{align}
      i \ket{ \Psi_{S^2}(R)}
    =\frac{1}{\sqrt{Z(2\pi\epsilon)}}
    \sum_{a,m} e^{\frac{-2 \pi^2 \epsilon}{l} (L_{0}+\bar{L_{0}})} \ket{R,\,a,\, m}_{V} \otimes \ket{R,\,a,\, \bar{m}}_{\bar{V}},
\end{align}
where $a=1,\cdots, \dim R$ labels the Kac-Moody zero modes.  This is a generalization of the co-product factorization in \eqref{eq:Ifact-2}. Moreover, the image of $i$ is exactly the entangling product \eqref{eq:fusion} with loop group of $\text{G}$ as the surface symmetry $\text{G}_{\mt{S}}$ \cite{Wong:2017pdm}.


\newpage

\bibliographystyle{JHEP-2}
\bibliography{3dgrav}



\end{document}